\documentclass[twocolumn]{aastex631}
\usepackage{amsmath}

\newcommand{\Wpms}{W\,m$^{-2}$} 
\newcommand{\WpmsK}{W\,m$^{-2}$\,K$^{-1}$}

\begin{document}

\title{Not Earth-like Yet Temperate? More Generic Climate Feedback Configurations Still Allow Temperate Climates in Habitable Zone Exo-Earth Candidates}

\author[0009-0007-4584-4417]{Chaucer Langbert}
\affiliation{Lunar and Planetary Laboratory, The University of Arizona, Tucson, AZ 85721, USA}

\author[0000-0003-3714-5855]{D\'aniel Apai}
\affiliation{Steward Observatory, The University of Arizona, Tucson, AZ 85721, USA}
\affiliation{Lunar and Planetary Laboratory, The University of Arizona, Tucson, AZ 85721, USA}
\affiliation{James C. Wyant College of Optical Sciences, The University of Arizona, AZ 85721, USA}
 
\begin{abstract}
Earth's climate is influenced by over a dozen feedbacks, but only three dominate its long-term climate behavior. Models of the exoplanet habitable zone (HZ) assume that this is similar for other Earth-like planets. We used dynamical simulations to study Earth-like planets with a fourth, (potentially strong) generalized climate feedback. Across over 20,000 climate simulations, we find that the addition of the fourth feedback produces novel behaviors, including runaway and chaotic climate trajectories, that are more diverse than one would expect based on Earth's climate configuration. Non-negligible fourth feedbacks -- if negative -- would not lessen the probability of planets with temperate climates. However, positive fourth feedbacks decrease the fraction of exo-Earth candidates that are long-term habitable. Therefore, strong fourth feedbacks will alter (and mostly shrink) the boundaries of the classical habitable zone. When combined with occurrence rates of Earth-sized planets around sun-like stars, our results imply that the fraction of stars hosting rocky planets with temperate climates may be substantially lower than classical estimates under Earth-like climate assumptions. Our results are subject to the validity of the model assumptions and not intended to represent conclusive predictions about exoplanet populations but rather to demonstrate the potential climate diversity that emerges from non-Earth-like model configurations.  Our conclusions provide context on sample sizes and science questions for next-generation exoplanet surveys. 
\end{abstract}

\keywords{}

\section{Introduction}

Broadly Earth-sized exoplanets show a large diversity in their physical properties (mass, density, orbital parameters, instellation levels). Multiple lines of evidence suggest that they are also chemically diverse: for example, studies of planet-forming disks show carbon-rich inner disks \citep{Pascucci2009,Long2025}; variations in refractory elements in stellar photospheres suggest variations in rocky planet bulk compositions, mineralogy, and geochemistry \citep{Bond2010,Putirka2021}; models of atmospheric photochemistry of super-Earths suggests a broad range of possible compositions \citep{Hu2014}; and early-phase magma oceans may fundamentally impact the thermal evolution and redox state of rocky planets \citep{Dorn2021}. 
Agnostic geochemical modeling showed that -- in addition to carbon-silicate weathering -- a broad range of exchange reactions between reservoirs in the Earth system could modulate carbon cycling and influence climate \citep{Kemeny2024}. 
These and other studies (e.g., refs. in \cite{Jontof-Hutter2019,Lichtenberg2025}) support the idea that habitable zone \citep{2013ApJ...765..131K} Earth-sized planets are likely geochemically diverse and, therefore, will likely have climates dominated by feedbacks cycles that are not identical to those of Earth. 

Paleoclimate records indicate that Earth's climate history was dominated by the operation of numerous interacting feedbacks across geological timescales. Beyond the well-known carbonate–silicate and ice–albedo mechanisms, feedbacks involving water vapor, lapse rate, clouds, and ice sheets have all exerted strong control on radiative balance and surface temperature \citep[e.g.,][]{Forster2021, 2010ppc..book.....P, 2005JGRD..110.1111P}. Biological and geochemical processes further expanded this network of feedbacks over time: oxygenic photosynthesis, organic carbon burial, and biogenic calcification reshaped atmospheric composition and strengthened climate stabilization \citep[e.g.,][]{Lenton2011, Mills2011, Krissansen-Totton2017}. These processes are also well documented across Earth's paleoclimate history, which exhibits repeated transitions among Icehouse, Coolhouse, Warmhouse, and Hothouse states driven by interacting feedbacks in CO$_2$, ice sheets, and radiative balance \citep[e.g.,][]{Lenton2008, Roe2006, Westerhold2020}. Collectively, these physical and biological feedbacks—some positive, others strongly negative—have allowed Earth's climate to remain within habitable bounds despite large perturbations and secular increases in solar luminosity \citep[e.g.,][]{1981JGR....86.9776W, 1993Icar..101..108K, 2015E&PSL.429...20M}.

The long-term stability of the Earth's climate is thought to depend heavily on the carbonate–silicate cycle, which regulates atmospheric CO$_2$ on geological timescales. Rainwater dissolves CO$_2$ to form a weak carbonic acid that weathers silicate rocks, producing bicarbonate and cations that are carried to the ocean. These products are eventually subducted and recycled into the mantle, with volcanic activity returning CO$_2$ to the atmosphere \citep{2010ppc..book.....P}. This negative feedback is strongly coupled to climate: higher surface temperatures or elevated pCO$_2$ enhance weathering and CO$_2$ drawdown, while cooler climates suppress weathering, allowing volcanic outgassing to replenish CO$_2$ \citep{1981JGR....86.9776W, 1997Icar..129..254W, 2000AREPS..28..611K, 2010ppc..book.....P}. This long-term stabilization is also supported by carbon-cycle reconstructions and geochemical modeling that quantify weathering feedback strength and CO$_2$ buffering over Phanerozoic time \citep{Berner2001, Berner1991, Sleep2001, Krissansen-Totton2017}.

In tandem with this stabilizing process, Earth's climate is strongly shaped by the ice–albedo feedback. Cooling increases surface ice cover, raising planetary albedo and further reducing absorbed stellar flux, while warming melts ice and lowers albedo, amplifying temperature change. At sufficiently low instellations or CO$_2$ concentrations, the planet can tip into a globally glaciated ``snowball'' state \citep{1998Sci...281.1342H, 2005JGRD..110.1111P, 2015E&PSL.429...20M}. Classical energy balance models showed that snowball and warm states can coexist at the same stellar flux and CO$_2$ levels, requiring either higher instellation or a strong greenhouse forcing to escape glaciation \citep{1969Tell...21..611B}. The carbonate–silicate cycle couples to this instability: long weathering-driven CO$_2$ drawdown can trigger snowball conditions, while continued volcanic outgassing during glaciations can eventually build up CO$_2$ to force deglaciation, producing oscillatory ``limit cycles" of climate \citep{2015E&PSL.429...20M}. Geological and modeling evidence for such events includes low-latitude glacial deposits linked to Neoproterozoic Snowball Earth episodes and the potential for multiple stable climate equilibria \citep{2007EP&S...59..293T, Kirschvink1992, Hyde2000, Evans2000}. General circulation models confirm that such bistability and limit cycles are dynamically feasible under Earth-like conditions \citep{2005JGRD..110.1111P, 2010QJRMS.136....2L, 2011CliPa...7..249V}.

Several additional processes, such as water vapor, lapse rate, and atmospheric Planck feedbacks, jointly dominate the behavior of outgoing longwave radiation (OLR) in one-dimensional and energy-balance models of Earth. As a result, many well-established formulations for Earth-like atmospheres treat them collectively as an effective OLR feedback, which can be accurately captured using a linearized parameterization \citep{2018PNAS..11510293K, 2020RSPSA.47600303A}. This approach provides a useful approximation for exploring large-scale climate dynamics but necessarily simplifies the underlying complexity and overall temperature sensitivity.

Together, these processes form the canonical three-feedback climate system: outgoing longwave radiation (OLR; stabilizing), ice–albedo (destabilizing), and carbonate–silicate weathering (stabilizing). Climate frameworks commonly focus on the three dominant feedbacks \citep[e.g.,][]{2016ApJ...827..117A, 2018PNAS..11510293K, 2020RSPSA.47600303A, 2021ApJ...912L..14W}. Models that incorporate these three feedbacks can reproduce Earth’s long-term mean climate behavior, including bistability and Snowball Earth episodes, and are broadly consistent with higher-dimensional GCM assessments of feedback strength and sensitivity \citep{2017ApJ...848...33P, 2023JCli...36..547Z, Arnscheidt2022}.

\begin{figure*}[htbp!]
    \centering
    \includegraphics[width=0.9\linewidth]{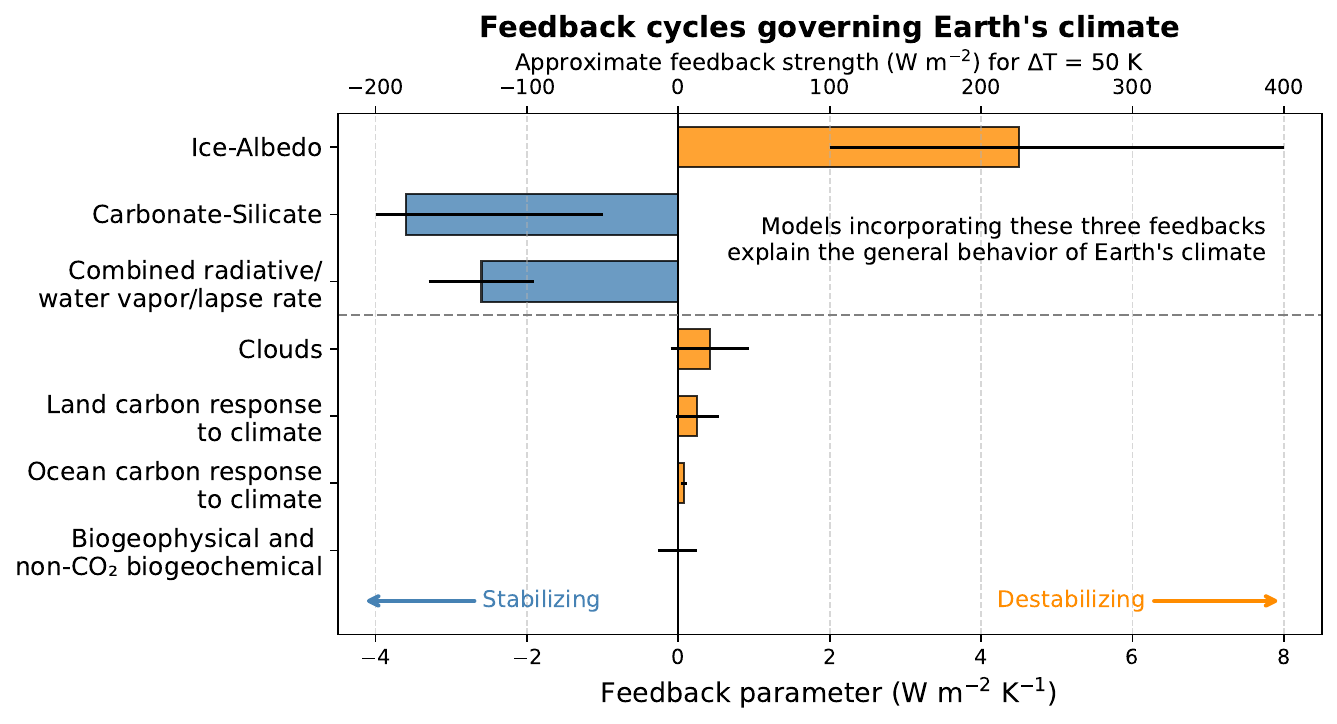}
    \caption{Strength of key long-term feedbacks governing Earth's climate system \citep{Forster2021, 2020RSPSA.47600303A, 2016ApJ...827..117A, 2018PNAS..11510293K}. Each bar represents the feedback parameter in \WpmsK, estimated from Earth system model assessments and paleoclimate constraints. The upper axis shows the approximate feedback strength associated with a 50 K climate perturbation, representative of transitions between major states (e.g., snowball to temperate). Feedbacks are sorted by magnitude. Negative values (blue) represent stabilizing feedbacks that damp temperature changes, while positive values (orange) indicate destabilizing feedbacks that amplify them. Outgoing longwave radiation and the carbonate–silicate cycle are among the strongest stabilizing processes, while the ice–albedo feedback exerts a strong destabilizing influence. Models incorporating the three dominant feedbacks explain the general behavior of Earth's long-term mean climate.}
    \label{fig:feed_schem}
\end{figure*}

This framework rests on the interplay of OLR, ice-albedo, and carbonate-silicate feedbacks (the strongest feedbacks in Figure \ref{fig:feed_schem}) \citep{2016ApJ...827..117A, 2018PNAS..11510293K, 2020RSPSA.47600303A}. Figure \ref{fig:feed_schem} displays the most dominant feedback cycles that affect Earth's long-term climate system and whether they act to stabilize or destabilize the climate. The top three strongest are the most commonly used feedbacks to explain Earth's long term behavior, while the other feedbacks are sufficiently small as to be often neglected in such models.

The Habitable Zone (HZ) concept \citep{2013ApJ...765..131K}, traditionally defined under the assumption that Earth-like climates are regulated by three dominant feedback cycles, has been foundational in shaping the goals of flagship exoplanet missions. Mission concepts such as HabEx and LUVOIR explicitly used the HZ to define science requirements and target selection \citep{Gaudi2020, TheLUVOIRTeam2019}, while the Astro2020 Decadal Survey emphasized the HZ as the central organizing principle for the Habitable Worlds Observatory \citep{NationalAcademiesofSciences2021}. 

However, given the likely very broad geochemical/geophysical  diversity of Earth-sized potentially habitable planets, it is compelling to explore planetary climate evolutions that are shaped by more complex planetary feedback configurations. Without understanding how additional feedbacks alter climate dynamics and habitability outcomes, current models may miss the diversity and stability of surface environments on rocky exoplanets. Approaches based on comprehensive general circulation models (e.g., SAMOSA model intercomparison,  \citealt{Haqq-Misra2022} and references in it) provide powerful insights into the behavior of strategically selected climate configurations. Naturally, due to the high dimensionality of the parameter space and the complexity of the processes involved, the comprehensive GCM-based approach is not readily scalable for sweeping studies of the parameter space.  Our focus here is complementary: by examining how additional long-term feedbacks reshape the structure of climate trajectories and stability regimes, we provide a broader dynamical context within which such focused, comprehensive models can be applied and interpreted.

While previous work has studied the maximally simple three-feedback model with stochastic variability \citep[e.g.,][]{2021ApJ...912L..14W} and run climate simulations of thousands of planets that assigned randomly generated climate feedback configurations on billion-year timescales \citep{Tyrrell2020}, here we narrow our focus on planets presumed to (a) follow known processes like OLR, (b) possess liquid water on their surface, which enables the ice-albedo feedback, (c) possess a stabilizing mechanism (carbonate-silicate weathering) which we presume to operate on Earth-like exoplanets in the HZ, and (d) have another (currently unknown) feedback process operating on its long-term mean climate. 

In our study, we developed a climate framework to explore climate evolution and stability for diverse climate feedback configurations. In particular, we focus on the importance of the strength and sign of a fourth climate feedback on the climate trajectories of broadly Earth-like planets, i.e., what climate diversity is likely represented among habitable zone (HZ) Earth-sized exoplanets.

\section{A dynamical model for multi-feedback climates}

We constructed a minimal, zero-dimensional dynamical system to explore how extending Earth-motivated climate models to include a fourth, general feedback term changes the prevalence of long-term climate states and the persistence of habitable surface conditions (defined here as the planetary surface being capable of supporting currently known, metabolically active bacterial life, discussed further in Section~\ref{subsec:hab}). The model is intentionally simple and general, focusing on the three dominant feedbacks while excluding known secondary or state-dependent processes on Earth whose roles in Earth-like climates remain uncertain or likely small in their global-mean influence. We aim to study general model behaviors rather than to make precise predictions, so we follow previous studies \citep{2016ApJ...827..117A, 2018PNAS..11510293K, 2020RSPSA.47600303A} and linearize the combined OLR feedback around a stable warm state with albedo and weathering as smoothly varying functions of temperature (for more details, see the Appendix). While different atmospheric compositions or water inventories could qualitatively shift OLR, weathering, and ice-albedo feedbacks, they are unlikely to alter the qualitative effects of introducing a fourth feedback.

To isolate the dynamical effects of introducing an additional feedback, we adopt a simplified Earth-like climate baseline rather than an exact Earth analog. Specifically, we assume circular orbits ($e=0$), Earth-like obliquity and rotation rate, and a present-day ocean inventory. Setting $e=0$ suppresses seasonal and insolation variability, allowing the effects of the additional climate feedback to be examined without conflating them with orbital forcing, while retaining the operation of ice–albedo and silicate weathering feedbacks. This configuration reflects standard assumptions used in low-order climate feedback models \citep[e.g.,][]{1981JGR....86.9776W, 2016ApJ...827..117A, 2018PNAS..11510293K, 2021ApJ...912L..14W}. Although limit cycles and bistability can depend on orbital and rotational parameters \citep[e.g.,][]{Lucarini2013, Linsenmeier2015, Checlair2017, Checlair2019}, our goal here is to understand how the addition of a fourth feedback modifies long-term climate trajectories under otherwise simplified Earth-like conditions. Exploring non-Earth orbital configurations, such as high obliquity, eccentricity, or asynchronous rotation, represents an important future extension of this framework.

Our model evolves three state variables: surface temperature $T$, atmospheric CO$_2$ partial pressure $P$, and a fourth, generalized feedback term $f$, in response to stellar and volcanic forcings, albedo changes, radiative losses, and geochemical cycling. The climate is governed by three coupled differential equations (Equations \ref{model1}-\ref{model3}).
\begin{figure*}[ht!]
\begin{align}
    C \dot{T} &= \frac{S(t)}{4}(1 - \alpha(T)) - \frac{S_0}{4}(1 - \alpha_0) - a(T - T_0) + b \ln\left(\frac{P}{P_0}\right) + c f \label{model1} \\
    \dot{P} &= V - W(T) e^{k(T - T_0)} \left(\frac{P}{P_0}\right)^{\beta} \label{model2} \\
    \dot{f} &= -\gamma_f \left[ f - \kappa \tanh(\delta_f(T - T_f)) \right] \label{model3}
\end{align}
\end{figure*}

$T$ is the average planetary surface temperature $(\text{K})$, with $\dot{T}$ its time derivative, and $C$ the effective surface–atmosphere heat capacity $(\text{J m}^{-2}\,\text{K}^{-1})$. $S$ is the stellar flux at the top of the atmosphere and $S_0$ a reference solar constant $(\text{W m}^{-2})$; $\alpha(T)$ and $\alpha_0$ are the temperature-dependent and reference albedos (dimensionless). The OLR is linearized around a reference warm state with coefficients $a$ $(\text{W m}^{-2}\,\text{K}^{-1})$ for temperature sensitivity and $b$ $(\text{W m}^{-2})$ for logarithmic CO$_2$ sensitivity. $P$ is the atmospheric CO$_2$ partial pressure with reference value $P_0$ $(\text{bar})$. Volcanic outgassing enters through $V$ $(\text{bar Gyr}^{-1})$, while silicate weathering is parameterized by a baseline rate $W(T)$ $(\text{bar Gyr}^{-1})$, an exponential temperature sensitivity $k$ $(\text{K}^{-1})$, and a CO$_2$ dependence exponent $\beta$ (dimensionless). 

The generalized fourth feedback is represented by a dimensionless state variable $f$, bounded between $-1$ and $+1$, scaled by an amplitude $c$ $(\text{W m}^{-2})$ in the energy balance. $f$ captures hypothetical processes, such as long-term biospheric or geochemical regulation, not included in current models but potentially relevant for planetary habitability. It has tunable strength $c$, relaxation rate $\gamma_f$, and activation temperature $T_f$, and is implemented as a smooth sigmoidal function that activates near $T_f$, with steepness set by $\delta_f$ and saturation at $\pm c$. Its temporal evolution follows a first-order relaxation with timescale $\gamma_f^{-1}$. A sigmoidal form is appropriate because many Earth-system couplings are thresholded, where they are weak away from their main zone of activation but strengthen as the system nears biologically or chemically favorable ranges, all while remaining bounded to avoid unphysical growth. This choice captures thresholded yet smooth responses seen in idealized climate models and avoids discontinuities, so it is both physically motivated and an agnostic regulating process. We implemented this climate model in a Python-based numerical integration framework (Eqs. \ref{model1}–\ref{model3}), solved using a backward differentiation formula (BDF) scheme. 

Expanding on \cite{2020RSPSA.47600303A}, we adopt expanded model bounds (T = 200–400 K; pCO$_2$ = 0–1000 bar) to allow trajectories to allow the system to explore transient excursions, bistability, and instability-driven transitions without being limited by edge effects. We expanded the ranges from \cite{2020RSPSA.47600303A} by 10-15\% without invoking specific chemical or physical limits to allow study of these behaviors. At extreme temperatures or pressures, additional processes (e.g., phase changes, atmospheric escape, or changes in dominant feedbacks) would require more complex modeling.

All simulations were performed on the University of Arizona High Performance Computing cluster using 94 CPU cores in parallel. The total computational cost was approximately 50,000 CPU hours ($\sim$1,000 node-hours). We leveraged multiprocessing to scale out our work, enabling thousands of simulations to be run simultaneously and drastically reducing computation time. This approach allowed us to comprehensively sample our high-dimensional parameter space while ensuring robust and reproducible results. Results were processed and visualized using a combination of custom scripts and pre-existing libraries in the Python programming ecosystem \citep{2013A&A...558A..33A}, enabling robust identification of attractors and long-term climate stability metrics.

\section{Results}

\subsection{Three-feedback model}
\label{sec: rothman repro}

\begin{figure*}[htbp!]
    \centering
    \includegraphics[width=\linewidth]{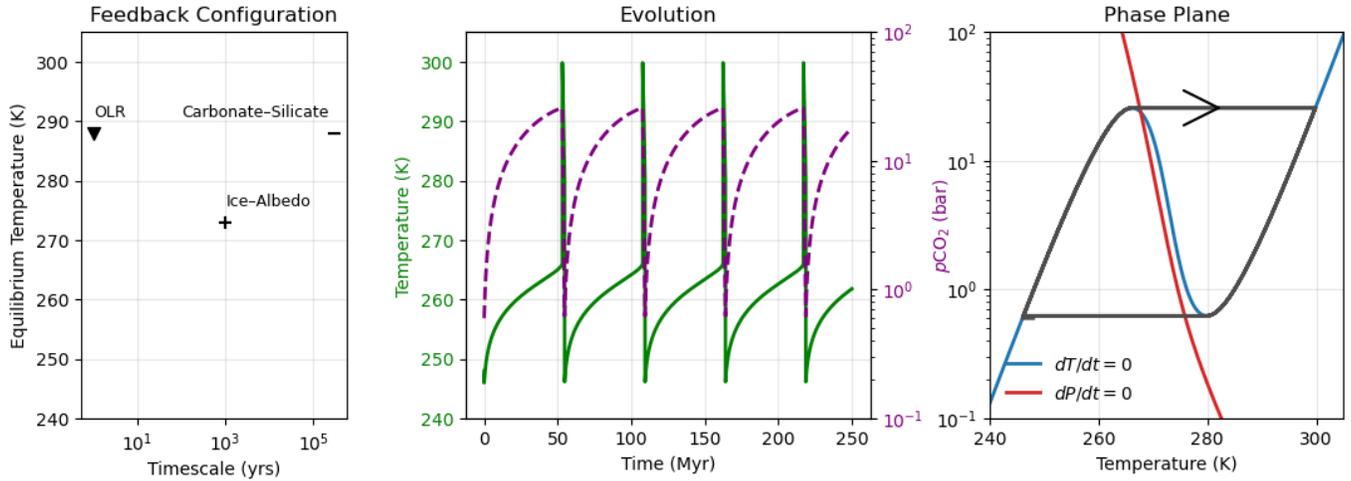}
    \caption{Dynamics of the planetary climate system with coupled ice-albedo and carbonate-silicate feedback mechanisms without a fourth feedback, consistent with \citet{2020RSPSA.47600303A}. Left: Feedback configuration showing the relationship between equilibrium temperature (K) and the timescale (years) for the three canonical feedbacks. Middle: T (green solid) and pCO$_2$ (purple dashed) time series. Right: Phase plane illustrating the trajectory of the system (black) in the T-pCO$_2$ phase space. The blue curve represents the temperature nullcline, while the red curve corresponds to the pCO$_2$ nullcline. The intersection of these nullclines denotes potential steady states, but with this particular configuration of $S = 1000$ W/m$^{2}$ and $V = 10 W_0$ we observe a limit cycle.}
    \label{fig:icealb_carbsil_null}
\end{figure*}

To verify our model, we first compare with previous results in the literature. The dynamical system described by \citet{2020RSPSA.47600303A} (in the absence of additional feedback cycles) has been shown to exhibit both steady states and limit cycles. In our model (equations \ref{model1}–\ref{model3}) with the feedback coefficient set to \( c=0 \) W/m$^{2}$, we recover these same qualitative behaviors (e.g., Figure \ref{fig:icealb_carbsil_null}). In the left column of Figure \ref{fig:icealb_carbsil_null} (``Feedback Configuration''), each marker's equilibrium temperature is the steady-state temperature of the feedback process—set to 288 K for OLR and carbonate–silicate) and 273 K for the water phase transition in the ice–albedo feedback. The timescales are prescribed from reference estimates (1 yr for OLR, $10^3$ yr for ice–albedo, and $2.4\times10^5$ yr for carbonate–silicate).

A typical phase-plane trajectory in the (T, pCO$_2$) phase plane for this parameter configuration (right panel of Figure \ref{fig:icealb_carbsil_null}) rapidly converges to the blue cubic temperature nullcline. It then slowly crawls along the nullcline until it reaches a knee, at which point it rapidly jumps to another branch. The resulting limit cycle between warm and glaciated states exhibits two widely separated timescales: a slow build-up followed by a fast transition.

These results demonstrate consistency with prior results. This agreement provides a validation check for our system in the baseline case, confirming that it captures the known dynamics of ice-albedo and carbonate-silicate cycling. Although minor numerical differences may exist due to model discretization and parameter choices, the overall dynamical structure remains unchanged. Our overall approach builds on the energy balance framework developed by \cite{2020RSPSA.47600303A}, which has been extensively validated against higher-order climate models and paleoclimate constraints \citep[e.g.,][]{2016ApJ...827..117A, 2015E&PSL.429...20M, North1983, 1969Tell...21..611B, 1969JApMe...8..392S}. This well-tested formulation provides a robust foundation for exploring nonlinear and chaotic behavior in Earth-like climate systems.

\subsection{Emergent climate behaviors of four-feedback model with fixed stellar luminosity} \label{sec:new behaviors}

We then systematically varied the strength and sign of the fourth climate feedback across more than 10,000 simulations, each integrated over 5 Gyr. This integration timescale was chosen to capture long-term climate behavior relevant for planetary habitability and to allow even the slowest feedbacks to reach equilibrium or exhibit their characteristic dynamics. Initial conditions for temperature ($T$), atmospheric pCO$_2$ ($P$), and feedback states ($f$) were randomized within physically plausible ranges to explore diverse climate trajectories. Importantly for the first 10,000 simulations, we neglected the role of stellar evolution where the star brightens by $\sim30$\% over five Gyr. Previous studies argue that various climate-relevant factors can change over geological time, and some of them could amplify or even outweigh the effects of stellar brightening, leading to a net change in the opposite direction and cooling the planet over time \citep{Tyrrell2020}. Examples include gradual increases in continental area over time that produce an increase in albedo, declining geothermal heating due to the decay of radioactive elements in the planetary interior, and progressive leakage out to space of low molecular mass gases which facilitates oxygenation of the atmosphere and removal of some greenhouse gases \citep[][and references within]{Tyrrell2020}. However, to create a more physically grounded model based on current conceptions of the HZ and to address the problem of the `Faint Young Sun Paradox,' we run an additional 10,000 simulations which incorporate this stellar forcing (Section \ref{sec:SE}).

\begin{figure*}[htbp!]
    \centering
    \includegraphics[width=\linewidth]{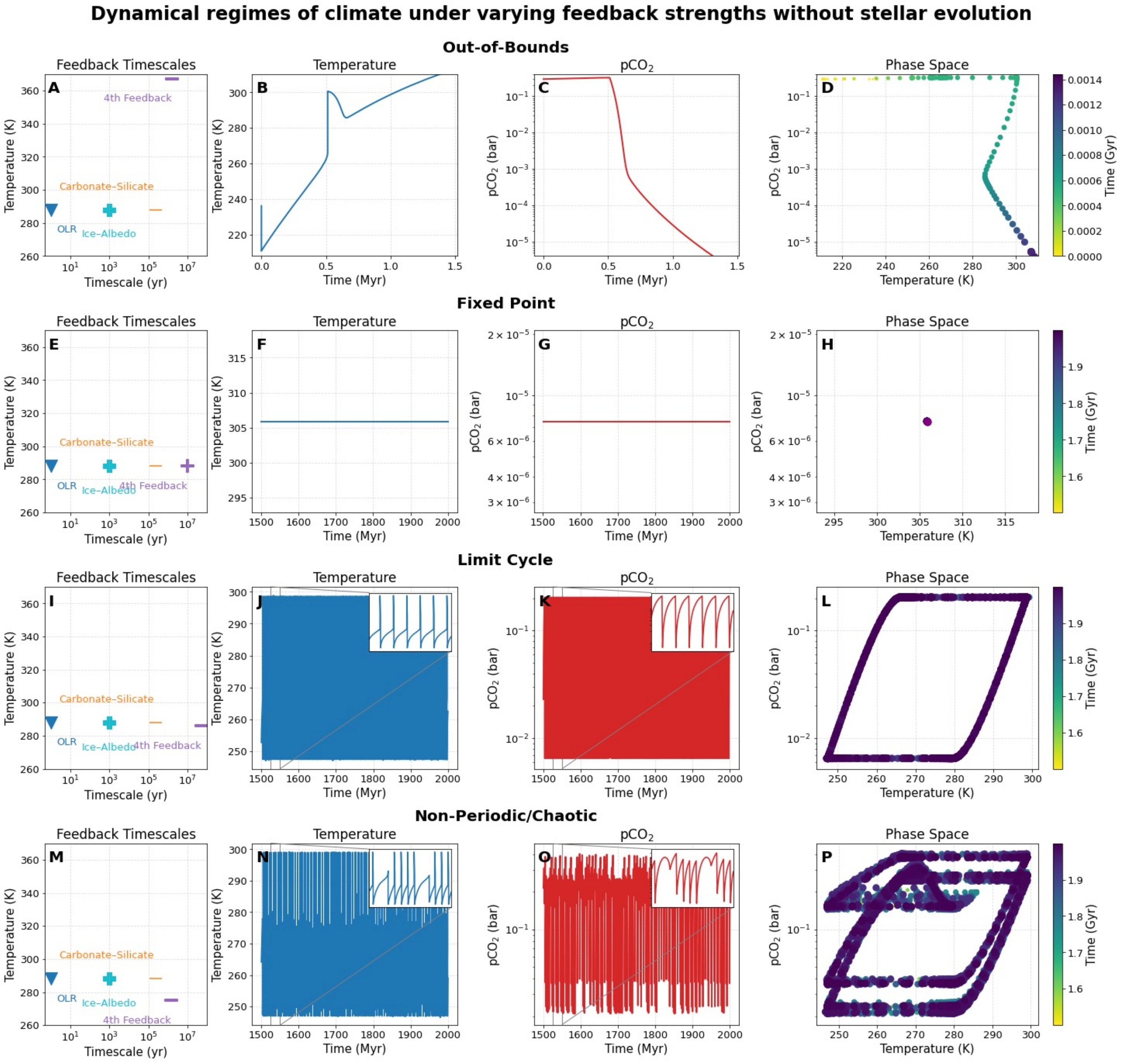}
    \caption{Representative examples of climate trajectory behaviors in a four-feedback Earth-like model, consistent with possible states of Earth but excluding the role of stellar evolution. Each row shows a different behavior: Out-of-Bounds, Fixed Point, Limit Cycle, and Non-Periodic/Chaotic. Left column: Feedback timescales and effective temperatures, with the fourth feedback shown in purple. Middle columns: Time evolution of temperature (blue) and pCO$_2$ (red, log scale) with zoomed-in series insets. Right column: Phase space plots of pCO$_2$ vs. temperature. Out-of-bounds runs leave the temperate regime before the end of the simulation; fixed points collapse to a single state; limit cycles trace loops; non-periodic/chaotic states show irregular structure.}
    \label{fig:model_exs}
\end{figure*}

Key physical parameters, such as stellar flux ($S$) and volcanic outgassing rate ($V$), were sampled from distributions tailored to represent plausible variations in instellation received at the top of the atmosphere and planetary geological activity, respectively. The ranges of parameters including the volcanic outgassing rate $V$ and the maximum amplitude of the generalized fourth feedback $c$ are intentionally broad because our objective is not to prescribe precise values, but rather to explore the range of qualitatively distinct climate behaviors that arise under plausible but uncertain process. For volcanic outgassing, published estimates for modern Earth span $\sim1-10$ bar Gyr$^{-1}$ \citep{Foley2016}, and geological constraints on Earth's early degassing rates extend even higher (tens of bar Gyr$^{-1}$) \citep{Muller2024}. Given this order-of-magnitude uncertainty, we adopt a wide exploratory range for $V$, allowing us to capture both Earth-like and non-Earth-like tectonic regimes. The system also incorporated smooth transition functions for albedo ($\alpha(T)$) and weathering ($W(T)$), ensuring continuity and avoiding numerical instabilities at critical thresholds. For further discussion of randomized parameters, see the Appendix.

For the maximum amplitude of the fourth feedback $c$, we explored a deliberately broad range of values ($|c| \leq 100$ \Wpms), which in our model corresponds to an effective feedback parameter of $\sim5-20$ \WpmsK. This is substantially stronger than Earth's known feedbacks, and is therefore not meant to represent physically constrained values from the Earth system. Rather, the intent is to bracket an upper bound on possible behaviors and to test whether our qualitative conclusions depend on the assumed feedback strength. 

In the initial case without stellar evolution, each simulation was automatically classified (see Appendix) into one of six categories based on its long-term climate trajectory: warm or cold fixed points (converging to a snowball or temperate stable equilibrium), limit cycles (bounded, periodic oscillations), non-periodic or chaotic states (irregular yet bounded oscillations), transient chaos (chaos sustained for fewer than 2.5 Gyr), or globally uninhabitable outcomes termed ``out-of-bounds'' (e.g., runaway greenhouse or snowball trajectories that exit the model validity regime). Out-of-bounds trajectories occur when either the temperature or pCO$_2$ values leave the validity range of the model, so we conservatively end the simulation and assume the climate remains uninhabitable for the rest of its evolution. We verified the robustness of this classification scheme by comparing automated labels with visual inspection of representative trajectories across parameter space. Figure~\ref{fig:model_exs} shows illustrative examples of each behavior (other than transient chaos, which is shown in Figure~\ref{fig:se_model_exs}). For each regime, we present (leftmost panels) the relative timescales and equilibrium temperatures of the contributing feedbacks, (second and third columns) the evolution of surface temperature and atmospheric pCO\textsubscript{2}, and (rightmost column) the corresponding phase space trajectories. These categories reflect qualitatively distinct dynamical regimes that can inform the interpretation of exoplanet observations and constrain expectations for Earth-like climate stability. Here, ``non-periodic/chaotic'' denotes trajectories that do not converge to a closed, time-invariant orbit and exhibit long-term sensitivity to initial conditions.

\begin{figure*}[htbp!]
    \centering
    \includegraphics[width=\linewidth]{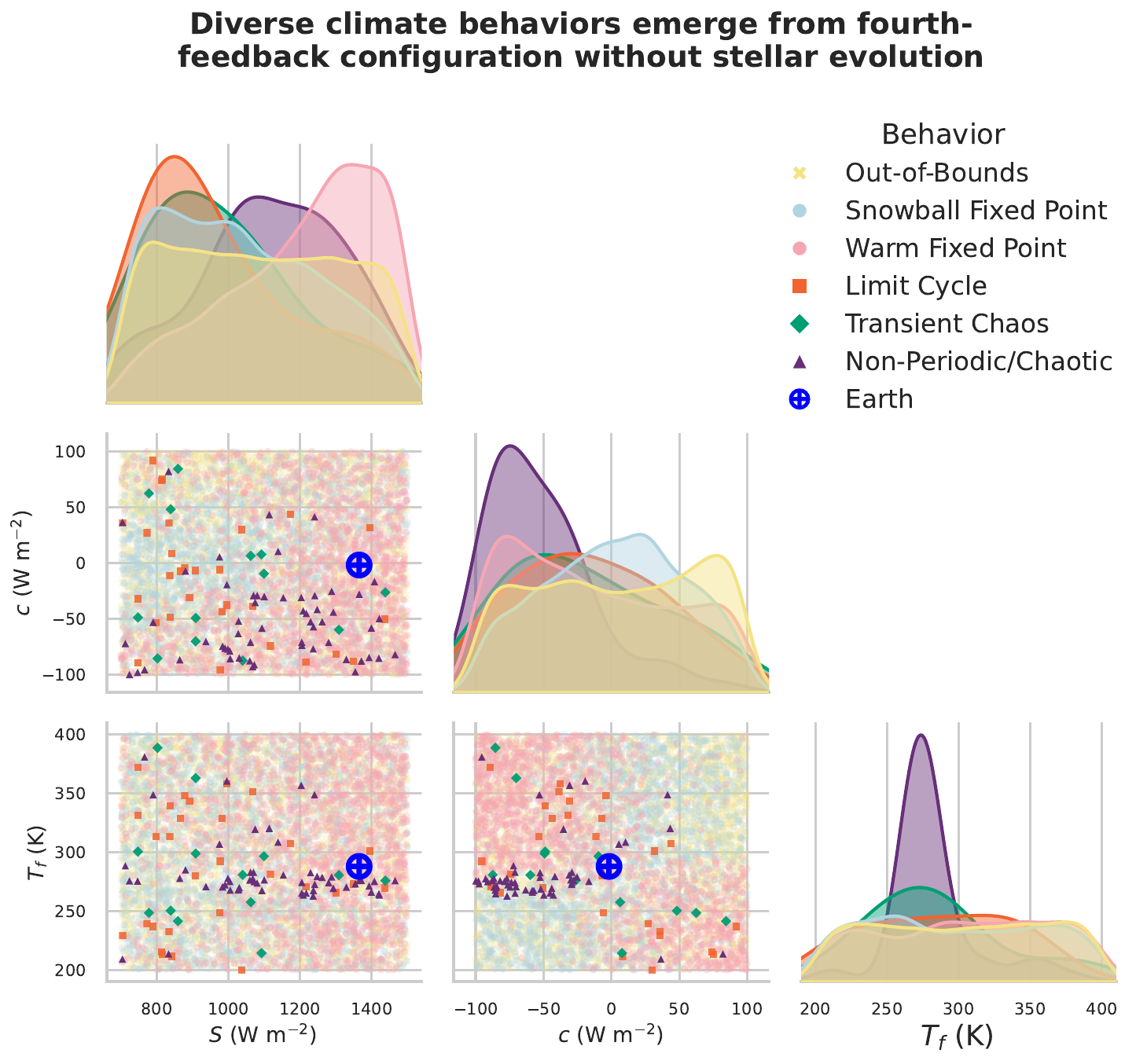}
    \caption{Climate behaviors emerge in distinct regions of planetary parameter space from a generalized Earth-motivated climate model that excludes stellar evolution. Each point represents the outcome of a single climate simulation sampled across stellar flux ($S$), fourth-feedback strength ($c$), and the feedback’s equilibrium temperature ($T_f$). Colors and marker shapes denote long-term climate behaviors, including warm and snowball fixed points, limit cycles, non-periodic/chaotic states, transient chaos, and out-of-bounds trajectories. The diagonal panels show the one-dimensional distributions of outcomes along each parameter, while the off-diagonal panels display pairwise projections of the sampled parameter space. Earth’s present-day parameters are shown for reference. Dynamical behaviors cluster cleanly within the parameter space, highlighting the substantial diversity of climate states produced by the four-feedback configuration.}
    \label{fig:behav_clusters}
\end{figure*}

The equilibrium temperature of each feedback is the steady-state temperature of the feedback process set at 288 K for OLR and carbonate–silicate, 273 K for the water phase transition in the ice–albedo feedback, and at the equilibrium temperature $T_f$ defined by the fourth-feedback parameterization. The timescales are either prescribed from reference estimates (1 yr for OLR, $10^3$ yr for ice–albedo, $2.4\times10^5$ yr for carbonate–silicate) or calculated directly as the inverse of the feedback decay rate, $\tau = 1/\gamma_f$, for the fourth feedback, consistent with its dynamical equation $df/dt = -\gamma_f \left[ f - \kappa \tanh(\delta_f(T - T_f)) \right]$.

The complete set of long-term climate behaviors for the first 10,000 simulations excluding stellar evolution is shown in Figure \ref{fig:behav_clusters}, where we can observe distinct clusters of climate behaviors in the parameter space. Out-of-bounds/runaway points dominated, comprising over half of simulations. A wide range of $c$ values, especially when $c \gg 0$ \Wpms, often triggered runaway states, consistent with the general expectation that sufficiently strong feedbacks can destabilize temperate climates~\citep{2016ApJ...827..117A, 2016ApJ...827..120H}. In contrast, more complex and long-term chaotic behaviors clustered in narrow regions of parameter space, particularly near moderately negative values of fourth-feedback strength ($c<0$) and activation temperatures of $T_f \sim 273$ K, where feedbacks interact to destabilize the climate (Fig. \ref{fig:behav_clusters}). We further observe transient chaos occurs within a similar but broader temperature range, as well as at lower instellations. This structure indicates that while chaotic or oscillatory dynamics are less probable, they emerge robustly in a distinct subset of feedback and system parameter configurations. The full landscape of climate behaviors reflects the interplay of multiple nonlinearities and highlights the potential for complex variability in Earth-like exoplanets.

\subsection{Stellar evolution expands climate diversity and maintains stability} \label{sec:SE}

\begin{figure*}[htbp!]
    \centering
    \includegraphics[width=\linewidth]{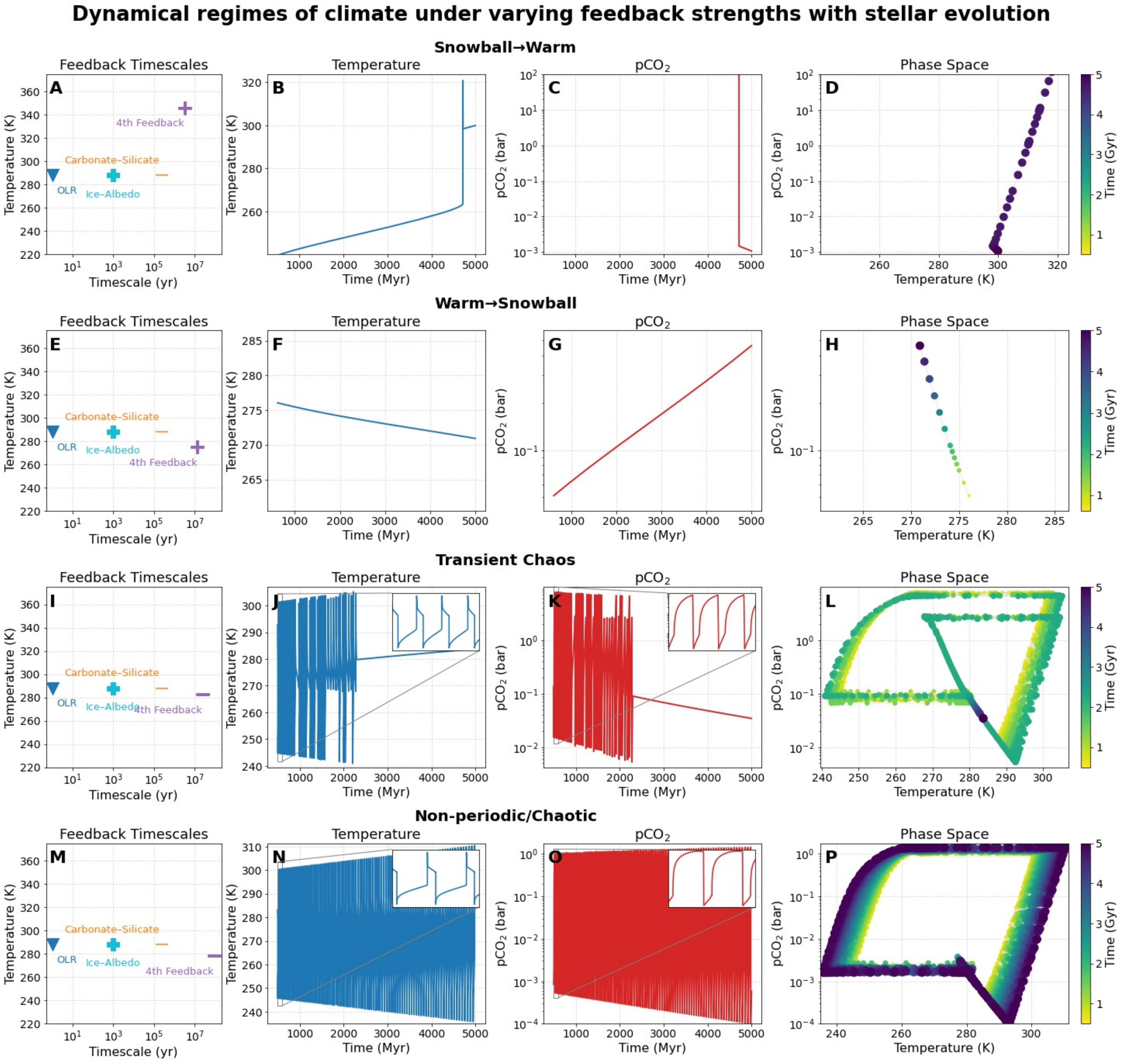}
    \caption{Representative examples of climate trajectory behaviors in a four-feedback Earth-like model under increasing stellar luminosity. Each row shows a different behavior: Snowball$\rightarrow$Warm transition, Warm$\rightarrow$Snowball transition, Transient Chaos, and Non-Periodic/Chaotic. Left column: Feedback timescales and equilibrium temperatures, with the fourth feedback shown in purple. Middle columns: Time evolution of temperature (blue) and pCO$_2$ (red, log scale) over the stellar evolution window, with insets highlighting oscillatory structure where present. Right column: Phase space plots of pCO$_2$ vs.\ temperature, colored by time. Snowball$\rightarrow$Warm runs cross from glaciated to temperate states; Warm$\rightarrow$Snowball runs collapse into permanent glaciation; transiently chaotic runs exhibit irregular excursions before settling; non-periodic/chaotic runs maintain sustained irregular variability.}
    \label{fig:se_model_exs}
\end{figure*}

We incorporated stellar evolution by following previous calculations that suggested that the early Sun was 25-30\% less luminous than it is today, and its luminosity has increased with time in an approximately linear manner \citep{Kasting1987}. Thus, for this second set of simulations, we introduce a linear equation $S(t)$: $S(t) = S_{\text{init}} * \left(1 + 0.3 * \left(\frac{t}{5\text{ Gyr}}\right)\right)$. We kept our randomization bounds the same as the previous 10,000 simulations without stellar evolution. 

The introduction of stellar evolution led to a slightly expanded set of qualitative behaviors (Figure~\ref{fig:se_model_exs}): (1) general temperature increases as the star brightens which (a) remain in a Snowball, (b) transition from a Snowball to Warm state, then continue to increase in temperature, or (c) remain in a Warm state; (2) get pushed out-of-bounds; (3) remain in non-periodic/chaotic trajectories; or (4) transition between non-periodic/chaotic and fixed point behaviors, which we will refer to hereafter as transient chaos if the chaotic behavior lasted for fewer than 2.5 Gyr. 

\begin{figure*}[htbp!]
    \centering
    \includegraphics[width=\linewidth]{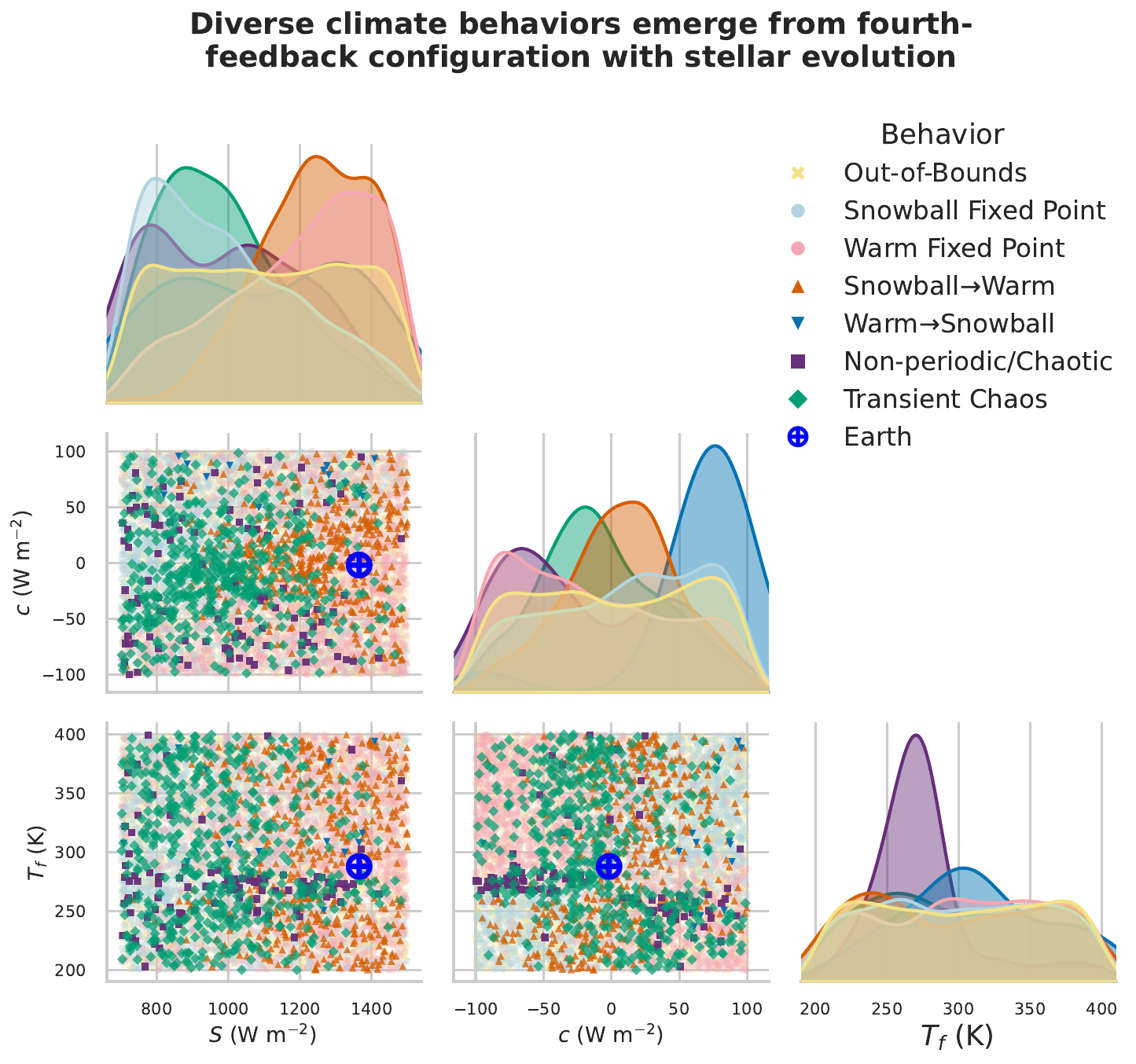}
    \caption{Diverse climate behaviors emerge in a four-feedback Earth-derived model that includes stellar evolution. Each point represents the outcome of a single climate simulation sampled across stellar flux ($S$), fourth-feedback strength ($c$), and the feedback’s equilibrium temperature ($T_f$). Colors and marker shapes denote long-term behaviors, including warm and snowball fixed points, snowball$\rightarrow$warm and warm$\rightarrow$snowball transitions, sustained non-periodic/chaotic states, transient chaos, and out-of-bounds trajectories. The diagonal panels show one-dimensional distributions of simulation outcomes along each parameter, while the off-diagonal panels display pairwise projections of the sampled parameter space. Earth’s present-day parameters are shown for reference. The inclusion of stellar evolution broadens the accessible climate trajectories and enables rare transitionary and chaotic regimes, which cluster in characteristic regions of the parameter space.}
    \label{fig:behav_clusters_SE}
\end{figure*}

The regime changes in the transitions between Snowball fixed points, warm fixed points, and non-periodic/chaotic behaviors reflect the underlying bifurcation structure of the climate system: increasing instellation erodes the basin of attraction until a fixed point loses stability and the climate is diverted to a new regime. Similar stability-loss and fold bifurcations have been identified in low-dimensional climate models \citep[e.g.,][]{2020RSPSA.47600303A, 2015E&PSL.429...20M, 1993Icar..101..108K}, and estimates of Earth-like feedback strengths \citep[e.g.,][]{Forster2021, Krissansen-Totton2017} indicate that relatively small forcing changes can trigger such tipping behavior.

The distribution of long-term climate behaviors when stellar evolution is included is shown in Figure \ref{fig:behav_clusters_SE} and can be studied in more detail in the Appendix. Compared to the fixed-luminosity case, the addition of stellar brightening reshapes the distributions of climate behaviors by reducing the dominance of out-of-bounds states and increasing the occurrence of transitionary and chaotic behaviors. In particular, we see new clusters of trajectories that undergo snowball to warm or rarer warm to snowball transitions, as well as more common instances of transient and sustained chaos. These chaotic behaviors emerge most clearly at intermediate values of the fourth-feedback strength $c$ and activation temperature $T_f \sim 273$ K, where the gradual increase in stellar flux interacts with internal feedbacks to destabilize the climate. At large $|c|$, stellar evolution continues to drive systems into runaway states. The resulting structure underscores that stellar evolution is a critical external driver that can both suppress stability and enable escape from chaotic regimes, expanding the diversity of long-term outcomes for Earth-like exoplanets.

\subsection{Complex climate evolution and limited habitability} \label{subsec:hab}

\begin{figure*}[htbp!]
    \centering
    \includegraphics[width=\textwidth]{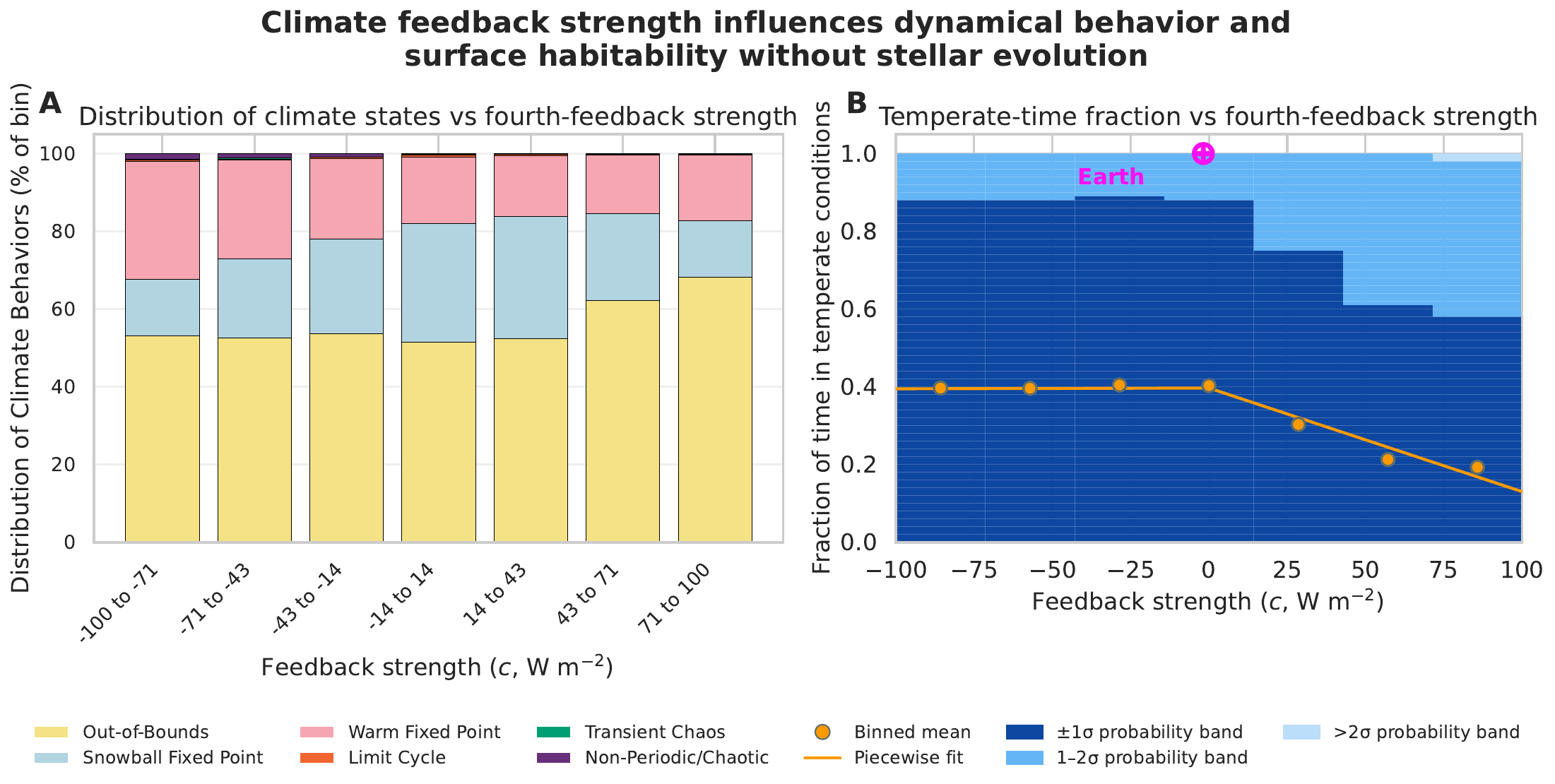} 
    \caption{Climate behavior and time spent in habitable surface conditions as a function of the fourth feedback strength $c$ (W/m$^2$) without stellar evolution. (\textbf{A}) Distribution of climate behaviors as a function of $c$, separated into 10 bins. (\textbf{B}) Average habitable fraction as a function of $c$ sorted into 10 bins, with our two-component linear fit overplotted. We conservatively assume that simulations that go out-of-bounds of the model remain there for the rest of the simulation. We show the data and fit (orange) as well as 1- and 2-$\sigma$ probability distributions in dark and light blue, respectively, with Earth for reference (pink). Climate behavior and time spent in temperate surface conditions are strongly dependent on the fourth-feedback strength. Earth's feedback configuration (weak fourth feedback) places it near the maximum probability for a long-term stable and habitable surface.}
    \label{fig:hist_c_hab} 
\end{figure*}

We now examine the role that this climate behavior diversity plays in the amount of time spent in temperate surface conditions suitable for known microbial life on Earth. We quantified microbial surface habitability as the fraction of time surface temperature remains within 253 to 395 K (-20 to 122$\,^\circ$C), the temperature range in which currently known bacterial Earth life is viable~\citep{doi:10.1128/AEM.70.1.550-557.2004, 2003AsBio...3..331G, 2008AGUFM.B51F..05T,Apai2025}. This threshold refers specifically to the temperature range for metabolically active bacterial life, rather than habitability in the broadest biological sense.We find that across our fixed-luminosity simulation ensemble, planets with weak to moderate negative fourth feedbacks ($c \lesssim 0$ \Wpms) were more likely to maintain temperate climates for a larger fraction of geologic time (Fig.~\ref{fig:hist_c_hab}). As $c$ increases above zero, the system becomes increasingly likely to be unstable, with trajectories exhibiting either prolonged glaciation or runaway warming. We found that the behavior is well-fit by a two-component linear trend:
\begin{equation}
f_{\text{hab}}(c) = \begin{cases}
0.4478 + 0.0005\,c, & \text{if } c < 0, \\
0.4478 - 0.0029\,c, & \text{if } c \ge 0.
\end{cases}
\end{equation}

The much shallower negative slope for $c \geq 0$ (-0.0005 versus +0.0029) indicates that amplifying (positive) feedbacks erode habitable conditions more rapidly than an equivalent-strength damping (negative) feedback, so strong positive feedbacks are disproportionately harmful to long‐term habitability. Maximal habitability occurs when $c \lesssim 0$, suggesting that long-term surface habitability is optimized by having no fourth feedback or a negative fourth feedback. These findings point to a narrowed window of additional feedback strengths compatible with sustained temperate conditions on rocky planets.

\subsection{The role of stellar evolution in long-term habitability}

\begin{figure*}[htbp!]
    \centering
    \includegraphics[width=\linewidth]{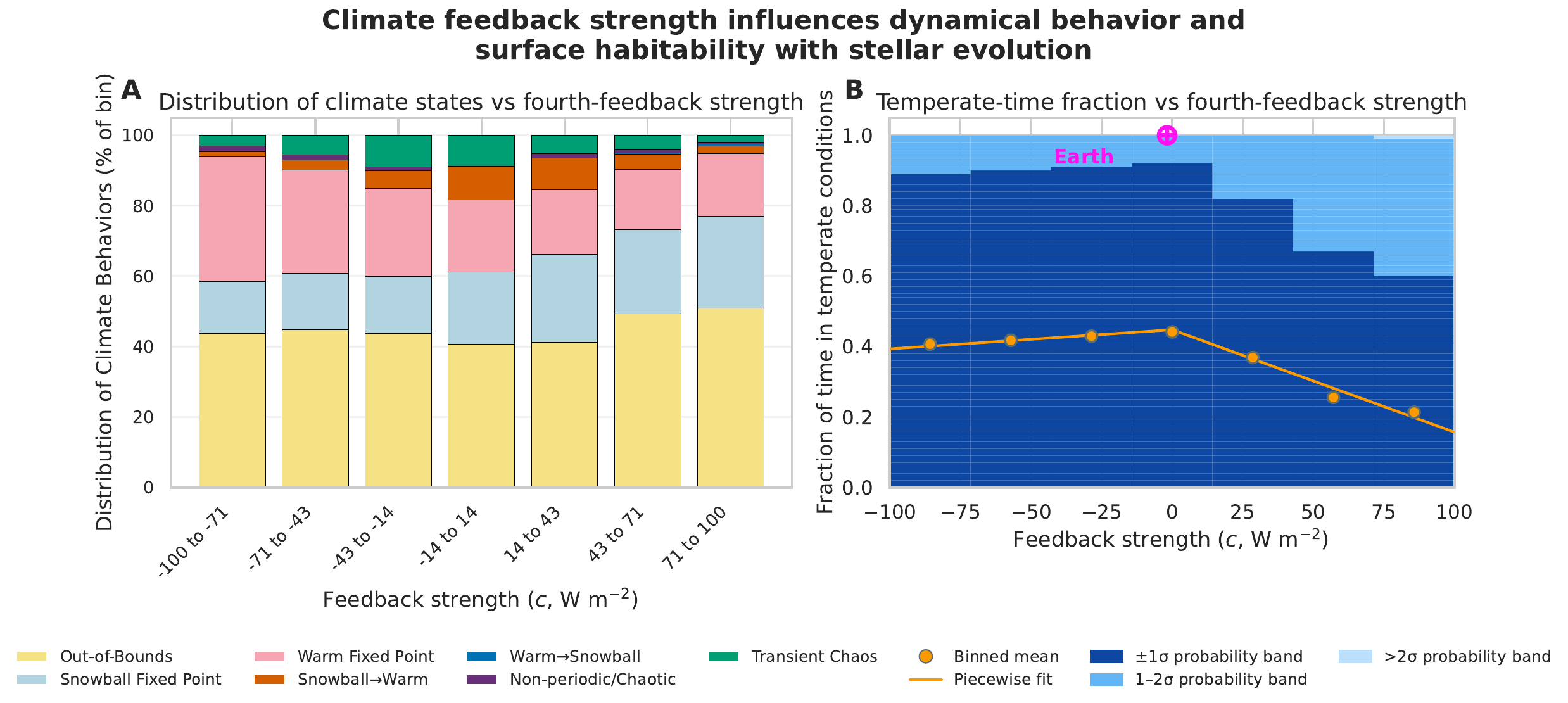}
    \caption{Climate behavior and time spent in habitable surface conditions as a function of the fourth feedback strength $c$ (W/m$^2$), now incorporating stellar evolution. (\textbf{A}) Relative occurrence of climate behaviors binned by fourth-feedback strength $c$. Stable fixed points dominate across most bins, but transitions (Snowball→Warm, Warm→Snowball) and rare chaotic states appear near intermediate $c$. Out-of-bounds trajectories (gray) are more common at large $|c|$. (\textbf{B}) Fraction of time spent in temperate conditions as a function of $c$. Orange circles mark binned means, with a linear fit in orange and probability bands ($\pm1\sigma$, $1$–$2\sigma$, $>2\sigma$) in blue shading. Earth’s present-day location is marked in magenta. Strong negative feedbacks increast the likelihood of stable climates, whereas strong positive feedbacks greatly decrease the fraction of exo-Earth candidates that are long-term habitable.}
    \label{fig:SE_hab}
\end{figure*}

When we account for the role of stellar evolution in our model, we find that the fraction of time that each planet spends in temperate surface conditions is fit by a similar two-component linear trend:
\begin{equation}
f_{\text{hab, SE}}(c) = \begin{cases}
    0.3964 + 0.0002\,c, & \text{if } c < 0, \\
    0.3964 - 0.0027\,c, & \text{if } c \ge 0.
\end{cases}
\end{equation}

This trend suggests that stellar brightening, combined with a strong stabilizing feedback, lead to slightly lower likelihoods of sustained habitable conditions, yet are more likely to support warm fixed points. On the other hand, the presence of a strong positive fourth feedback would greatly decrease the fraction of exo-Earth candidates that are habitable on billion-year timescales. Similarly to our previous fixed-luminosity findings, Earth-like (warm fixed point) climate trajectories may be more common when given a negative fourth feedback, but become increasingly rare with stronger positive fourth feedbacks. 

\subsection{Earth in the climate parameter space}

It is instructive to compare Earth with the distribution of planetary climates our model predicts. To plot Earth, we adopted the strength of the cloud feedback as its estimated strongest fourth feedback (after the ice-albedo, carbon-silicate, and the combined OLR feedback). The cloud feedback is estimated to have a typical strength around $+0.5$ to $+1.5$ \WpmsK, but is very uncertain and has values potentially ranging from slightly negative to positive depending on the model and cloud type \citep{Forster2021, 1997BAMS...78..197K, 2006JCli...19.3354S}. The pink Earth symbol in Figure \ref{fig:hist_c_hab}B shows that Earth is very near to the peak of the habitable fraction curve, i.e., its physical climate system falls close to the peak of long-term temperate conditions. This is, of course, consistent with Earth's actual climate behavior. 

We note, however, that our framework omits biogeochemical and biospheric feedbacks, which are known to play major roles in regulating Earth's present-day climate. These processes are highly complex, difficult to parameterize within a zero-dimensional model, and largely unconstrained for other Earth-like exoplanets. Consequently, we treat Earth here as a simplified physical reference point, rather than as a literal comparison to the modern biosphere-influenced Earth system.

There are no a priori reasons to believe that a fourth feedback on broadly Earth-like planets must be either absent or universally weak. It is therefore reasonable to expect that such feedbacks often have non-zero strengths—potentially weak, but not negligible. Even modest additional feedbacks can alter long-term climate stability, and our results show that the presence of a non-zero fourth feedback frequently shifts systems away from Earth-like equilibria. In this sense, Earth's current climate (with biological feedbacks omitted) may represent an unusually stable configuration among the broader diversity of possible temperate worlds.

In the broader dynamical landscape (Figs.~\ref{fig:behav_clusters},~\ref{fig:behav_clusters_SE}), Earth occupies a central region within the warm fixed-point domain. Here, climate trajectories are stable, bounded, and less sensitive to perturbations that would otherwise induce large-amplitude cycles or chaotic excursions. This position highlights Earth's location near a dynamical sweet spot: sufficiently regulated to avoid prolonged surface extremes, but not so tightly stabilized into a globally frozen state such a Snowball Earth, which-while not sterile-would have severely limited surface habitability \citep{2019JGRE..124.2087P}. Together, these results suggest that weak fourth feedbacks or stabilizing feedbacks in the presence of increasing stellar luminosity may offer robust pathways to temperate climate stability for Earth-like worlds on billion-year timescales.

\section{Discussion}

\subsection{Model verification}

Verification of the model's ice–albedo and carbonate–silicate dynamics (Section \ref{sec: rothman repro}) confirms its ability to reproduce established climate states and oscillations consistent with previous analytical and numerical studies \citep{2015E&PSL.429...20M, 2020RSPSA.47600303A, 2023MNRAS.521..690A}. These results corroborate prior findings that stable climate regulation naturally arises from coupled radiative and geochemical feedbacks \citep{2018MNRAS.477..727N, 2020MNRAS.492.2572A}.

\subsection{Model: Simplifications and Caveats}  \label{sec:caveats}

Our model is idealized in a number of important ways. The simplified, zero-dimensional treatment of climate described above is intentionally simple to apply to a broad swath of potential Earth-like exoplanet climates. Our treatment of weathering processes is also simple and ignores seafloor weathering and the mantle CO$_2$ cycle, consistent with previous studies \citep[e.g.,][and references within]{2015E&PSL.429...20M}. The model assumes Earth-like parameters and uniform parameter distributions unless otherwise specified.

Our simplified parameterization of albedo and weathering feedbacks may underestimate complexities highlighted by empirical studies, especially under extreme perturbations \citep{2016ApJ...827..117A, 2020RSPSA.47600303A}. Cloud microphysics, notably influenced by aerosols, also remains highly uncertain and could significantly alter radiative balances \citep{Forster2021, 2014BGD....11.8443Z, 2023JCli...36..547Z}. Further refinement in modeling cloud and aerosol interactions would enhance accuracy.

However, real-world paleoclimate records, such as abrupt glaciations and the mid-Pleistocene transition, indicate more complex nonlinear behaviors than captured by idealized models \citep{2004GPC....41...95R, 2013JCli...26.8289F}. Future work should examine the model’s sensitivity to additional stochastic forcings like orbital variations and internal ocean-atmosphere variability, known to influence planetary climates \citep{2019PhRvL.122o8701L, 2023JCli...36..547Z}.

Additionally, the model's assumptions primarily apply to Earth-like planets dominated by silicate weathering and CO$_2$-rich atmospheres. Expanding its applicability to planets with alternative volatile cycles (e.g., sulfur, methane) or diverse geochemical contexts would broaden its relevance \citep{2016ApJ...827..117A, 2020RSPSA.47600303A}. 

In this initial study, we do not consider any potential biological processes, such as organic carbon burial or potential evolutionary innovations that could wreak havoc on the climate (e.g., the Medea Hypothesis \citep{Ward2009}). While our fourth feedback is designed to remain agnostic and potentially capture similar processes, we do not assume in our model that Earth-like exoplanets must develop these or similar feedbacks.

There exist a plethora of other potential feedbacks (e.g., glacier-accelerated chemical weathering) that are not treated explicitly in our model, but rather can be incorporated as part of our generalized feedback framework. Before confidently assigning specific feedbacks to the model, work should be done to constrain the potential (bio)geochemistries present on other terrestrial exoplanets.

In summary, our model successfully reproduces fundamental climate behaviors, including long-term stability and nonlinear feedback impacts on stability, aligning well with previous theoretical and numerical findings. While future enhancements are needed, particularly in parameter distribution constraints, cloud feedbacks, climate forcings, irreversible transitions, and broader planetary conditions, this work provides a robust, simplified framework for exploring planetary climate dynamics and habitability.

\subsection{Impact of additional, generic feedbacks and time spent in temperate surface conditions}

The relevance of incorporating an additional, generic climate feedback should not be understated. Earth itself exhibits numerous feedback mechanisms \citep{2024PNAS..12116535K}, and it is likely that other terrestrial planets harbor additional or different geochemical feedbacks. Because the geochemical and radiative properties of exoplanets remain largely unconstrained, our modeling framework is designed to accommodate a wide range of climate parameters and feedback structures. By doing so, it allows us to map out the potential climates that may arise across a diverse array of planetary environments.

The inclusion of this fourth, additional feedback broadens the model’s dynamical repertoire (Section \ref{sec:new behaviors}). Simulations reveal the emergence of quasi-periodic and chaotic climate trajectories in response to variations in internal feedback parameters, emphasizing the importance of (fourth) feedback strengths and equilibrium temperatures in governing long-term behavior. 

Feedback strengths in the model should be interpreted as effective parameters that depend on atmospheric composition, surface inventory, and geological state. While calibrated to Earth-like conditions, varying these parameters allows the framework to capture both Earth’s climatic evolution and a broader diversity of exoplanet climates at a qualitative, dynamical level.

Through our work, feedback mechanisms emerge as important modulators of climate complexity. In the absence of stellar evolution, negative feedbacks would not lessen the probability of temperate conditions over long timescales (5 Gyr), which we interpret to show that Earth's current feedback configuration is stabilizing, and further ``stabilization'' is inefficient. However, stronger positive feedbacks (\( c \gg 0 \)) more often lead to unstable or uninhabitable outcomes (Figure \ref{fig:hist_c_hab}). Thus, only strong positive feedbacks would determine whether an Earth-like planet is more likely to be uninhabitable.

When stellar evolution is included, we observe the same general trends in the amount of time spent in temperate conditions (Figure \ref{fig:SE_hab}), but now see a greater diversity in climate behaviors and more common transitions between different states. The persistence of quasi-periodic, chaotic, and transiently chaotic trajectories suggests that episodic warming events, potentially driven by ice-albedo fluctuations, may play a key role in intermittently sustaining habitable conditions on otherwise frozen planets. The extent to which such planets could sustain liquid surface water intermittently over geological timescales remains an open question, but our results suggest that even small variations in feedback strength could have significant implications for long-term climate stability.

\subsection{Implications for geochemical models}

Our model predicts that a broad range of climate trajectories are likely for broadly Earth-like planets. Examples for strong fourth feedbacks may include coupling the cycling through the carbon reservoir to another second geochemical cycle (e.g., sulfur- or nitrogen cycle, \cite{Kemeny2024}) or to a different mode of plate recycling \citep[e.g.,][]{Affholder2025}. 
Our results motivate research into the most likely candidates for strong fourth feedbacks. Future research may also explore whether there are geophysical or geochemical processes that shape climate system feedbacks to be limited by a set of three feedbacks. If not, then Earth-like climate states are likely to be well-represented in the exoplanet population and would also mean that our long-term habitable Earth is still a commonplace outcome of climate evolution despite expanded climate behavior diversity.

\subsection{Climate diversity in exoplanet surveys}

Our simulations identify trends with the strengths of a fourth feedback. These trends carry important implications for the designs of exoplanet surveys, as they inform about the fraction of sun-like stars with HZ Earth-size planets that are broadly Earth-like (i.e., have fixed-point or limit cycle-type climate evolution allowing for long-term temperate surfaces). Our results suggest that the fraction of Earth-like, temperate climates ($f_{EL}$) will be lower than the total number of all rocky exoplanets. Planets, of course, may differ in many other ways from Earth (e.g., bulk composition, internal thermal evolution, atmospheric chemistry, presence/absence of life). Given these factors, our findings should be interpreted as conditional outcomes within the Earth-like parameter space explored here, rather than as representative population statistics. When combined with estimates of $\eta_\oplus$, the fraction sun-like stars that have Earth-sized HZ planets (e.g., ExoPAG SAG13 report\footnote{\url{https://exoplanets.nasa.gov/exep/exopag/sag/}},  \citealt{Pascucci2019,Bryson2020, 2022AJ....164..190B}), our results imply that under simplified Earth-like feedback assumptions, the fraction of Sun-like stars hosting broadly Earth-like (long-term temperate surface) planets would be even lower.

Our results show that including a strong fourth feedback shapes the expected distribution of climate outcomes for Earth-sized habitable zone planets. Compared to models that assume exact Earth-like feedbacks, fourth feedbacks generate a far greater climate diversity, introducing out-of-bounds and chaotic states that do not otherwise occur. The sign of the fourth feedback is important: if negative, the probability of sustained temperate surface conditions remains the same, whereas if positive, it reduces the fraction of planets that remain habitable over gigayear timescales. In either case, strong fourth feedbacks will shift the inner and outer boundaries of the classical habitable zone.

\begin{figure*}[htbp!]
    \centering
    \includegraphics[width=0.49\textwidth]{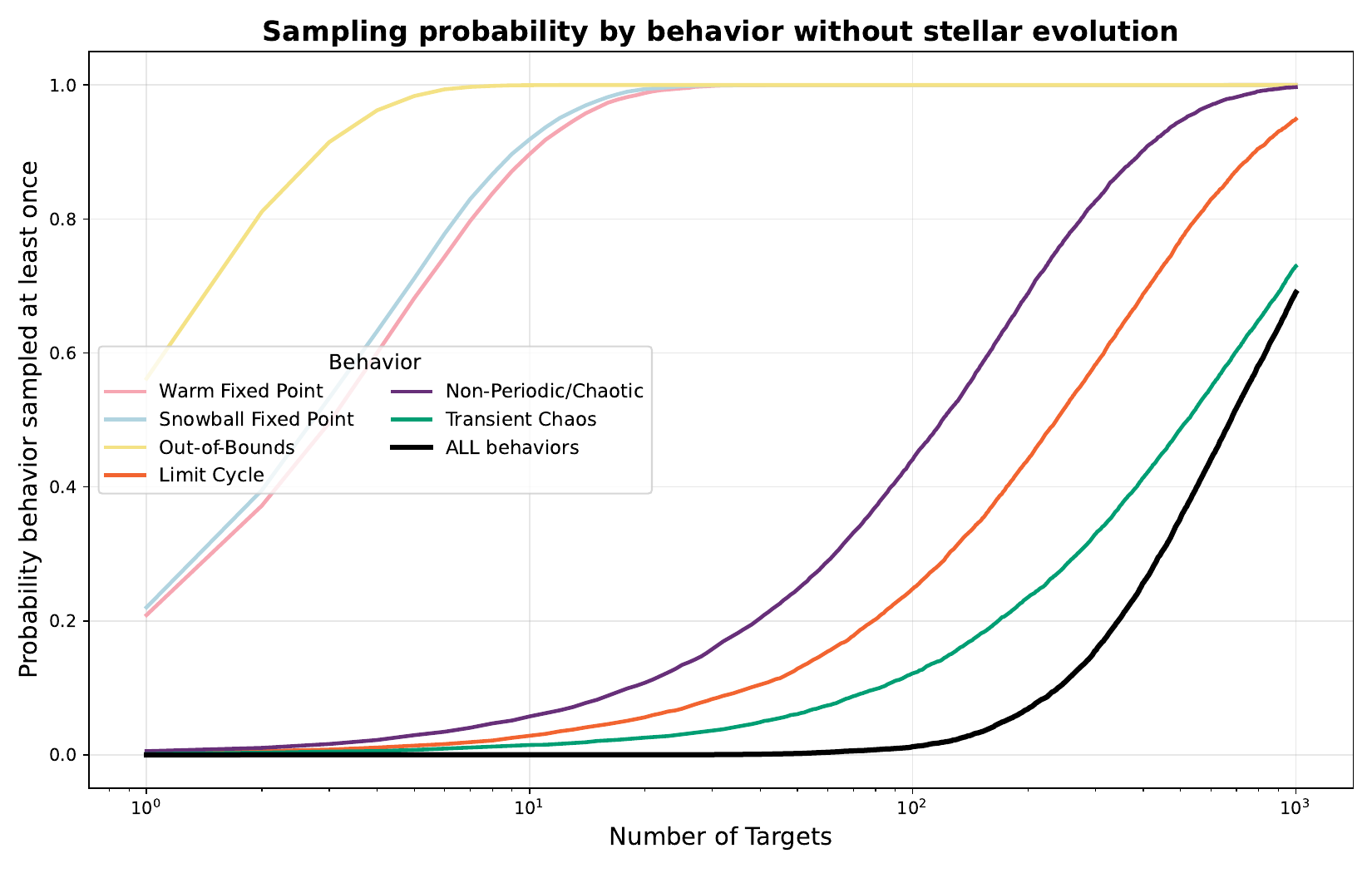}
    \includegraphics[width=0.49\textwidth]{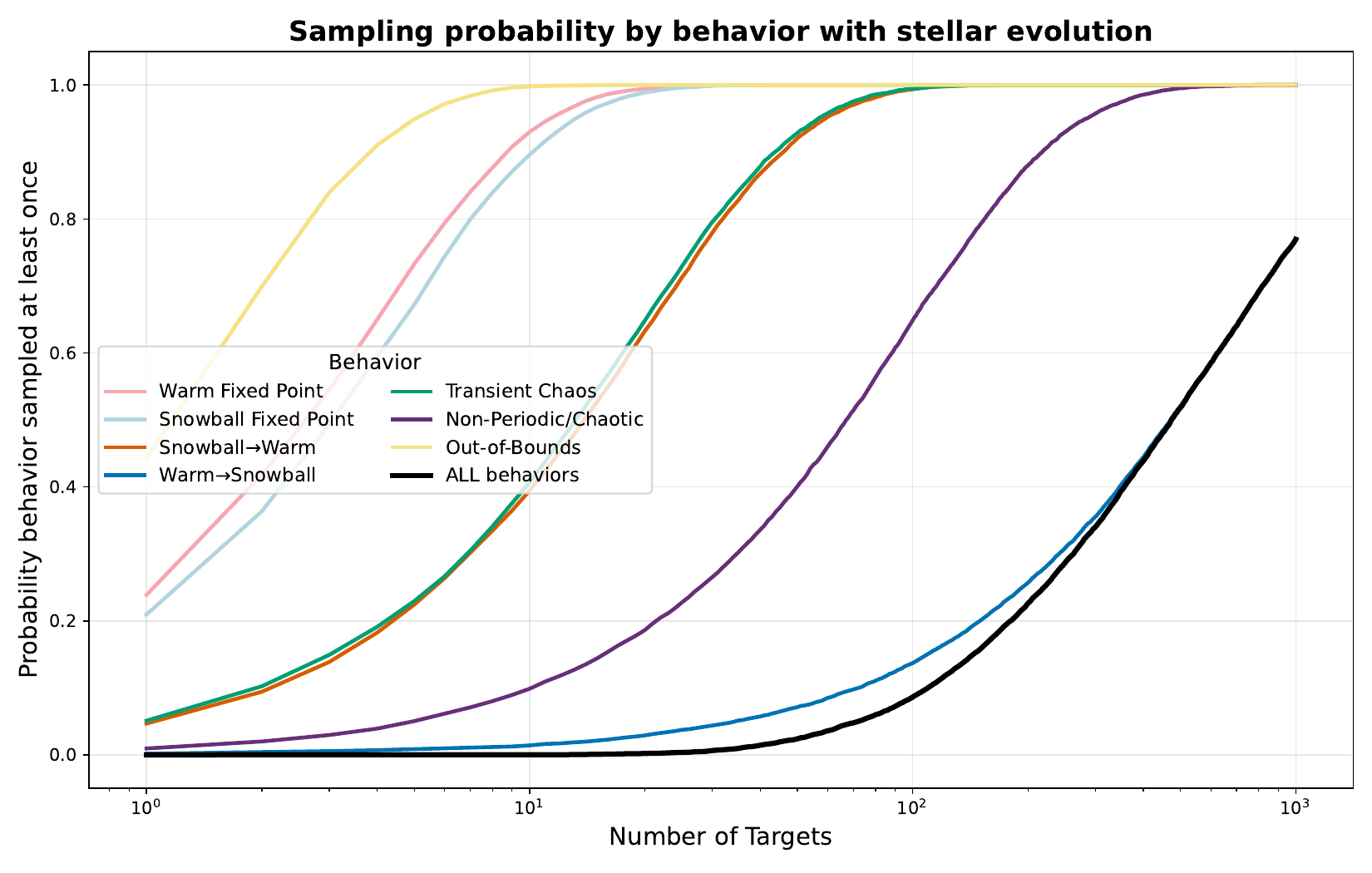}
    \caption{Monte Carlo sampling probability for climate behaviors as a function of the number of exoplanet targets observed (logarithmic $x$–axis), subject to model assumptions (Section \ref{sec:caveats}). Each curve shows the probability that at least one example of a given behavior has been sampled after $N$ observations, while the black curve tracks the probability that all behaviors have been observed at least once. 
    Left: simulations without stellar evolution. Warm and snowball fixed points, along with out-of-bounds cases, are recovered with only a handful to tens of observations, while limit cycles and non-periodic/chaotic states require much larger samples. 
    Right: simulations including stellar evolution. Stellar brightening introduces additional behaviors (e.g., snowball$\rightarrow$warm), which appear at intermediate sample sizes, while non-periodic/chaotic states remain rare and require large observational samples. Stellar evolution thus increases behavioral diversity and delays the point at which all behaviors are likely to be sampled.}
    \label{fig:samp_prob}
\end{figure*}

The Monte Carlo sampling experiment (Figure \ref{fig:samp_prob}) highlights how survey size controls the likelihood of recovering the full diversity of climate outcomes predicted by our model with and without incorporated stellar evolution. Fixed-point behaviors (both warm and snowball) and out-of-bounds trajectories are sampled quickly, with high probabilities of detection after only a handful of observations. Surveys with the next-generation ``extremely large'' ground-based telescopes will be limited to small samples \citep{Hardegree-Ullman2023,Currie2023,Hardegree-Ullman2025}. We find that with our assumed planet distribution and fourth feedback distribution (see Section \ref{sec:caveats}), these samples would likely be dominated by fixed-point climates (warm and snowball) and out-of-bounds climates. Similarly, space-based direct imaging surveys (e.g., \citealt{TheLUVOIRTeam2019} and \citealt{Gaudi2020}) and space-based interferometry (e.g., LIFE: \citealt{Quanz2022}) are realistically limited to smaller (O(10)) sample sizes and will likely probe these two types of climates. This is broadly consistent with insights from the SAMOSA intercomparison \citep{Haqq-Misra2022}, which showed that sparse samples preferentially recover dominant climate regimes when the underlying parameter space is structured by strong attractors. In our model, this leads small surveys to overwhelmingly detect fixed points and out-of-bounds states. For example, the proposed Habitable Worlds Observatory \citep[e.g.,][]{Feinberg2024} aims to study $\sim$25 habitable zone Earth-sized planets \citep{Stark2024,Tuchow2024}. Our study suggests (Fig.~\ref{fig:samp_prob}) that the planet population probed by HWO will most likely consist of a mix of fixed-point (some habitable and some not) and some out-of-bounds (runaway) climates, even within the habitable zone. A summary of the predicted outcome of missions is captured in Table \ref{tab:exoearth-yields}.

\begin{deluxetable*}{lcll}
    \tablecaption{Projected Exo-Earth candidate yields and likely behaviors sampled for exoplanet surveys, given our model assumptions (see Section \ref{sec:caveats}).\label{tab:exoearth-yields}}
    \tablehead{
        \colhead{Project} & \colhead{\# Exo-Earth Candidates} & \colhead{Refs.} & \colhead{Behaviors Sampled (given model)}
    }
    \startdata
    LFAST & $\sim$3 & \cite{Bender2022, Hardegree-Ullman2023}  & Fixed point; runaway \\
    GMT & $\sim$10 & \cite{Hardegree-Ullman2025} & Fixed point; runaway \\
    ELT & $\sim$30 & \cite{Hardegree-Ullman2025} & Fixed point; runaway \\
    HWO & $\sim$25 & \cite{Stark2024,Tuchow2024} & Fixed point; runaway \\
    LIFE & $\sim$10 & \cite{Quanz2022,Glauser2024} & Fixed point; runaway \\
    Nautilus & $\sim$1000 & \cite{Apai2019} & Full diversity; rare behaviors likely \\
    \enddata
\end{deluxetable*}
In contrast, less common climate outcomes, such as limit cycles, non-periodic/chaotic, and transient chaotic states require surveys with tens to hundreds of targets before they are likely found. Importantly, the ``all behaviors'' curves in each case (Fig.~\ref{fig:samp_prob}) show that dozens to hundreds of observed exoplanets are needed before a survey can reliably capture the full spectrum of climate regimes from a randomly-drawn exoplanet population. When stellar evolution is incorporated, even more planets will need to be observed before we can expect to sample the full diversity of climates that our model predicts. 

Sampling very large samples ($\sim$1,000 exo-earth candidates) in the foreseeable future may only be practically possible via transit spectroscopy. This method does not rely on spatially separating star and planet light and it is thus less sensitive to the distance of the targets. The Nautilus Space Observatory \citep{Apai2019} concept aims to study such large samples \citep{2022SPIE12221E..0CA}. This result reflects the intrinsic rarity of dynamic attractors (limit cycles and chaotic trajectories) in parameter space compared to more common fixed points or runaway behaviors, underscoring that their detection in exoplanet surveys is likely to be observationally expensive.

These findings suggest a clear tradeoff for future missions: small samples will efficiently constrain the prevalence of Earth-like fixed points (habitable vs. non-habitable) but will be much less sensitive to rarer dynamical regimes. Conversely, surveys randomly targeting hundreds of HZ Earth-sized planets are required to test the hypothesis that strong fourth-feedback processes drive planets into chaotic or cycling states. Thus, the demographic scale of an observing program directly determines whether it can distinguish between a habitable-zone population dominated by Earth-like stability or one shaped by more exotic dynamical attractors. In this sense, the rarity of Earth-like temperate fixed points becomes testable not through individual observations, but through population-level completeness in sampling the climate diversity landscape.

\begin{figure*}[htbp!]
    \centering
    \includegraphics[width=\linewidth]{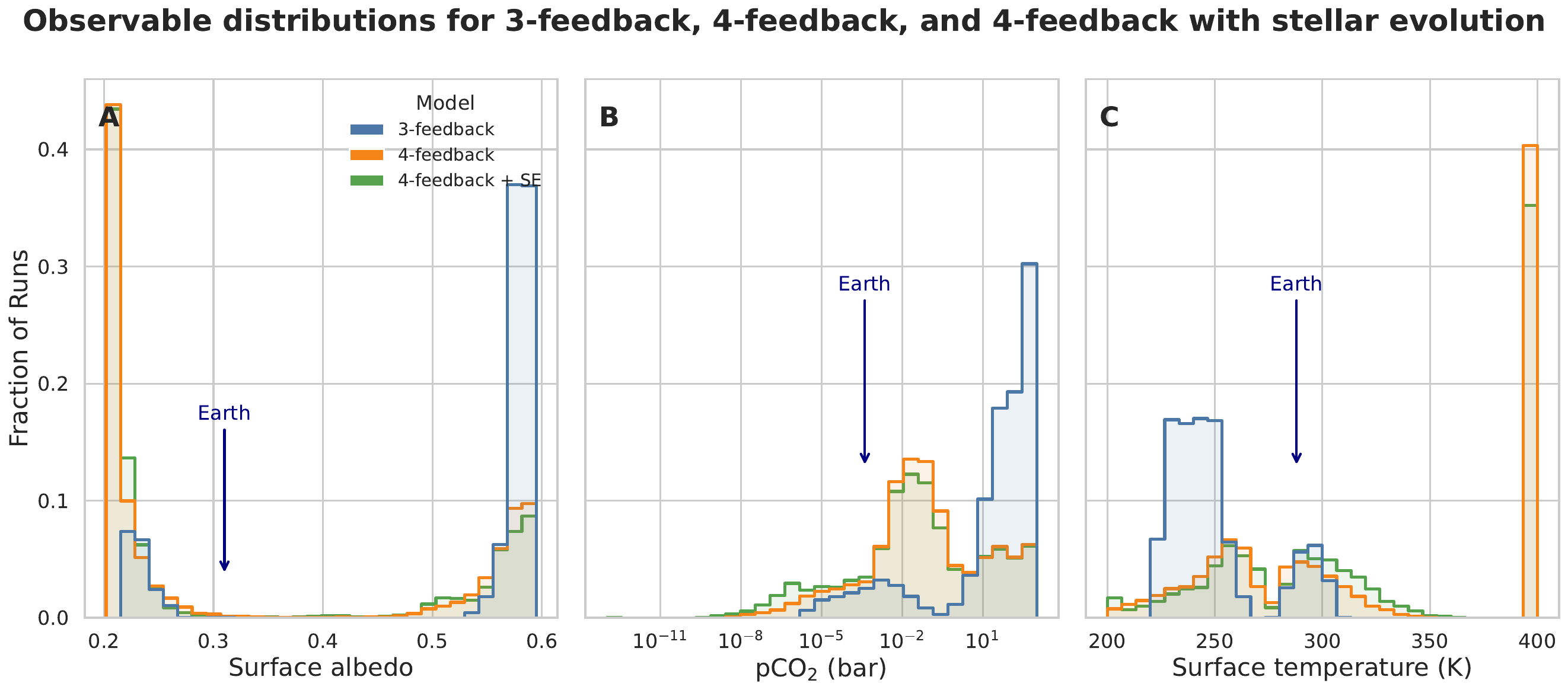}
    \caption{Observable distributions from the climate ensembles given our model assumptions (see Section \ref{sec:caveats}). Panels \textbf{A–C} show histograms of representative (median) post-transient values for surface albedo, atmospheric pCO$_2$ (log axis), and surface temperature. Colors denote model families in the order analyzed here: 3-feedback (blue), 4-feedback (orange), and 4-feedback with stellar evolution (green). Bars give the fraction of simulations in each bin. Vertical arrows mark present-day Earth for reference. Relative to the 3-feedback baseline, adding a fourth feedback broadens and shifts the distributions, and including stellar evolution further spreads outcomes—extending both low- and high-$p\mathrm{CO}_2$ and temperature tails—illustrating the increased climatic diversity expected when additional feedbacks and stellar brightening are considered.}
    \label{fig:obs_dist}
\end{figure*}

While our model is intentionally simplified and not designed to predict quantitative population statistics, it provides a framework for reasoning about how added feedback complexity would reshape the relative distributions of observable climate properties under otherwise Earth-like assumptions. Many degrees of freedom, such as atmospheric mass and composition, land fraction, rotation rate, aerosols, and photochemistry, remain unconstrained and could further broaden real distributions. Additionally, our model is largely built around Earth, and we selected randomized distributions for parameters including volcanic outgassing and fourth feedbacks since they are not currently constrained in the exoplanet population. Future work should be done to examine the effects of other planet parameter distributions and fourth feedback configurations to make more precise predictions.

Nevertheless, if fourth feedbacks are common, their presence would statistically modify the shapes of observable distributions (e.g., surface temperature, albedo, pCO$_2$). This reasoning follows previous conceptual studies \citep[e.g.,][]{Underwood2025}, which treat observable distributions as a statistical test of underlying feedback structure rather than as population forecasts. Thus, our analysis should be viewed as hypothesis-generating rather than predictive, motivating future statistical tests of climate diversity to provide more quantitative yield estimates.

Testing this hypothesis and its prediction requires large exoplanet samples or carefully-treated sampling techniques, e.g. \cite{Haqq-Misra2022}, and precise measurements of climate-related observables, such as atmospheric composition \citep{1993Icar..101..108K, 2018SciA....4.5747K}, surface mineralogy \citep{2019A&A...628A..12T, 2019JGRE..124.2087P, 2011ApJ...733L..48W}, albedo \citep{2011ApJ...731...76C, 2018AJ....155..230G,Tuchow2025}, and atmospheric retention \citep{2015AsBio..15..119L, 2017ApJ...843..122Z, 2017ApJ...841L..24B}. These indicators, when aggregated across populations, can provide statistical constraints on planetary climate diversity. While distinguishing individual climate states for single exoplanets is infeasible, ensemble trends in observables may reveal the statistical imprint of feedback complexity \citep[e.g.,][]{Underwood2025}.

Thus, while it is difficult to directly detect chaos in individual systems, population-level trends may reveal it indirectly: chaotic climates are expected to show broader or multimodal distributions in observables compared to the more regular patterns of fixed points or limit cycles (Figure \ref{fig:obs_dist}). For example, statistical outliers or multimodal distributions in retrieved albedo or phase curve variability could reflect transitions between multiple climate attractors. Spectral retrievals from next-generation observatories might constrain such distributions and reveal the imprint of underlying dynamical regimes. This demographic approach offers a promising path to empirically validate the dynamical regimes identified in our model.

Future missions could incorporate this insight by designing observations to capture the distribution of climate behaviors across the HZ. By statistically characterizing variations in observables, next-generation space telescopes such as the proposed Habitable Worlds Observatory (HWO) \citep{Feinberg2024, Stark2024,Tuchow2024}, and concepts like Nautilus \citep{Apai2019,2022SPIE12221E..0CA} and LIFE \citep{Quanz2022}, could test predictions of feedback-driven dynamical regimes. Such surveys may place population-level constraints on the configuration of planetary climate systems.

\section{Summary}

We developed a physically-motivated multi-feedback climate model for planetary climates in the HZ, incorporating ice–albedo, carbonate–silicate, OLR, and additional feedbacks, to explore the potential range of long-term climate behaviors of Earth-like planets around Sun-like stars. Our simulations reveal that both external forcings and internal feedbacks are critical in dictating long-term climate behavior.

Our key results are as follows:
\begin{enumerate}
    \item The addition of an extra (fourth) feedback to the classical three-feedback energy balance model expands the range of possible behaviors beyond simple fixed points and limit cycles, introducing quasi-periodic, chaotic, and transiently chaotic states (Figures \ref{fig:hist_c_hab}, \ref{fig:SE_hab}). These additional feedback mechanisms create more complex interactions within the climate system, allowing for oscillatory patterns that do not repeat exactly and, in some cases, sensitive dependence on initial conditions. 
    \item We find that quasi-periodic and chaotic climate trajectories occur due to the interplay of negative feedback strengths and specific feedback equilibrium temperature ranges, indicating complex but bounded climate variability. While previous studies have examined the role of feedbacks in exoplanet climates, our analysis highlights how quasi-periodic, chaotic, and transient chaotic trajectories can emerge in Earth-like habitable zone climates within a physically grounded multi-feedback framework.
    \item When we incorporate stellar evolution into our four feedback models, we find that negative fourth feedbacks (weak or strong) would not significantly impact the fraction of exo-Earth candidates that support long-term temperate surface conditions.
    \item In contrast, we find that strong positive fourth feedbacks would decrease the fraction of planets in long-term temperature states. 
    \item Assuming our underlying planet and feedback distributions hold, surveys with O(10) randomly-drawn exoplanets would likely sample fixed point (warm or snowball) and runaway (out-of-bounds) climates. 
    \item Given our model assumptions, our study suggests that sampling the diversity of climate states in the observable exoplanet population would require the characterization of on the order of 100 exo-Earth candidates. 
\end{enumerate}

These demographic insights have significant implications for future exoplanet surveys focused on habitability. Recognizing the parameter conditions that lead to rarer climate trajectories helps guide observational strategies aimed at sampling the full diversity of climate states.


\begin{acknowledgments}
    We thank A. Affholder, T. Barman, R. Ferri\`ere, R. Malhotra, M. Marley, and I. Pascucci for helpful and constructive comments. This material is based upon work supported by the National Aeronautics and Space Administration under Agreement No. 80NSSC21K0593 for the program ``Alien Earths". This publication was partly funded by the Heising-Simons Foundation through grant \#2024-5688. The results reported herein benefited from collaborations and/or information exchange within NASA’s Nexus for Exoplanet System Science (NExSS) research coordination network sponsored by NASA's Science Mission Directorate.

    This material is based upon High Performance Computing (HPC) resources supported by the University of Arizona TRIF, UITS, and Research, Innovation, and Impact (RII) and maintained by the UArizona Research Technologies department. This research has made use of NASA's Astrophysics Data System.
\end{acknowledgments}

\appendix

\section{Model Parameters and Formulation} \label{sec:params}

\begin{figure}[htbp!]
    \centering
    \includegraphics[width=0.9\linewidth]{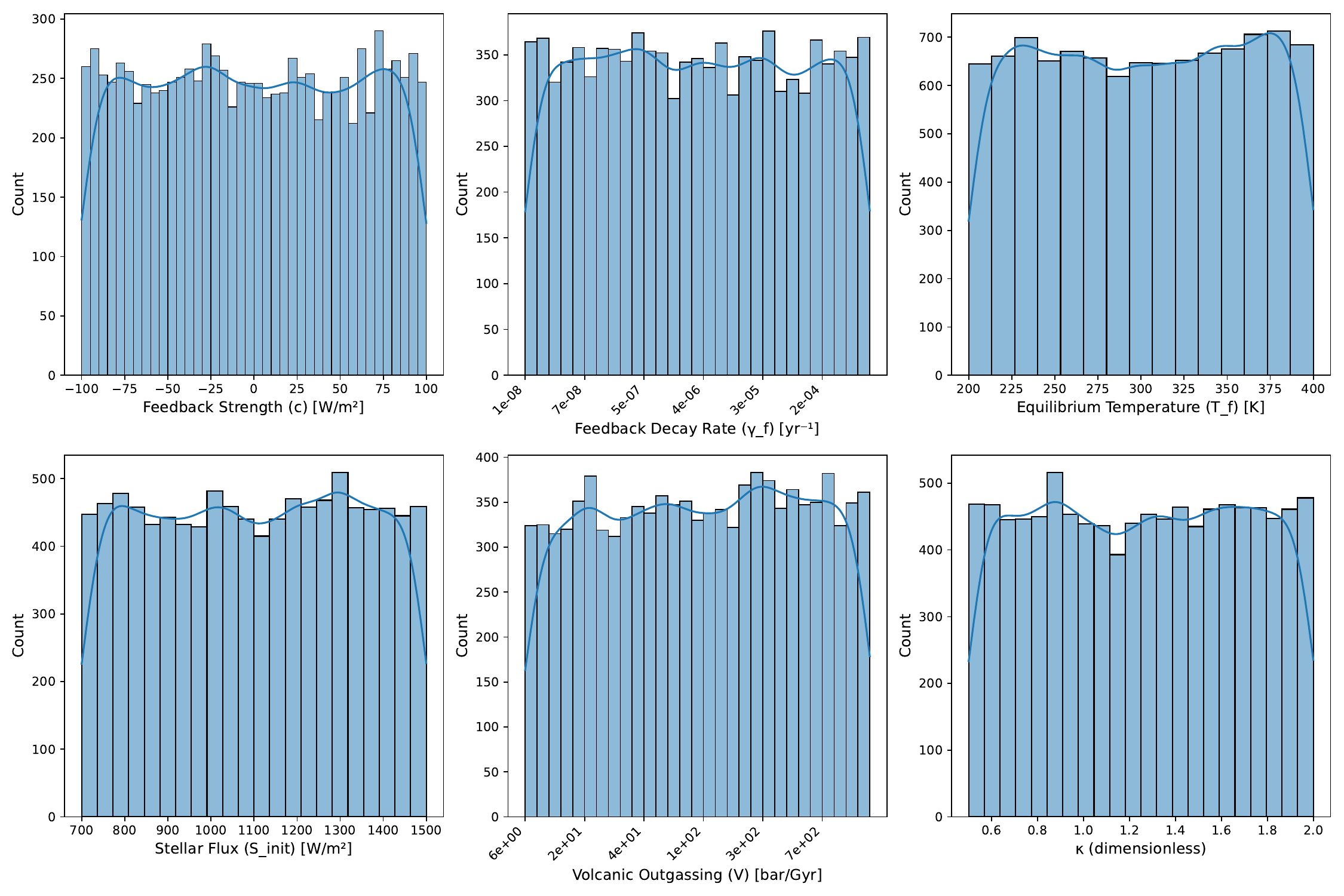}    \caption{Distributions of sampled parameters in the climate feedback configuration without stellar evolution. (Top row) Histograms of feedback strength (\( c \)), feedback decay rate (\( \gamma_f \)), and equilibrium temperature (\( T_f \)). (Bottom row) Histograms of stellar flux (\( S \)), volcanic outgassing (\( V \)), and activation strength ($\kappa$). The parameters \( \gamma_f \) and \( V \) are log-scaled to enhance visualization, as their values span several orders of magnitude. Overlaid kernel density estimates (KDEs) provide a smoothed representation of each distribution. These parameter ranges were chosen at random to systematically explore a broad spectrum of climate dynamics, ensuring coverage of both stable and dynamically evolving regimes.}
    \label{fig:config}
\end{figure}

\begin{figure}[htbp!]
    \centering
    \includegraphics[width=0.9\linewidth]{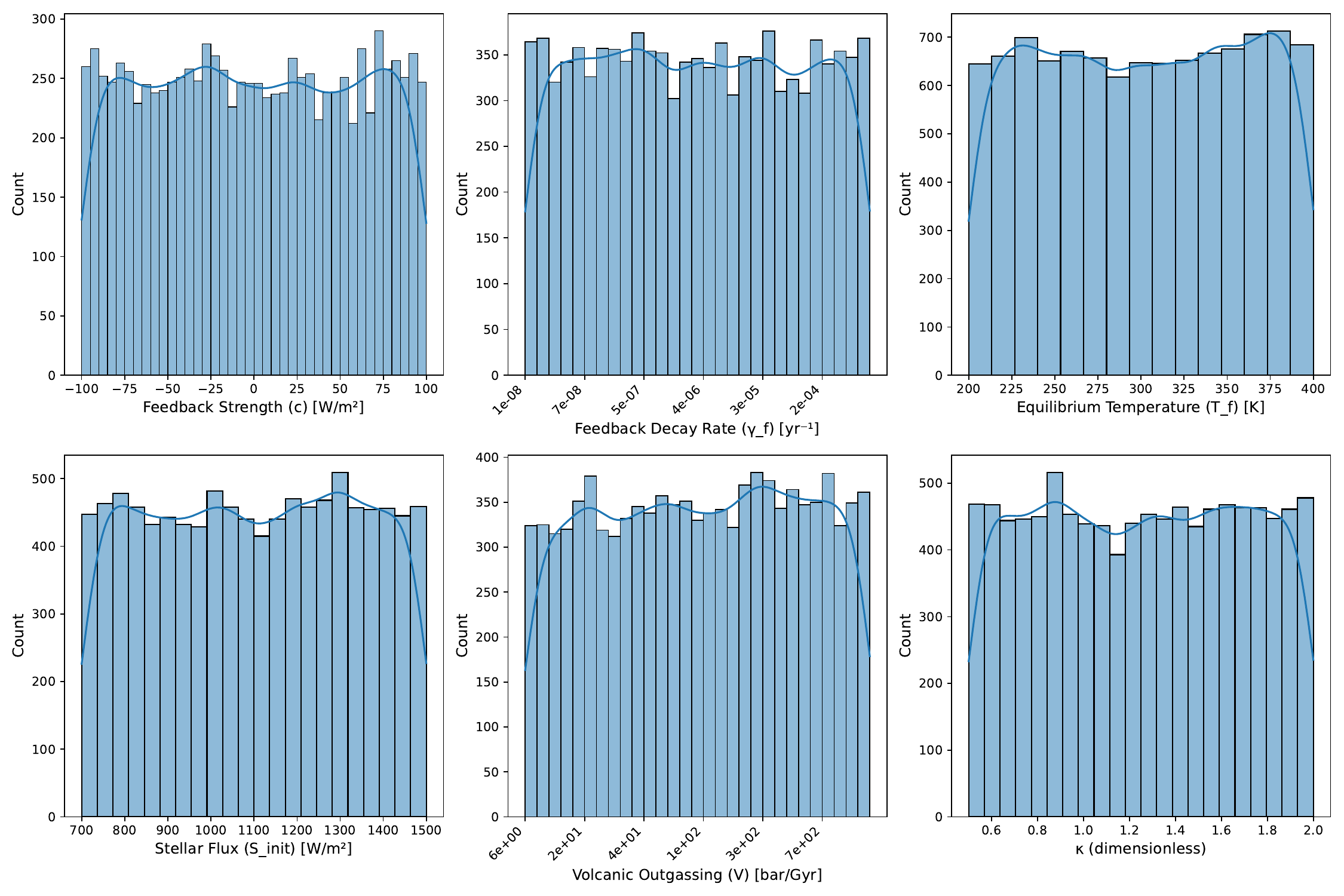}    \caption{Distributions of sampled parameters in the climate feedback configuration including stellar evolution. (Top row) Histograms of feedback strength (\( c \)), feedback decay rate (\( \gamma_f \)), and equilibrium temperature (\( T_f \)). (Bottom row) Histograms of stellar flux (\( S \)), volcanic outgassing (\( V \)), and activation strength ($\kappa$). The parameters \( \gamma_f \) and \( V \) are log-scaled to enhance visualization, as their values span several orders of magnitude. Overlaid kernel density estimates (KDEs) provide a smoothed representation of each distribution. These parameter ranges were chosen at random to systematically explore a broad spectrum of climate dynamics, ensuring coverage of both stable and dynamically evolving regimes.}
    \label{fig:config_SE}
\end{figure}

To enable a comprehensive exploration of climate feedback interactions, we selected parameter ranges that encompass both well-constrained physical values and more extreme cases that could reveal emergent behaviors. Figures \ref{fig:config} and \ref{fig:config_SE} display the resulting distributions of these sampled parameters for simulations excluding and including the role of stellar evolution, respectively. The feedback strength \( c \) was sampled uniformly around zero, allowing absolute values that could reach the strengths of both the carbonate-silicate and ice-albedo feedbacks. The feedback decay rate \( \gamma_f \) and volcanic outgassing rate \( V \) span multiple orders of magnitude, and their ranges capture diverse climate response timescales and atmospheric evolution scenarios. Stellar flux \( S \) and equilibrium temperature \( T_f \) were sampled more uniformly to ensure systematic coverage of planetary conditions. Kernel density estimates (KDEs) overlaid on histograms highlight key trends to aid interpretation. This parameter selection enables a comprehensive investigation of equilibrium states, oscillations, and chaotic climate dynamics, ensuring robust analysis of feedback interactions and long-term stability. The model parameters and standard values are given in Table \ref{tab:default_parameters}.

\begin{table}[ht!]
    \centering
    \caption{\textbf{Model Parameters, Descriptions, and Standard Values.}}
    \begin{tabular}{|l|l|l|}
        \hline
        \textbf{Parameter} & \textbf{Description} & \textbf{Standard Value} \\
        \hline
        \( S_0 \) & Solar constant & 1365 W m$^{-2}$ \\
        \( S \) & Stellar flux & Variable (W m$^{-2}$) \\
        \( T_0 \) & Warm-state temperature & 288 K \\
        \( P_0 \) & Atmospheric pCO\(_2\) in warm state & \(3 \times 10^{-4}\) bar \\
        \( a \) & OLR temperature coefficient & 2.2 W m\(^{-2}\) K\(^{-1}\) \\
        \( b \) & OLR CO\(_2\) sensitivity coefficient & 8 W m\(^{-2}\) \\
        \( \alpha_0 \) & Stable warm-state albedo & 0.241 \\
        \( \alpha_c \) & Cold-state albedo limit & 0.6 \\
        \( \alpha_w \) & Warm-state albedo limit & 0.2 \\
        \( C \) & Planetary heat capacity & \(2 \times 10^8\) J m\(^{-2}\) K\(^{-1}\) \\
        \( T_i \) & Transitional temperature for albedo/weathering & 273 K \\
        \( W_w \) & Maximum weathering rate (warm state) & 70 bar Gyr$^{-1}$ \\
        \( k \) & Weathering temperature sensitivity & 0.1 K\(^{-1}\) \\
        \( \beta \) & CO\(_2\) sensitivity exponent in weathering & 0.5 \\
        \( V \) & Volcanic outgassing rate & Variable (bar Gyr$^{-1}$) \\
        \( c \) & Strength of additional feedback & Variable (W m\(^{-2}\)) \\
        \( f \) & Feedback amplitude & Variable \\
        \( \delta_{\rm f} \) & Feedback activation sensitivity  & 0.1 K\(^{-1}\) \\
        \( \gamma_f \) & Feedback decay rate  & Variable (yr$^{-1}$) \\
        \( \kappa \) & Feedback activation rate & Variable \\
        \( T_f \) & Equilibrium temperature for fourth feedback & Variable (K) \\
        \hline
    \end{tabular}
    \label{tab:default_parameters}
\end{table}

We adopt albedo and weathering functions from \cite{2020RSPSA.47600303A}. Albedo parameterization goes as

\begin{equation}
    \alpha(T) = \alpha_c - (\alpha_c - \alpha_w)(0.5 + \arctan{((T - T_i) / \gamma)} / \pi)
\end{equation}

Here, $\alpha_c$ is the limit of the albedo in a cold state, and $\alpha_w$ the limit of the albedo in a warm state. $T_i$ denotes the transitional temperature at which the albedo is halfway between $\alpha_c$ and $\alpha_w$, and $\gamma$ sets the temperature scale over which this transition occurs. Assuming that weathering must go to zero with decreasing $T$ similar to the way in which $\alpha(T)$ goes towards its maximum value $\alpha_c$ with decreasing $T$, we follow the parameterization of \cite{2020RSPSA.47600303A} for $W(T)$:

\begin{equation}
    W(T) = W_w (0.5 + \arctan({(T - T_i) / \gamma}) / \pi)
\end{equation}

where $W_w$ represents the limit of $W(T)$ in a warm state.

\section{Fourth feedback formulation and assumptions}

\begin{figure}[htbp!]
    \centering
    \includegraphics[width=0.9\linewidth]{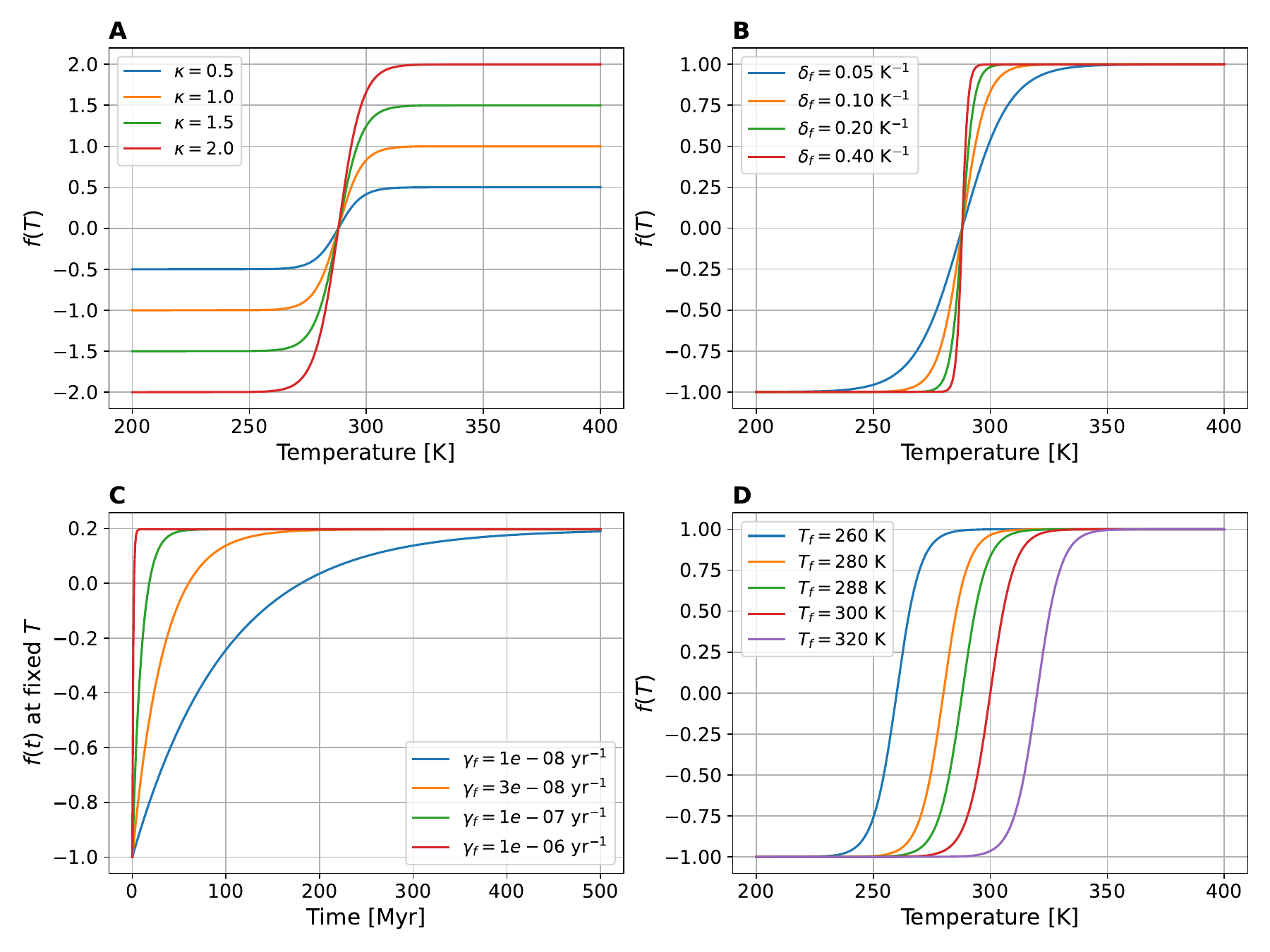}
    \caption{Exploring the sensitivity of the generalized climate feedback variable $f$ to feedback parameters. (\textbf{A}) Steady-state values of the fourth feedback $f(T)$ as a function of temperature for different feedback activation rates $\kappa$, which scale the amplitude of the internal response. Higher $\kappa$ increases the saturation of the feedback. (\textbf{B}) Effect of the feedback activation sensitivity $\delta_f$ on the transition behavior of $f(T)$. Larger $\delta_f$ produces a steeper, more threshold-like response around the feedback equilibrium temperature $T_f$. (\textbf{C}) Time evolution of $f(t)$ toward equilibrium at a fixed temperature $T = 290$\,K, for several values of the decay rate $\gamma_f$. Faster decay rates lead to quicker relaxation toward the equilibrium value. (\textbf{D}) Influence of the feedback equilibrium temperature $T_f$ on the location of the sigmoid transition in $f(T)$. Higher values of $T_f$ shift the feedback onset to warmer temperatures.}
    \label{fig:sens_analysis}
\end{figure}

It is important to understand how the behavior of the generalized climate feedback variable $f$, described in the main text, depends on our assumptions for its governing parameters. We are particularly interested in how the fourth-feedback function $\Dot{f}(T) = -\gamma_f \left[f - \kappa \tanh(\delta_f (T - T_f))\right]$ responds to changes in each of the three key control parameters: the amplitude $\kappa$, the sharpness $\delta_f$, and the feedback equilibrium temperature $T_f$. As shown in fig. \ref{fig:sens_analysis}, each of these parameters modulates a different aspect of the feedback behavior. The amplitude $\kappa$ sets the maximum possible feedback strength and directly determines the forcing contribution of the internal feedback at saturation. The sharpness parameter $\delta_f$ controls the steepness of the temperature transition: small values lead to a gradual, extended response across a broad range of temperatures, while larger values yield a rapid, threshold-like transition. Finally, the pivot temperature $T_f$ sets the point at which the feedback switches sign, determining whether the feedback activates in cold, temperate, or warm climates.

Our analysis shows that the fourth feedback is robustly tunable across physically plausible values of $\kappa$, $\delta_f$, and $T_f$, allowing the climate system to exhibit a wide variety of dynamical responses. This flexibility is critical when considering exoplanetary climates, where internal feedbacks may not be Earth-like in structure or strength. For example, increasing $\delta_f$ could model abrupt biogeochemical thresholds, such as sudden methane release or vegetation collapse, while varying $T_f$ could represent shifts in the climatic regime where such thresholds activate. Likewise, a stronger $\kappa$ may represent a more efficient biological or chemical amplifier operating within the system. Although our choice of parameter ranges is motivated by conceptual model simplicity, we find that the qualitative structure of $f(T)$ and its relaxation timescale remains well-behaved across the entire space. Future work could explore how different combinations of feedback shape and placement affect multistability, hysteresis, and the persistence of quasi-periodic or chaotic trajectories, especially in coupled models with multiple competing feedbacks.

\section{Climate trajectory classification}

Trajectories from the climate simulations were classified by analyzing their long-term dynamical behaviors. Each simulation provided time series of temperature, partial pressure of CO$_2$, and feedback amplitude. Stable fixed points were identified when both temperature and CO$_2$ gradients were negligible over the final simulation window, indicating no significant change with time. Simulations where temperature or CO$_2$ exceeded physically meaningful thresholds were classified as out-of-bounds. 

If gradients indicated persistent variability, further classification distinguished between limit cycles (regular periodic behavior), quasi-periodic orbits (multiple overlapping cycles), and chaotic trajectories (irregular and unpredictable dynamics), using a combination of spectral analysis \citep{1976Ap&SS..39..447L, 1982ApJ...263..835S}, fractal dimension calculations \citep{1983PhRvL..50..346G}, and Lyapunov exponents \citep{1986PhRvA..34.4971E}. 

\begin{figure}[ht!] 
    \centering 
    \includegraphics[width=0.6\linewidth]{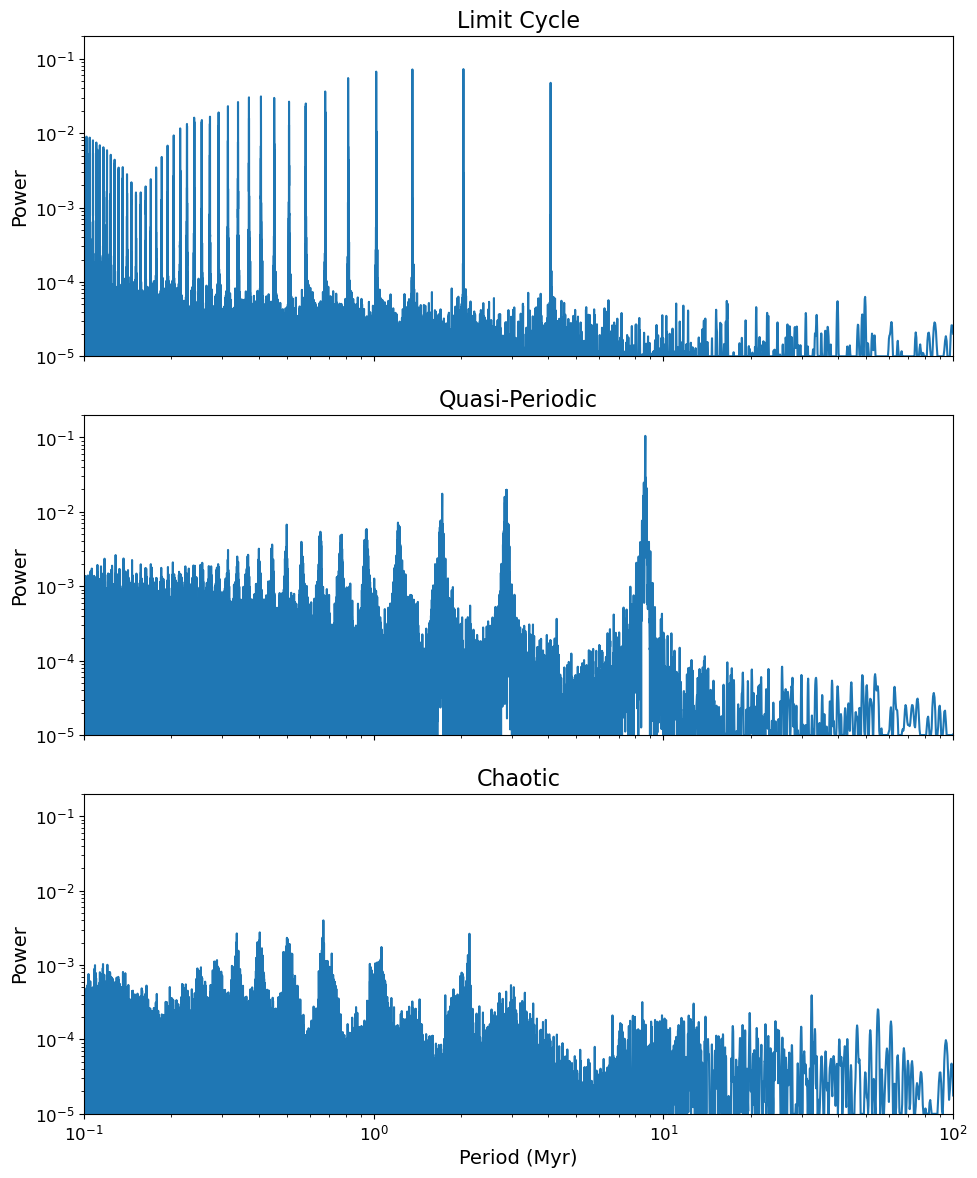} \caption{Temperature time series power spectra for three representative climate trajectories using a Lomb-Scargle periodogram. A noise floor of $10^{-5}$ has been applied to suppress numerical artifacts. Top: Periodic state, characterized by sharp peaks at regularly spaced intervals. Middle: Quasi-periodic behavior showing multiple sizable peaks with less regular structure and no single fundamental frequency dominating. Bottom: Chaotic trajectory with no narrow peaks and a broader, flatter power distribution.} \label{fig:power_spectra} 
\end{figure}

The distinction between periodic, quasiperiodic, and chaotic states is further examined using power spectra (Figure \ref{fig:power_spectra}). A Lomb-Scargle periodogram is applied to each time series to identify the dominant frequencies and assess the presence of broadband noise, which is a hallmark of chaotic systems \citep{1987ZPhyB..68..251S, 1980NYASA.357..453F}. To prevent numerical artifacts, a noise floor of $10^{-6}$ is imposed, ensuring that only meaningful spectral features are retained.

The results follow expected trends. The limit cycle case exhibits discrete peaks at integer multiples of a fundamental frequency, indicative of purely periodic motion. The quasiperiodic case retains several dominant peaks, but additional frequency components introduce complexity, suggesting an approach toward chaos. The chaotic case, in contrast, exhibits a broad distribution of power across frequencies, confirming the presence of aperiodic fluctuations. Such broadband noise, as described in earlier studies of chaotic systems, is characteristic of sensitive dependence on initial conditions and an underlying fractal structure of the attractor \citep{1987ZPhyB..68..251S, 1980NYASA.357..453F}.

Beyond spectral analysis, we examine the geometry of the attractor using correlation dimension calculations and recurrence quantification analysis (RQA) \citep{2007PhR...438..237M}. Unlike simple periodic or quasiperiodic attractors, chaotic attractors exhibit a fractal structure, occupying a non-integer dimension in state space. The correlation dimension, \( v \), quantifies this property by assessing how the number of neighboring trajectory points scales with distance. In addition, Lyapunov exponents provide a complementary measure of trajectory stability \citep{1981ZNatA..36...80S}, while recurrence statistics capture the underlying temporal structure. The results are summarized in Table \ref{tab:corr_dim}.

\begin{table*}[ht!]
    \centering
    \begin{tabular}{c|c|c|c}
        Metric & Limit Cycle & Quasiperiodic & Chaotic \\
        \hline
        $v$ & 0.636 & 0.579 & 0.762 \\
        $\lambda_1$ & 0.0667 & 0.0614 & 0.0714 \\
        $\lambda_2$ & 0.0665 & 0.0613 & 0.0712 \\
        Determinism & 0.918 & 0.872 & 0.853 \\
    \end{tabular}
    \caption{
    Metrics to distinguish between behaviors include correlation dimension \( v \), Lyapunov exponents \( \lambda \), and determinism for the three example climate states shown in Figure \ref{fig:power_spectra}.
    }
    \label{tab:corr_dim}
\end{table*}

Lyapunov exponents provide a dynamical criterion for distinguishing between stable, quasiperiodic, and chaotic regimes \citep{1985PhyD...16..285W}. The largest Lyapunov exponent, \(\lambda_1\), determines whether nearby trajectories diverge over time, with positive values indicating chaos and values near zero corresponding to quasiperiodic or periodic motion. The second exponent, \(\lambda_2\), reflects the presence of contraction in phase space, which in chaotic systems is typically negative due to contraction or dissipation.

Recurrence quantification analysis (RQA) offers additional insight into trajectory complexity \citep{2007PhR...438..237M}. The determinism (DET) metric quantifies the fraction of recurrence points that form structured diagonal lines in the recurrence plot, with high values (close to one) indicating periodic or quasiperiodic behavior.

Table \ref{tab:corr_dim} summarizes the key metrics used to distinguish among the dynamical regimes. In the limit cycle example, a relatively low correlation dimension (\(v = 0.635\)) is observed alongside nearly identical Lyapunov exponents (\(\lambda_1 = 0.0642\) and \(\lambda_2 = 0.0638\)) and very high determinism (DET = 0.918), consistent with a strongly recurrent, periodic attractor. By contrast, the quasiperiodic state exhibits a slightly lower correlation dimension (\(v = 0.575\)) and a modest reduction in both \(\lambda_1\) (\(0.0578\)) and \(\lambda_2\) (\(0.0571\)), with determinism reduced to 0.872. This suggests a broader range of recurrence times, indicative of a toroidal attractor. Finally, the chaotic case is marked by a notably higher correlation dimension (\(v = 0.744\)) and the highest Lyapunov exponents (\(\lambda_1 = 0.071\) and \(\lambda_2 = 0.0701\)) paired with the lowest determinism (DET = 0.853), reflecting a more complex, fragmented attractor geometry.

Taken together, these metrics demonstrate that while limit cycles and quasiperiodic regimes are characterized by strongly structured, low-dimensional recurrences, the transition to chaos is accompanied by increased attractor dimensionality and reduced recurrence structure, marking a subtle shift toward higher dynamical complexity.

\section{Layered views of behavior clusters with and without stellar evolution}\label{app:se_layers}

\begin{figure*}[htbp!]
  \gridline{\fig{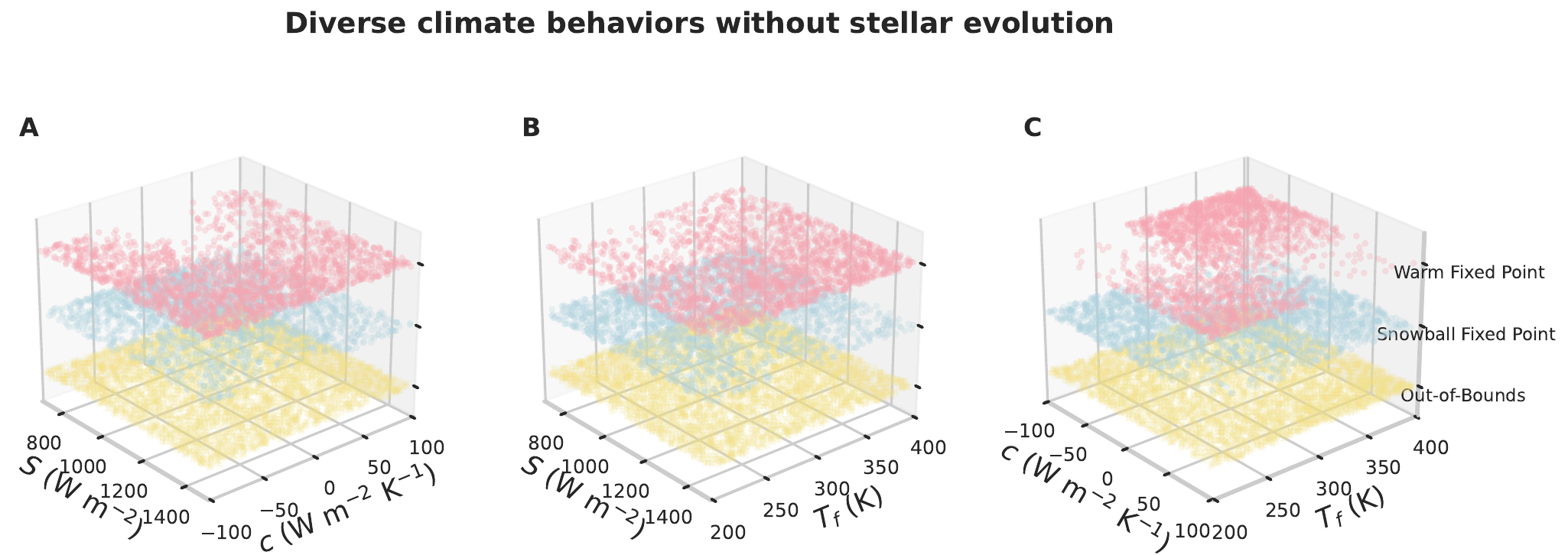}{0.98\textwidth}{}}
  \vspace{-4pt}
  \gridline{\fig{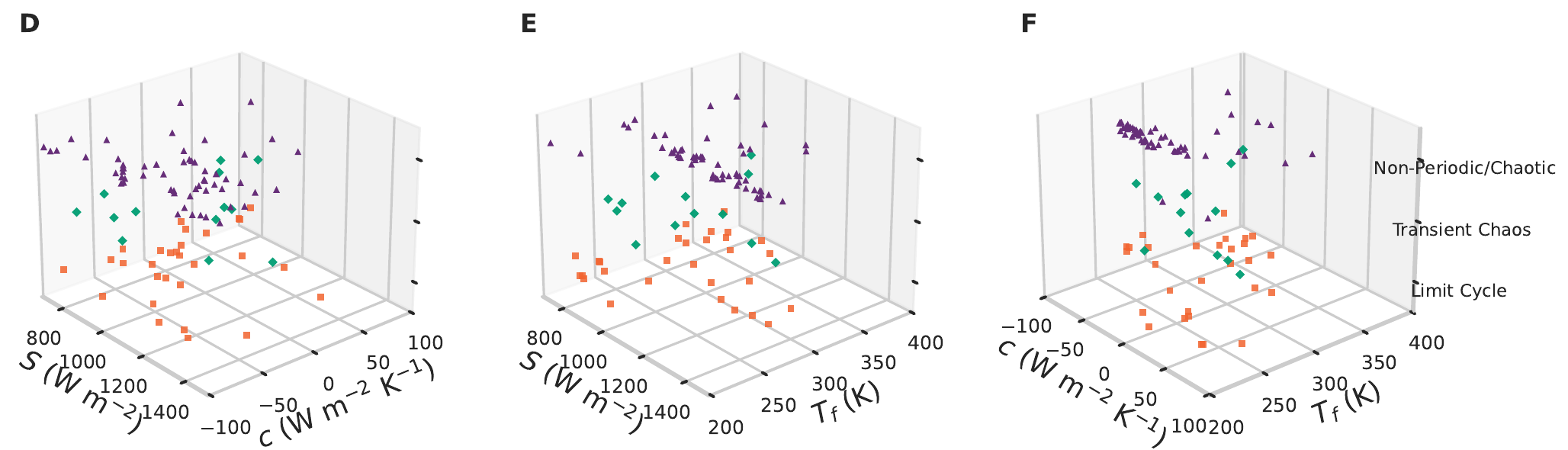}{0.98\textwidth}{}}
  \vspace{-6pt}
  \caption{
  Layered 3D triptychs showing how climate behaviors populate parameter space when stellar evolution is excluded. Each row contains three identical projections—(\textbf{A},\textbf{D}) $S$ vs.\ $c$, (\textbf{B},\textbf{E}) $S$ vs.\ $T_f$, and (\textbf{C},\textbf{F}) $c$ vs.\ $T_f$—with the third dimension used only to separate behaviors and avoid overplotting. The $z$‐axis is labeled by behavior name and is shared across the three panels in each row.
  \emph{Top row (A–C):} Out-of-Bounds (yellow), Snowball Fixed Point (light blue), Warm Fixed Point (pink).
  \emph{Bottom row (D–F):} Snowball$\rightarrow$Warm (red), Warm$\rightarrow$Snowball (dark blue), Non-periodic/Chaotic (purple), and Transient Chaos (green). Layering the categories in $z$ reveals their full spatial distributions that would otherwise be hidden in 2D projections.
  }
  \label{fig:layers_3D}
\end{figure*}

\begin{figure*}[htbp!]
  \gridline{\fig{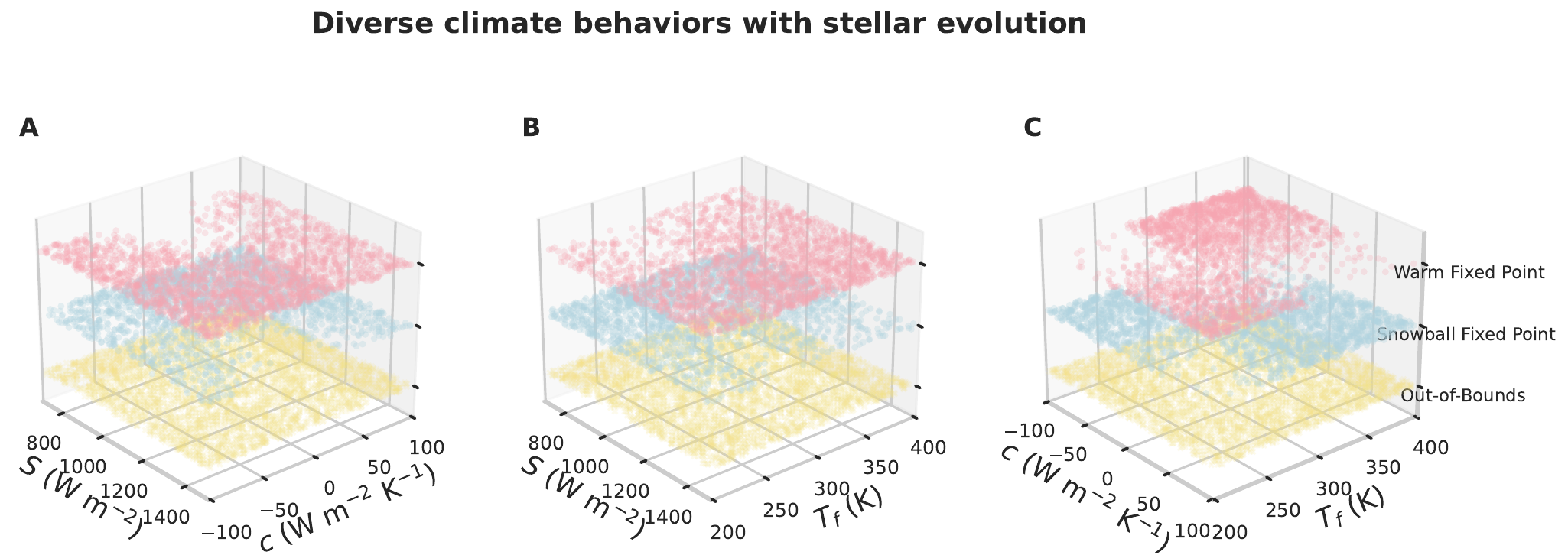}{0.98\textwidth}{}}
  \vspace{-4pt}
  \gridline{\fig{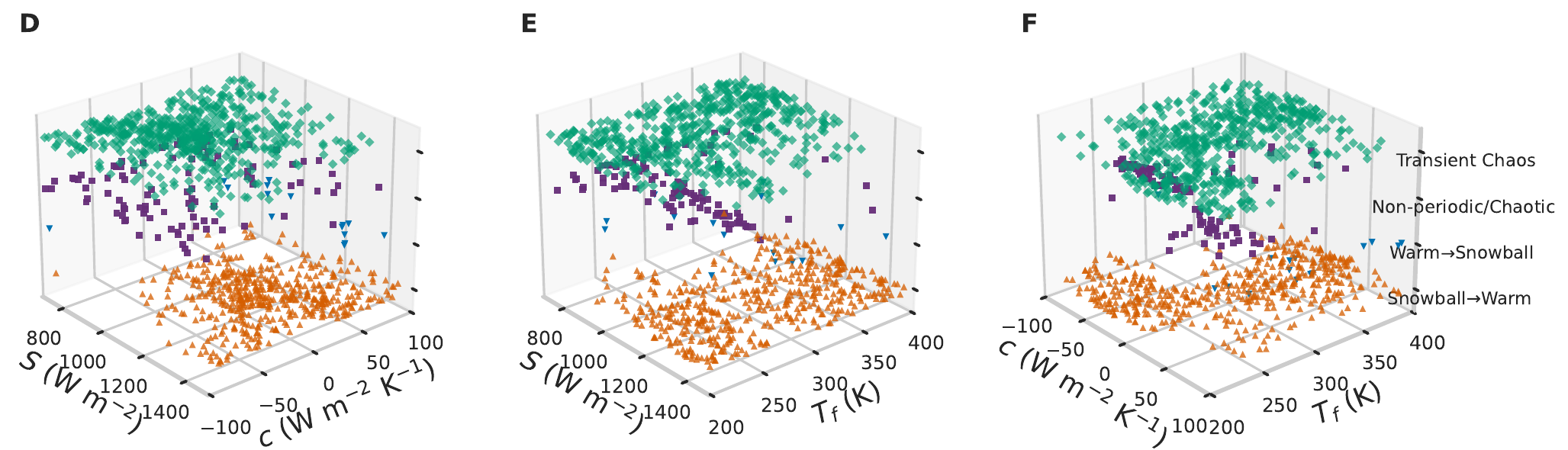}{0.98\textwidth}{}}
  \vspace{-6pt}
  \caption{
  Layered 3D triptychs showing how climate behaviors populate parameter space when stellar evolution is included. Each row contains three identical projections—(\textbf{A},\textbf{D}) $S$ vs.\ $c$, (\textbf{B},\textbf{E}) $S$ vs.\ $T_f$, and (\textbf{C},\textbf{F}) $c$ vs.\ $T_f$—with the third dimension used only to separate behaviors and avoid overplotting. The $z$‐axis is labeled by behavior name and is shared across the three panels in each row.
  \emph{Top row (A–C):} Out-of-Bounds (yellow), Snowball Fixed Point (light blue), Warm Fixed Point (pink).
  \emph{Bottom row (D–F):} Snowball$\rightarrow$Warm (red), Warm$\rightarrow$Snowball (dark blue), Non-periodic/Chaotic (purple), and Transient Chaos (green). Layering the categories in $z$ reveals their full spatial distributions that would otherwise be hidden in 2D projections.
  }
  \label{fig:se_layers_3D}
\end{figure*}

Figures~\ref{fig:layers_3D} and \ref{fig:se_layers_3D} present two complementary, layered (3D) triptychs that visualize how simulated climate behaviors populate the parameter space when stellar evolution is excluded and is included, respectively, as complementary visuals to Figures~\ref{fig:behav_clusters} and \ref{fig:behav_clusters_SE}. In each triptych the three panels show the same projections— (\textbf{A},\textbf{D}) stellar flux $S$ versus fourth–feedback strength $c$, (\textbf{B},\textbf{E}) $S$ versus the activation/equilibrium temperature $T_f$, and (\textbf{C},\textbf{F}) $c$ versus $T_f$—but the third dimension is used only to separate categories of behavior. The $z$–axis is categorical, labeled with the behavior name and shared across the three panels in each row, so the reader can see the full distribution of each behavior without overplotting.

Each point is one simulation, classified by its long–term trajectory. We split the behaviors across two rows to reduce visual crowding. The plotting limits are identical between rows, enabling direct, visual comparisons across panels. 

Within a given panel, the horizontal and vertical axes are the physical parameters being varied ($S$, $c$, $T_f$); the vertical ``Behavior'' axis simply stacks categories on shallow, evenly spaced layers.  Because the layers are separated in $z$ but viewed at an inclination, the shape and extent of each category's footprint in the $(S,c)$, $(S,T_f)$, and $(c,T_f)$ planes are visible simultaneously.  Apparent density gradients arise from point overlap and should be interpreted qualitatively; for quantitative incidence and temperate-time statistics, see the main-text histograms and probability summaries.

\bibliography{main}{}

@ARTICLE{Long2025,
       author = {{Long}, Feng and {Pascucci}, Ilaria and {Houge}, Adrien and {Banzatti}, Andrea and {Pontoppidan}, Klaus M. and {Najita}, Joan and {Krijt}, Sebastiaan and {Xie}, Chengyan and {Williams}, Joe and {Herczeg}, Gregory J. and {Andrews}, Sean M. and {Bergin}, Edwin and {Blake}, Geoffrey A. and {Colmenares}, Mar{\'\i}a Jos{\'e} and {Harsono}, Daniel and {Romero-Mirza}, Carlos E. and {Li}, Rixin and {Lu}, Cicero X. and {Pinilla}, Paola and {Wilner}, David J. and {Vioque}, Miguel and {Zhang}, Ke and {JDISCS Collaboration}},
        title = "{The First JWST View of a 30-Myr-old Protoplanetary Disk Reveals a Late-stage Carbon-rich Phase}",
      journal = {Astrophys. J. Letters},
         year = 2025,
        month = jan,
       volume = {978},
       number = {2},
          eid = {L30},
        pages = {L30},
          doi = {10.3847/2041-8213/ad99d2},
archivePrefix = {arXiv},
       eprint = {2412.05535},
 primaryClass = {astro-ph.EP},
       adsurl = {https://ui.adsabs.harvard.edu/abs/2025ApJ...978L..30L}
}

@ARTICLE{Currie2023,
       author = {{Currie}, Miles H. and {Meadows}, Victoria S. and {Rasmussen}, Kaitlin C.},
        title = "{There's More to Life than O$_{2}$: Simulating the Detectability of a Range of Molecules for Ground-based, High-resolution Spectroscopy of Transiting Terrestrial Exoplanets}",
      journal = {\psj},
         year = 2023,
        month = may,
       volume = {4},
       number = {5},
          eid = {83},
        pages = {83},
          doi = {10.3847/PSJ/accf86},
archivePrefix = {arXiv},
       eprint = {2304.10683},
 primaryClass = {astro-ph.EP},
       adsurl = {https://ui.adsabs.harvard.edu/abs/2023PSJ.....4...83C}
}

@INPROCEEDINGS{Glauser2024,
       author = {{Glauser}, Adrian M. and {Quanz}, Sascha P. and {Hansen}, Jonah and {Dannert}, Felix and {Ireland}, Michael and {Linz}, Hendrik and {Absil}, Olivier and {Alei}, Eleonora and {Angerhausen}, Daniel and {Birbacher}, Thomas and {Defr{\`e}re}, Denis and {Fortier}, Andrea and {Huber}, Philipp A. and {Kammerer}, Jens and {Laugier}, Romain and {Lichtenberg}, Tim and {Noack}, Lena and {Ranganathan}, Mohanakrishna and {Rugheimer}, Sarah and {Airapetian}, Vladimir and {Alibert}, Yann and {Amado}, Pedro J. and {Anger}, Marius and {Anugu}, Narsireddy and {Aragon}, Max and {Armstrong}, David J. and {Balbi}, Amedeo and {Balsalobre-Ruza}, Olga and {Banik}, Deepayan and {Beck}, Mathias and {Bhattarai}, Surendra and {Biren}, Jonas and {Bottoni}, Jacopo and {Braam}, Marrick and {Brandeker}, Alexis and {Buchhave}, Lars A. and {Caballero}, Jos{\'e} A. and {Cabrera}, Juan and {Carone}, Ludmila and {Carri{\'o}n-Gonz{\'a}lez}, {\'O}scar and {Castro-Gonz{\'a}lez}, Amadeo and {Chan}, Kenny and {Coelho}, Ligia F. and {Constantinou}, Tereza and {Cowan}, Nicolas and {Danchi}, William and {Dandumont}, Colin and {Davoult}, Jeanne and {Dawn}, Arjun and {de Vera}, Jean-Pierre P. and {de Visser}, Pieter J. and {Dorn}, Caroline and {Duque Lara}, Juan A. and {Elowitz}, Mark and {Ertel}, Steve and {Fang}, Yuedong and {Felix}, Simon and {Fortney}, Jonathan and {Fridlund}, Malcolm and {Garc{\'\i}a Mu{\~n}oz}, Antonio and {Gillmann}, Cedric and {Golabek}, Gregor and {Grenfell}, John Lee and {Guidi}, Greta and {Guilera}, Octavio and {Hagelberg}, Janis and {Hansen}, Janina and {Haqq-Misra}, Jacob and {Hara}, Nathan and {Helled}, Ravit and {Herbst}, Konstantin and {Hernitschek}, Nina and {Hinkley}, Sasha and {Ito}, Takahiro and {Itoh}, Satoshi and {Ivanovski}, Stavro and {Janson}, Markus and {Johansen}, Anders and {Jones}, Hugh and {Kane}, Stephen and {Kitzmann}, Daniel and {Kovacevic}, Andjelka B. and {Kraus}, Stefan and {Krause}, Oliver and {Kruijssen}, J.~M. Diederik and {Kuiper}, Rolf and {Kuriakose}, Alen and {Labadie}, Lucas and {Lacour}, Sylvestre and {Lanza}, Antonino F. and {Leedj{\"a}rv}, Laurits and {Lendl}, Monika and {Leung}, Michaela and {Lillo-Box}, Jorge and {Loicq}, J{\'e}r{\^o}me and {Luque}, Rafael and {Mahadevan}, Suvrath and {Majumdar}, Liton and {Malbet}, Fabien and {Mallia}, Franco and {Mathew}, Joice and {Matsuo}, Taro and {Matthews}, Elisabeth and {Meadows}, Victoria and {Mennesson}, Bertrand and {Meyer}, Michael R. and {Molaverdikhani}, Karan and {Molli{\`e}re}, Paul and {Monnier}, John and {Navarro}, Ramon and {Nsamba}, Benard and {Oguri}, Kenshiro and {Oza}, Apurva and {Palle}, Enric and {Persson}, Carina and {Pitman}, Joe and {Pl{\'a}valov{\'a}}, Eva and {Pozuelos}, Francisco J. and {Quirrenbach}, Andreas and {Ramirez}, Ramses and {Reiners}, Ansgar and {Ribas}, Ignasi and {Rice}, Malena and {Ricketti}, Berke Vow and {Roelfsema}, Peter and {Romagnolo}, Amedeo and {Ronco}, Mar{\'\i}a. Paula and {Schlecker}, Martin and {Schonhut-Stasik}, Jessica and {Schwieterman}, Edward and {Sefilian}, Antranik A. and {Serabyn}, Eugene and {Shahi}, Chinmay and {Sharma}, Siddhant and {Silva}, Laura and {Singh}, Swapnil and {Sneed}, Evan L. and {Spencer}, Locke and {Squicciarini}, Vito and {Staguhn}, Johannes and {Stapelfeldt}, Karl and {Stassun}, Keivan and {Tamura}, Motohide and {Taysum}, Benjamin and {van der Tak}, Floris and {van Kempen}, Tim A. and {Vasisht}, Gautam and {Wang}, Haiyang S. and {Wordsworth}, Robin and {Wyatt}, Mark},
        title = "{The Large Interferometer For Exoplanets (LIFE): a space mission for mid-infrared nulling interferometry}",
    booktitle = {Optical and Infrared Interferometry and Imaging IX},
         year = 2024,
       editor = {{Kammerer}, Jens and {Sallum}, Stephanie and {Sanchez-Bermudez}, Joel},
       series = {Society of Photo-Optical Instrumentation Engineers (SPIE) Conference Series},
       volume = {13095},
        month = aug,
          eid = {130951D},
        pages = {130951D},
          doi = {10.1117/12.3019090},
       adsurl = {https://ui.adsabs.harvard.edu/abs/2024SPIE13095E..1DG}
}

@INPROCEEDINGS{Feinberg2024,
       author = {{Feinberg}, Lee and {Ziemer}, John and {Ansdell}, Megan and {Crooke}, Julie and {Dressing}, Courtney and {Mennesson}, Bertrand and {O'Meara}, John and {Pepper}, Joshua and {Roberge}, Aki},
        title = "{The Habitable Worlds Observatory engineering view: status, plans, and opportunities}",
    booktitle = {Space Telescopes and Instrumentation 2024: Optical, Infrared, and Millimeter Wave},
         year = 2024,
       editor = {{Coyle}, Laura E. and {Matsuura}, Shuji and {Perrin}, Marshall D.},
       series = {Society of Photo-Optical Instrumentation Engineers (SPIE) Conference Series},
       volume = {13092},
        month = aug,
          eid = {130921N},
        pages = {130921N},
          doi = {10.1117/12.3018328},
       adsurl = {https://ui.adsabs.harvard.edu/abs/2024SPIE13092E..1NF}
}

@ARTICLE{Gaudi2020,
       author = {{Gaudi}, B. Scott and {Seager}, Sara and {Mennesson}, Bertrand and {Kiessling}, Alina and {Warfield}, Keith and {Cahoy}, Kerri and {Clarke}, John T. and {Domagal-Goldman}, Shawn and {Feinberg}, Lee and {Guyon}, Olivier and {Kasdin}, Jeremy and {Mawet}, Dimitri and {Plavchan}, Peter and {Robinson}, Tyler and {Rogers}, Leslie and {Scowen}, Paul and {Somerville}, Rachel and {Stapelfeldt}, Karl and {Stark}, Christopher and {Stern}, Daniel and {Turnbull}, Margaret and {Amini}, Rashied and {Kuan}, Gary and {Martin}, Stefan and {Morgan}, Rhonda and {Redding}, David and {Stahl}, H. Philip and {Webb}, Ryan and {Alvarez-Salazar}, Oscar and {Arnold}, William L. and {Arya}, Manan and {Balasubramanian}, Bala and {Baysinger}, Mike and {Bell}, Ray and {Below}, Chris and {Benson}, Jonathan and {Blais}, Lindsey and {Booth}, Jeff and {Bourgeois}, Robert and {Bradford}, Case and {Brewer}, Alden and {Brooks}, Thomas and {Cady}, Eric and {Caldwell}, Mary and {Calvet}, Rob and {Carr}, Steven and {Chan}, Derek and {Cormarkovic}, Velibor and {Coste}, Keith and {Cox}, Charlie and {Danner}, Rolf and {Davis}, Jacqueline and {Dewell}, Larry and {Dorsett}, Lisa and {Dunn}, Daniel and {East}, Matthew and {Effinger}, Michael and {Eng}, Ron and {Freebury}, Greg and {Garcia}, Jay and {Gaskin}, Jonathan and {Greene}, Suzan and {Hennessy}, John and {Hilgemann}, Evan and {Hood}, Brad and {Holota}, Wolfgang and {Howe}, Scott and {Huang}, Pei and {Hull}, Tony and {Hunt}, Ron and {Hurd}, Kevin and {Johnson}, Sandra and {Kissil}, Andrew and {Knight}, Brent and {Kolenz}, Daniel and {Kraus}, Oliver and {Krist}, John and {Li}, Mary and {Lisman}, Doug and {Mandic}, Milan and {Mann}, John and {Marchen}, Luis and {Marrese-Reading}, Colleen and {McCready}, Jonathan and {McGown}, Jim and {Missun}, Jessica and {Miyaguchi}, Andrew and {Moore}, Bradley and {Nemati}, Bijan and {Nikzad}, Shouleh and {Nissen}, Joel and {Novicki}, Megan and {Perrine}, Todd and {Pineda}, Claudia and {Polanco}, Otto and {Putnam}, Dustin and {Qureshi}, Atif and {Richards}, Michael and {Eldorado Riggs}, A.~J. and {Rodgers}, Michael and {Rud}, Mike and {Saini}, Navtej and {Scalisi}, Dan and {Scharf}, Dan and {Schulz}, Kevin and {Serabyn}, Gene and {Sigrist}, Norbert and {Sikkia}, Glory and {Singleton}, Andrew and {Shaklan}, Stuart and {Smith}, Scott and {Southerd}, Bart and {Stahl}, Mark and {Steeves}, John and {Sturges}, Brian and {Sullivan}, Chris and {Tang}, Hao and {Taras}, Neil and {Tesch}, Jonathan and {Therrell}, Melissa and {Tseng}, Howard and {Valente}, Marty and {Van Buren}, David and {Villalvazo}, Juan and {Warwick}, Steve and {Webb}, David and {Westerhoff}, Thomas and {Wofford}, Rush and {Wu}, Gordon and {Woo}, Jahning and {Wood}, Milana and {Ziemer}, John and {Arney}, Giada and {Anderson}, Jay and {Ma{\'\i}z-Apell{\'a}niz}, Jes{\'u}s and {Bartlett}, James and {Belikov}, Ruslan and {Bendek}, Eduardo and {Cenko}, Brad and {Douglas}, Ewan and {Dulz}, Shannon and {Evans}, Chris and {Faramaz}, Virginie and {Feng}, Y. Katherina and {Ferguson}, Harry and {Follette}, Kate and {Ford}, Saavik and {Garc{\'\i}a}, Miriam and {Geha}, Marla and {Gelino}, Dawn and {G{\"o}tberg}, Ylva and {Hildebrandt}, Sergi and {Hu}, Renyu and {Jahnke}, Knud and {Kennedy}, Grant and {Kreidberg}, Laura and {Isella}, Andrea and {Lopez}, Eric and {Marchis}, Franck and {Macri}, Lucas and {Marley}, Mark and {Matzko}, William and {Mazoyer}, Johan and {McCandliss}, Stephan and {Meshkat}, Tiffany and {Mordasini}, Christoph and {Morris}, Patrick and {Nielsen}, Eric and {Newman}, Patrick and {Petigura}, Erik and {Postman}, Marc and {Reines}, Amy and {Roberge}, Aki and {Roederer}, Ian and {Ruane}, Garreth and {Schwieterman}, Edouard and {Sirbu}, Dan and {Spalding}, Christopher and {Teplitz}, Harry and {Tumlinson}, Jason and {Turner}, Neal and {Werk}, Jessica and {Wofford}, Aida and {Wyatt}, Mark and {Young}, Amber and {Zellem}, Rob},
        title = "{The Habitable Exoplanet Observatory (HabEx) Mission Concept Study Final Report}",
      journal = {arXiv e-prints},
         year = 2020,
        month = jan,
          eid = {arXiv:2001.06683},
        pages = {arXiv:2001.06683},
          doi = {10.48550/arXiv.2001.06683},
archivePrefix = {arXiv},
       eprint = {2001.06683},
 primaryClass = {astro-ph.IM},
       adsurl = {https://ui.adsabs.harvard.edu/abs/2020arXiv200106683G}
}

@ARTICLE{Bryson2020,
       author = {{Bryson}, S. and {Coughlin}, J. and {Batalha}, N.~M. and {Berger}, T. and {Huber}, D. and {Burke}, C. and {Dotson}, J. and {Mullally}, S.~E.},
        title = "{A Probabilistic Approach to Kepler Completeness and Reliability for Exoplanet Occurrence Rates}",
      journal = {\aj},
         year = 2020,
        month = jun,
       volume = {159},
       number = {6},
          eid = {279},
        pages = {279},
          doi = {10.3847/1538-3881/ab8a30},
archivePrefix = {arXiv},
       eprint = {1906.03575},
 primaryClass = {astro-ph.EP},
       adsurl = {https://ui.adsabs.harvard.edu/abs/2020AJ....159..279B}
}

@ARTICLE{Tuchow2025,
       author = {{Tuchow}, Noah W. and {Stark}, Christopher C. and {Apai}, Daniel and {Schlecker}, Martin and {Hardegree-Ullman}, Kevin K.},
        title = "{Bioverse: Assessing the Ability of Direct Imaging Surveys to Empirically Constrain the Habitable Zone via Trends in Albedo}",
      journal = {arXiv e-prints},
         year = 2025,
        month = sep,
          eid = {arXiv:2509.07297},
        pages = {arXiv:2509.07297},
          doi = {10.48550/arXiv.2509.07297},
archivePrefix = {arXiv},
       eprint = {2509.07297},
 primaryClass = {astro-ph.EP},
       adsurl = {https://ui.adsabs.harvard.edu/abs/2025arXiv250907297T}
}

@ARTICLE{Apai2019,
       author = {{Apai}, D{\'a}niel and {Milster}, Tom D. and {Kim}, Dae Wook and {Bixel}, Alex and {Schneider}, Glenn and {Liang}, Ronguang and {Arenberg}, Jonathan},
        title = "{A Thousand Earths: A Very Large Aperture, Ultralight Space Telescope Array for Atmospheric Biosignature Surveys}",
      journal = {\aj},
         year = 2019,
        month = aug,
       volume = {158},
       number = {2},
          eid = {83},
        pages = {83},
          doi = {10.3847/1538-3881/ab2631},
archivePrefix = {arXiv},
       eprint = {1906.05079},
 primaryClass = {astro-ph.IM},
       adsurl = {https://ui.adsabs.harvard.edu/abs/2019AJ....158...83A}
}

@ARTICLE{Stark2024,
       author = {{Stark}, Christopher C. and {Mennesson}, Bertrand and {Bryson}, Steve and {Ford}, Eric B. and {Robinson}, Tyler D. and {Belikov}, Ruslan and {Bolcar}, Matthew R. and {Feinberg}, Lee D. and {Guyon}, Olivier and {Latouf}, Natasha and {Mandell}, Avi M. and {Rauscher}, Bernard J. and {Sirbu}, Dan and {Tuchow}, Noah W.},
        title = "{Paths to robust exoplanet science yield margin for the Habitable Worlds Observatory}",
      journal = {Journal of Astronomical Telescopes, Instruments, and Systems},
         year = 2024,
        month = jul,
       volume = {10},
          eid = {034006},
        pages = {034006},
          doi = {10.1117/1.JATIS.10.3.034006},
archivePrefix = {arXiv},
       eprint = {2405.19418},
 primaryClass = {astro-ph.EP},
       adsurl = {https://ui.adsabs.harvard.edu/abs/2024JATIS..10c4006S}
}

@ARTICLE{TheLUVOIRTeam2019,
       author = {{The LUVOIR Team}},
        title = "{The LUVOIR Mission Concept Study Final Report}",
      journal = {arXiv e-prints},
         year = 2019,
        month = dec,
          eid = {arXiv:1912.06219},
        pages = {arXiv:1912.06219},
          doi = {10.48550/arXiv.1912.06219},
archivePrefix = {arXiv},
       eprint = {1912.06219},
 primaryClass = {astro-ph.IM},
       adsurl = {https://ui.adsabs.harvard.edu/abs/2019arXiv191206219T}
}

@ARTICLE{Hardegree-Ullman2023,
       author = {{Hardegree-Ullman}, Kevin K. and {Apai}, D{\'a}niel and {Bergsten}, Galen J. and {Pascucci}, Ilaria and {L{\'o}pez-Morales}, Mercedes},
        title = "{Bioverse: A Comprehensive Assessment of the Capabilities of Extremely Large Telescopes to Probe Earth-like O$_{2}$ Levels in Nearby Transiting Habitable-zone Exoplanets}",
      journal = {\aj},
         year = 2023,
        month = jun,
       volume = {165},
       number = {6},
          eid = {267},
        pages = {267},
          doi = {10.3847/1538-3881/acd1ec},
archivePrefix = {arXiv},
       eprint = {2304.12490},
 primaryClass = {astro-ph.EP},
       adsurl = {https://ui.adsabs.harvard.edu/abs/2023AJ....165..267H}
}

@ARTICLE{Hardegree-Ullman2025,
       author = {{Hardegree-Ullman}, Kevin K. and {Apai}, D{\'a}niel and {Haffert}, Sebastiaan Y. and {Schlecker}, Martin and {Kasper}, Markus and {Kammerer}, Jens and {Wagner}, Kevin},
        title = "{Bioverse: Giant Magellan Telescope and Extremely Large Telescope Direct Imaging and High-resolution Spectroscopy Assessment{\textemdash}Surveying Exo-Earth O$_{2}$ and Testing the Habitable Zone Oxygen Hypothesis}",
      journal = {\aj},
         year = 2025,
        month = mar,
       volume = {169},
       number = {3},
          eid = {171},
        pages = {171},
          doi = {10.3847/1538-3881/adb02f},
archivePrefix = {arXiv},
       eprint = {2405.11423},
 primaryClass = {astro-ph.EP},
       adsurl = {https://ui.adsabs.harvard.edu/abs/2025AJ....169..171H}
}

@ARTICLE{Pascucci2019,
       author = {{Pascucci}, Ilaria and {Mulders}, Gijs D. and {Lopez}, Eric},
        title = "{The Impact of Stripped Cores on the Frequency of Earth-size Planets in the Habitable Zone}",
      journal = {\apjl},
         year = 2019,
        month = sep,
       volume = {883},
       number = {1},
          eid = {L15},
        pages = {L15},
          doi = {10.3847/2041-8213/ab3dac},
archivePrefix = {arXiv},
       eprint = {1908.06192},
 primaryClass = {astro-ph.EP},
       adsurl = {https://ui.adsabs.harvard.edu/abs/2019ApJ...883L..15P}
}

@ARTICLE{Apai2025,
       author = {{Apai}, D{\'a}niel and {Barnes}, Rory and {Murphy}, Matthew M. and {Lichtenberg}, Tim and {Tuchow}, Noah and {Ferri{\`e}re}, R{\'e}gis and {Wagner}, Kevin and {Affholder}, Antonin and {Malhotra}, Renu and {Journaux}, Baptiste and {Vazan}, Allona and {Ramirez}, Ramses and {M{\'e}ndez}, Abel and {Kane}, Stephen R. and {Klawender}, Veronica H. and {NExSS Quantitative Habitability Science Working Group}},
        title = "{A Terminology and Quantitative Framework for Assessing the Habitability of Solar System and Extraterrestrial Worlds}",
      journal = {\psj},
         year = 2025,
        month = jul,
       volume = {6},
       number = {7},
          eid = {165},
        pages = {165},
          doi = {10.3847/PSJ/addda8},
archivePrefix = {arXiv},
       eprint = {2505.22808},
 primaryClass = {astro-ph.EP},
       adsurl = {https://ui.adsabs.harvard.edu/abs/2025PSJ.....6..165A}
}

@ARTICLE{Bond2010,
       author = {{Bond}, Jade C. and {O'Brien}, David P. and {Lauretta}, Dante S.},
        title = "{The Compositional Diversity of Extrasolar Terrestrial Planets. I. In Situ Simulations}",
      journal = {Astrophys. J. },
         year = 2010,
        month = jun,
       volume = {715},
       number = {2},
        pages = {1050-1070},
          doi = {10.1088/0004-637X/715/2/1050},
archivePrefix = {arXiv},
       eprint = {1004.0971},
 primaryClass = {astro-ph.EP},
       adsurl = {https://ui.adsabs.harvard.edu/abs/2010ApJ...715.1050B}
}

@ARTICLE{Jontof-Hutter2019,
       author = {{Jontof-Hutter}, Daniel},
        title = "{The Compositional Diversity of Low-Mass Exoplanets}",
      journal = {Annual Review of Earth and Planetary Sciences},
         year = 2019,
        month = may,
       volume = {47},
        pages = {141-171},
          doi = {10.1146/annurev-earth-053018-060352},
archivePrefix = {arXiv},
       eprint = {1911.04598},
 primaryClass = {astro-ph.EP},
       adsurl = {https://ui.adsabs.harvard.edu/abs/2019AREPS..47..141J}
}

@ARTICLE{Lichtenberg2025,
       author = {{Lichtenberg}, Tim and {Miguel}, Yamila},
        title = "{Super-Earths and Earth-like Exoplanets}",
      journal = {Treatise on Geochemistry},
         year = 2025,
        month = jan,
       volume = {7},
        pages = {51-112},
          doi = {10.1016/B978-0-323-99762-1.00122-4},
archivePrefix = {arXiv},
       eprint = {2405.04057},
 primaryClass = {astro-ph.EP},
       adsurl = {https://ui.adsabs.harvard.edu/abs/2025TrGeo...7...51L}
}

@ARTICLE{Affholder2025,
       author = {{Affholder}, Antonin and {Mazevet}, St{\'e}phane and {Sauterey}, Boris and {Apai}, D{\'a}niel and {Ferri{\`e}re}, R{\'e}gis},
        title = "{Interior Convection Regime, Host Star Luminosity, and Predicted Atmospheric CO$_{2}$ Abundance in Terrestrial Exoplanets}",
      journal = {Astron. J.},
         year = 2025,
        month = mar,
       volume = {169},
       number = {3},
          eid = {125},
        pages = {125},
          doi = {10.3847/1538-3881/ada384},
archivePrefix = {arXiv},
       eprint = {2406.16104},
 primaryClass = {astro-ph.EP},
       adsurl = {https://ui.adsabs.harvard.edu/abs/2025AJ....169..125A}
}

@ARTICLE{Putirka2021,
       author = {{Putirka}, Keith D. and {Dorn}, Caroline and {Hinkel}, Natalie R. and {Unterborn}, Cayman T.},
        title = "{Compositional Diversity of Rocky Exoplanets}",
      journal = {Elements},
         year = 2021,
        month = aug,
       volume = {17},
       number = {4},
        pages = {235},
          doi = {10.2138/gselements.17.4.235},
archivePrefix = {arXiv},
       eprint = {2108.08383},
 primaryClass = {astro-ph.EP},
       adsurl = {https://ui.adsabs.harvard.edu/abs/2021Eleme..17..235P}
}

@ARTICLE{Dorn2021,
       author = {{Dorn}, Caroline and {Lichtenberg}, Tim},
        title = "{Hidden Water in Magma Ocean Exoplanets}",
      journal = {Astrophys. J. Letters},
         year = 2021,
        month = nov,
       volume = {922},
       number = {1},
          eid = {L4},
        pages = {L4},
          doi = {10.3847/2041-8213/ac33af},
archivePrefix = {arXiv},
       eprint = {2110.15069},
 primaryClass = {astro-ph.EP},
       adsurl = {https://ui.adsabs.harvard.edu/abs/2021ApJ...922L...4D}
}

@ARTICLE{Hu2014,
       author = {{Hu}, Renyu and {Seager}, Sara},
        title = "{Photochemistry in Terrestrial Exoplanet Atmospheres. III. Photochemistry and Thermochemistry in Thick Atmospheres on Super Earths and Mini Neptunes}",
      journal = {Astrophys. J. },
         year = 2014,
        month = mar,
       volume = {784},
       number = {1},
          eid = {63},
        pages = {63},
          doi = {10.1088/0004-637X/784/1/63},
archivePrefix = {arXiv},
       eprint = {1401.0948},
 primaryClass = {astro-ph.EP},
       adsurl = {https://ui.adsabs.harvard.edu/abs/2014ApJ...784...63H}
}

@ARTICLE{Kemeny2024,
       author = {{Kemeny}, Preston Cosslett and {Torres}, Mark A. and {Fischer}, Woodward W. and {Bl{\"a}ttler}, Clara L.},
        title = "{Balance and imbalance in biogeochemical cycles reflect the operation of closed, exchange, and open sets}",
      journal = {Proceedings of the National Academy of Science},
         year = 2024,
        month = mar,
       volume = {121},
       number = {12},
          eid = {e2316535121},
        pages = {e2316535121},
          doi = {10.1073/pnas.2316535121},
       adsurl = {https://ui.adsabs.harvard.edu/abs/2024PNAS..12116535K}
}

@ARTICLE{Pascucci2009,
       author = {{Pascucci}, I. and {Apai}, D. and {Luhman}, K. and {Henning}, Th. and {Bouwman}, J. and {Meyer}, M.~R. and {Lahuis}, F. and {Natta}, A.},
        title = "{The Different Evolution of Gas and Dust in Disks around Sun-Like and Cool Stars}",
      journal = {Astrophys. J. },
         year = 2009,
        month = may,
       volume = {696},
       number = {1},
        pages = {143-159},
          doi = {10.1088/0004-637X/696/1/143},
archivePrefix = {arXiv},
       eprint = {0810.2552},
 primaryClass = {astro-ph},
       adsurl = {https://ui.adsabs.harvard.edu/abs/2009ApJ...696..143P}
}

@ARTICLE{1969JApMe...8..392S,
       author = {{Sellers}, William D.},
        title = "{A Global Climatic Model Based on the Energy Balance of the Earth-Atmosphere System.}",
      journal = {Journal of Applied Meteorology},
         year = 1969,
        month = jun,
       volume = {8},
       number = {3},
        pages = {392-400},
          doi = {10.1175/1520-0450(1969)008<0392:AGCMBO>2.0.CO;2},
       adsurl = {https://ui.adsabs.harvard.edu/abs/1969JApMe...8..392S}
}

@ARTICLE{Arnscheidt2022,
       author = {{Arnscheidt}, Constantin W. and {Rothman}, Daniel H.},
        title = "{Presence or absence of stabilizing Earth system feedbacks on different time scales}",
      journal = {Science Advances},
         year = 2022,
        month = nov,
       volume = {8},
       number = {46},
          eid = {eadc9241},
        pages = {eadc9241},
          doi = {10.1126/sciadv.adc9241},
       adsurl = {https://ui.adsabs.harvard.edu/abs/2022SciA....8C9241A}
}

@ARTICLE{2018PNAS..11510293K,
       author = {{Koll}, Daniel D.~B. and {Cronin}, Timothy W.},
        title = "{Earth's outgoing longwave radiation linear due to H2O greenhouse effect}",
      journal = {Proceedings of the National Academy of Science},
         year = 2018,
        month = oct,
       volume = {115},
       number = {41},
        pages = {10293-10298},
          doi = {10.1073/pnas.1809868115},
       adsurl = {https://ui.adsabs.harvard.edu/abs/2018PNAS..11510293K}
}

@ARTICLE{2018MNRAS.477..727N,
       author = {{Nicholson}, Arwen E. and {Wilkinson}, David M. and {Williams}, Hywel T.~P. and {Lenton}, Timothy M.},
        title = "{Gaian bottlenecks and planetary habitability maintained by evolving model biospheres: the ExoGaia model}",
      journal = {\mnras},
         year = 2018,
        month = jun,
       volume = {477},
       number = {1},
        pages = {727-740},
          doi = {10.1093/mnras/sty658},
archivePrefix = {arXiv},
       eprint = {1803.08063},
 primaryClass = {astro-ph.EP},
       adsurl = {https://ui.adsabs.harvard.edu/abs/2018MNRAS.477..727N}
}

@ARTICLE{2020MNRAS.492.2572A,
       author = {{Alcabes}, Olivia D.~N. and {Olson}, Stephanie and {Abbot}, Dorian S.},
        title = "{Robustness of Gaian feedbacks to climate perturbations}",
      journal = {\mnras},
         year = 2020,
        month = feb,
       volume = {492},
       number = {2},
        pages = {2572-2577},
          doi = {10.1093/mnras/staa055},
archivePrefix = {arXiv},
       eprint = {1906.01112},
 primaryClass = {astro-ph.EP},
       adsurl = {https://ui.adsabs.harvard.edu/abs/2020MNRAS.492.2572A}
}

@ARTICLE{2014BGD....11.8443Z,
       author = {{Zuluaga}, J.~I. and {Salazar}, J.~F. and {Cuartas-Restrepo}, P. and {Poveda}, G.},
        title = "{The Habitable Zone of Inhabited Planets}",
      journal = {Biogeosciences Discussions},
         year = 2014,
        month = jun,
       volume = {11},
       number = {6},
        pages = {8443-8483},
          doi = {10.5194/bgd-11-8443-2014},
archivePrefix = {arXiv},
       eprint = {1405.4576},
 primaryClass = {astro-ph.EP},
       adsurl = {https://ui.adsabs.harvard.edu/abs/2014BGD....11.8443Z}
}

@ARTICLE{2023MNRAS.521..690A,
       author = {{Arthur}, Rudy and {Nicholson}, Arwen},
        title = "{A Gaian habitable zone}",
      journal = {\mnras},
         year = 2023,
        month = may,
       volume = {521},
       number = {1},
        pages = {690-707},
          doi = {10.1093/mnras/stad547},
archivePrefix = {arXiv},
       eprint = {2301.02150},
 primaryClass = {astro-ph.EP},
       adsurl = {https://ui.adsabs.harvard.edu/abs/2023MNRAS.521..690A}
}

@ARTICLE{2023JCli...36..547Z,
       author = {{Zhu}, Fangze and {Rose}, Brian E.~J.},
        title = "{Multiple Equilibria in a Coupled Climate-Carbon Model}",
      journal = {Journal of Climate},
         year = 2023,
        month = jan,
       volume = {36},
       number = {2},
        pages = {547-564},
          doi = {10.1175/JCLI-D-21-0984.1},
       adsurl = {https://ui.adsabs.harvard.edu/abs/2023JCli...36..547Z}
}

@ARTICLE{2020RSPSA.47600303A,
       author = {{Arnscheidt}, Constantin W. and {Rothman}, Daniel H.},
        title = "{Routes to global glaciation}",
      journal = {Proceedings of the Royal Society of London Series A},
         year = 2020,
        month = jul,
       volume = {476},
       number = {2239},
          eid = {20200303},
        pages = {20200303},
          doi = {10.1098/rspa.2020.0303},
       adsurl = {https://ui.adsabs.harvard.edu/abs/2020RSPSA.47600303A}
}

@ARTICLE{2013JCli...26.8289F,
       author = {{Feldl}, Nicole and {Roe}, Gerard H.},
        title = "{The Nonlinear and Nonlocal Nature of Climate Feedbacks}",
      journal = {Journal of Climate},
         year = 2013,
        month = nov,
       volume = {26},
       number = {21},
        pages = {8289-8304},
          doi = {10.1175/JCLI-D-12-00631.1},
       adsurl = {https://ui.adsabs.harvard.edu/abs/2013JCli...26.8289F}
}

@ARTICLE{2004GPC....41...95R,
       author = {{Rial}, J.~A.},
        title = "{Abrupt climate change: chaos and order at orbital and millennial scales}",
      journal = {Global and Planetary Change},
         year = 2004,
        month = apr,
       volume = {41},
       number = {2},
        pages = {95-109},
          doi = {10.1016/j.gloplacha.2003.10.004},
       adsurl = {https://ui.adsabs.harvard.edu/abs/2004GPC....41...95R}
}

@ARTICLE{2015E&PSL.429...20M,
       author = {{Menou}, Kristen},
        title = "{Climate stability of habitable Earth-like planets}",
      journal = {Earth and Planetary Science Letters},
         year = 2015,
        month = nov,
       volume = {429},
        pages = {20-24},
          doi = {10.1016/j.epsl.2015.07.046},
archivePrefix = {arXiv},
       eprint = {1411.5564},
 primaryClass = {astro-ph.EP},
       adsurl = {https://ui.adsabs.harvard.edu/abs/2015E&PSL.429...20M}
}

@ARTICLE{2016ApJ...827..117A,
       author = {{Abbot}, Dorian S.},
        title = "{Analytical Investigation of the Decrease in the Size of the Habitable Zone Due to a Limited CO$_{2}$ Outgassing Rate}",
      journal = {\apj},
         year = 2016,
        month = aug,
       volume = {827},
       number = {2},
          eid = {117},
        pages = {117},
          doi = {10.3847/0004-637X/827/2/117},
archivePrefix = {arXiv},
       eprint = {1606.03030},
 primaryClass = {astro-ph.EP},
       adsurl = {https://ui.adsabs.harvard.edu/abs/2016ApJ...827..117A}
}

@ARTICLE{2024PNAS..12116535K,
       author = {{Kemeny}, Preston Cosslett and {Torres}, Mark A. and {Fischer}, Woodward W. and {Bl{\"a}ttler}, Clara L.},
        title = "{Balance and imbalance in biogeochemical cycles reflect the operation of closed, exchange, and open sets}",
      journal = {Proceedings of the National Academy of Science},
         year = 2024,
        month = mar,
       volume = {121},
       number = {12},
          eid = {e2316535121},
        pages = {e2316535121},
          doi = {10.1073/pnas.2316535121},
       adsurl = {https://ui.adsabs.harvard.edu/abs/2024PNAS..12116535K}
}

@ARTICLE{1987ZPhyB..68..251S,
       author = {{Schuster}, H.~G.},
        title = "{Estimating the strength of chaos from the power spectrum}",
      journal = {Zeitschrift fur Physik B Condensed Matter},
         year = 1987,
        month = jun,
       volume = {68},
       number = {2-3},
        pages = {251-252},
          doi = {10.1007/BF01304235},
       adsurl = {https://ui.adsabs.harvard.edu/abs/1987ZPhyB..68..251S}
}

@ARTICLE{1980NYASA.357..453F,
       author = {{Farmer}, D. and {Crutchfield}, J. and {Frochling}, H. and {Packard}, N. and {Shaw}, R.},
        title = "{Power spectra and mixing properties of strange attractors}",
      journal = {Annals of the New York Academy of Sciences},
         year = 1980,
        month = dec,
       volume = {357},
        pages = {453-472},
          doi = {10.1111/j.1749-6632.1980.tb29710.x},
       adsurl = {https://ui.adsabs.harvard.edu/abs/1980NYASA.357..453F}
}

@ARTICLE{2013A&A...558A..33A,
       author = {{Astropy Collaboration} and {Robitaille}, Thomas P. and {Tollerud}, Erik J. and {Greenfield}, Perry and {Droettboom}, Michael and {Bray}, Erik and {Aldcroft}, Tom and {Davis}, Matt and {Ginsburg}, Adam and {Price-Whelan}, Adrian M. and {Kerzendorf}, Wolfgang E. and {Conley}, Alexander and {Crighton}, Neil and {Barbary}, Kyle and {Muna}, Demitri and {Ferguson}, Henry and {Grollier}, Fr{\'e}d{\'e}ric and {Parikh}, Madhura M. and {Nair}, Prasanth H. and {Unther}, Hans M. and {Deil}, Christoph and {Woillez}, Julien and {Conseil}, Simon and {Kramer}, Roban and {Turner}, James E.~H. and {Singer}, Leo and {Fox}, Ryan and {Weaver}, Benjamin A. and {Zabalza}, Victor and {Edwards}, Zachary I. and {Azalee Bostroem}, K. and {Burke}, D.~J. and {Casey}, Andrew R. and {Crawford}, Steven M. and {Dencheva}, Nadia and {Ely}, Justin and {Jenness}, Tim and {Labrie}, Kathleen and {Lim}, Pey Lian and {Pierfederici}, Francesco and {Pontzen}, Andrew and {Ptak}, Andy and {Refsdal}, Brian and {Servillat}, Mathieu and {Streicher}, Ole},
        title = "{Astropy: A community Python package for astronomy}",
      journal = {\aap},
         year = 2013,
        month = oct,
       volume = {558},
          eid = {A33},
        pages = {A33},
          doi = {10.1051/0004-6361/201322068},
archivePrefix = {arXiv},
       eprint = {1307.6212},
 primaryClass = {astro-ph.IM},
       adsurl = {https://ui.adsabs.harvard.edu/abs/2013A&A...558A..33A}
}

@ARTICLE{2003AsBio...3..331G,
       author = {{Gilichinsky}, D. and {Rivkina}, E. and {Shcherbakova}, V. and {Laurinavichuis}, K. and {Tiedje}, J.},
        title = "{Supercooled Water Brines Within Permafrost-An Unknown Ecological Niche for Microorganisms: A Model for Astrobiology}",
      journal = {Astrobiology},
         year = 2003,
        month = jun,
       volume = {3},
       number = {2},
        pages = {331-341},
          doi = {10.1089/153110703769016424},
       adsurl = {https://ui.adsabs.harvard.edu/abs/2003AsBio...3..331G}
}

@article{doi:10.1128/AEM.70.1.550-557.2004,
author = {Karen Junge and Hajo Eicken and Jody W. Deming},
title = {Bacterial Activity at −2 to −20°C in Arctic Wintertime Sea Ice},
journal = {Applied and Environmental Microbiology},
volume = {70},
number = {1},
pages = {550-557},
year = {2004},
doi = {10.1128/AEM.70.1.550-557.2004},

URL = {https://journals.asm.org/doi/abs/10.1128/aem.70.1.550-557.2004},
eprint = {https://journals.asm.org/doi/pdf/10.1128/aem.70.1.550-557.2004}
}

@INPROCEEDINGS{2008AGUFM.B51F..05T,
       author = {{Takai}, K. and {Nakamura}, K. and {Toki}, T. and {Tsunogai}, U.},
        title = "{Methanogenesis in the hot and deep: implication for the deep biosphere}",
    booktitle = {AGU Fall Meeting Abstracts},
         year = 2008,
       volume = {2008},
        month = dec,
          eid = {B51F-05},
        pages = {B51F-05},
       adsurl = {https://ui.adsabs.harvard.edu/abs/2008AGUFM.B51F..05T}
}

@INPROCEEDINGS{2022SPIE12221E..0CA,
       author = {{Apai}, D{\'a}niel and {Milster}, Tom D. and {Kim}, Daewook and {Kim}, Youngsik and {Schneider}, Glenn and {Rackham}, Benjamin V. and {Arenberg}, Jonathan and {Choi}, Heejoo and {Esparza}, Marcos A. and {Wang}, Zichan and {Zhang}, Yingying and {Bixel}, Alex},
        title = "{Nautilus Space Observatory: a very large aperture space telescope constellation enabled by scalable optical manufacturing technologies}",
    booktitle = {Optical Manufacturing and Testing XIV},
         year = 2022,
       editor = {{Kim}, Daewook and {Choi}, Heejoo and {Ottevaere}, Heidi and {Rascher}, Rolf},
       series = {Society of Photo-Optical Instrumentation Engineers (SPIE) Conference Series},
       volume = {12221},
        month = oct,
          eid = {122210C},
        pages = {122210C},
          doi = {10.1117/12.2633184},
       adsurl = {https://ui.adsabs.harvard.edu/abs/2022SPIE12221E..0CA}
}

@ARTICLE{1976Ap&SS..39..447L,
       author = {{Lomb}, N.~R.},
        title = "{Least-Squares Frequency Analysis of Unequally Spaced Data}",
      journal = {\apss},
         year = 1976,
        month = feb,
       volume = {39},
       number = {2},
        pages = {447-462},
          doi = {10.1007/BF00648343},
       adsurl = {https://ui.adsabs.harvard.edu/abs/1976Ap&SS..39..447L}
}

@ARTICLE{1982ApJ...263..835S,
       author = {{Scargle}, J.~D.},
        title = "{Studies in astronomical time series analysis. II. Statistical aspects of spectral analysis of unevenly spaced data.}",
      journal = {\apj},
         year = 1982,
        month = dec,
       volume = {263},
        pages = {835-853},
          doi = {10.1086/160554},
       adsurl = {https://ui.adsabs.harvard.edu/abs/1982ApJ...263..835S}
}

@ARTICLE{1983PhRvL..50..346G,
       author = {{Grassberger}, Peter and {Procaccia}, Itamar},
        title = "{Characterization of strange attractors}",
      journal = {\prl},
         year = 1983,
        month = jan,
       volume = {50},
       number = {5},
        pages = {346-349},
          doi = {10.1103/PhysRevLett.50.346},
       adsurl = {https://ui.adsabs.harvard.edu/abs/1983PhRvL..50..346G}
}

@ARTICLE{1986PhRvA..34.4971E,
       author = {{Eckmann}, J. -P. and {Oliffson Kamphorst}, S. and {Ruelle}, D. and {Ciliberto}, S.},
        title = "{Liapunov exponents from time series}",
      journal = {\pra},
         year = 1986,
        month = dec,
       volume = {34},
       number = {6},
        pages = {4971-4979},
          doi = {10.1103/PhysRevA.34.4971},
       adsurl = {https://ui.adsabs.harvard.edu/abs/1986PhRvA..34.4971E}
}

@ARTICLE{2007PhR...438..237M,
       author = {{Marwan}, Norbert and {Carmen Romano}, M. and {Thiel}, Marco and {Kurths}, J{\"u}rgen},
        title = "{Recurrence plots for the analysis of complex systems}",
      journal = {\physrep},
         year = 2007,
        month = jan,
       volume = {438},
       number = {5-6},
        pages = {237-329},
          doi = {10.1016/j.physrep.2006.11.001},
archivePrefix = {arXiv},
       eprint = {2501.13933},
 primaryClass = {nlin.CD},
       adsurl = {https://ui.adsabs.harvard.edu/abs/2007PhR...438..237M}
}

@ARTICLE{1981ZNatA..36...80S,
       author = {{Shaw}, Robert},
        title = "{Strange Attractors, Chaotic Begavior, and Information Flow}",
      journal = {Zeitschrift Naturforschung Teil A},
         year = 1981,
        month = jan,
       volume = {36},
       number = {1},
        pages = {80-112},
          doi = {10.1515/zna-1981-0115},
       adsurl = {https://ui.adsabs.harvard.edu/abs/1981ZNatA..36...80S}
}

@ARTICLE{1985PhyD...16..285W,
       author = {{Wolf}, Alan and {Swift}, Jack B. and {Swinney}, Harry L. and {Vastano}, John A.},
        title = "{Determining Lyapunov exponents from a time series}",
      journal = {Physica D Nonlinear Phenomena},
         year = 1985,
        month = jul,
       volume = {16},
       number = {3},
        pages = {285-317},
          doi = {10.1016/0167-2789(85)90011-9},
       adsurl = {https://ui.adsabs.harvard.edu/abs/1985PhyD...16..285W}
}

@ARTICLE{2021ApJ...912L..14W,
       author = {{Wordsworth}, R.},
        title = "{How Likely Are Snowball Episodes Near the Inner Edge of the Habitable Zone?}",
      journal = {\apjl},
         year = 2021,
        month = may,
       volume = {912},
       number = {1},
          eid = {L14},
        pages = {L14},
          doi = {10.3847/2041-8213/abf7c7},
archivePrefix = {arXiv},
       eprint = {2104.06216},
 primaryClass = {astro-ph.EP},
       adsurl = {https://ui.adsabs.harvard.edu/abs/2021ApJ...912L..14W}
}

@ARTICLE{1993Icar..101..108K,
       author = {{Kasting}, James F. and {Whitmire}, Daniel P. and {Reynolds}, Ray T.},
        title = "{Habitable Zones around Main Sequence Stars}",
      journal = {\icarus},
         year = 1993,
        month = jan,
       volume = {101},
       number = {1},
        pages = {108-128},
          doi = {10.1006/icar.1993.1010},
       adsurl = {https://ui.adsabs.harvard.edu/abs/1993Icar..101..108K}
}

@ARTICLE{2018SciA....4.5747K,
       author = {{Krissansen-Totton}, Joshua and {Olson}, Stephanie and {Catling}, David C.},
        title = "{Disequilibrium biosignatures over Earth history and implications for detecting exoplanet life}",
      journal = {Science Advances},
         year = 2018,
        month = jan,
       volume = {4},
       number = {1},
        pages = {eaao5747},
          doi = {10.1126/sciadv.aao5747},
archivePrefix = {arXiv},
       eprint = {1801.08211},
 primaryClass = {astro-ph.EP},
       adsurl = {https://ui.adsabs.harvard.edu/abs/2018SciA....4.5747K}
}

@ARTICLE{2019A&A...628A..12T,
       author = {{Turbet}, Martin and {Ehrenreich}, David and {Lovis}, Christophe and {Bolmont}, Emeline and {Fauchez}, Thomas},
        title = "{The runaway greenhouse radius inflation effect. An observational diagnostic to probe water on Earth-sized planets and test the habitable zone concept}",
      journal = {\aap},
         year = 2019,
        month = aug,
       volume = {628},
          eid = {A12},
        pages = {A12},
          doi = {10.1051/0004-6361/201935585},
archivePrefix = {arXiv},
       eprint = {1906.03527},
 primaryClass = {astro-ph.EP},
       adsurl = {https://ui.adsabs.harvard.edu/abs/2019A&A...628A..12T}
}

@ARTICLE{2019JGRE..124.2087P,
       author = {{Paradise}, Adiv and {Menou}, Kristen and {Valencia}, Diana and {Lee}, Christopher},
        title = "{Habitable Snowballs: Temperate Land Conditions, Liquid Water, and Implications for CO$_{2}$ Weathering}",
      journal = {Journal of Geophysical Research (Planets)},
         year = 2019,
        month = aug,
       volume = {124},
       number = {8},
        pages = {2087-2100},
          doi = {10.1029/2019JE005917},
archivePrefix = {arXiv},
       eprint = {1803.00511},
 primaryClass = {astro-ph.EP},
       adsurl = {https://ui.adsabs.harvard.edu/abs/2019JGRE..124.2087P}
}

@ARTICLE{2011ApJ...733L..48W,
       author = {{Wordsworth}, Robin D. and {Forget}, Fran{\c{c}}ois and {Selsis}, Franck and {Millour}, Ehouarn and {Charnay}, Benjamin and {Madeleine}, Jean-Baptiste},
        title = "{Gliese 581d is the First Discovered Terrestrial-mass Exoplanet in the Habitable Zone}",
      journal = {\apjl},
         year = 2011,
        month = jun,
       volume = {733},
       number = {2},
          eid = {L48},
        pages = {L48},
          doi = {10.1088/2041-8205/733/2/L48},
archivePrefix = {arXiv},
       eprint = {1105.1031},
 primaryClass = {astro-ph.EP},
       adsurl = {https://ui.adsabs.harvard.edu/abs/2011ApJ...733L..48W}
}

@ARTICLE{2015AsBio..15..119L,
       author = {{Luger}, R. and {Barnes}, R.},
        title = "{Extreme Water Loss and Abiotic O2Buildup on Planets Throughout the Habitable Zones of M Dwarfs}",
      journal = {Astrobiology},
         year = 2015,
        month = feb,
       volume = {15},
       number = {2},
        pages = {119-143},
          doi = {10.1089/ast.2014.1231},
archivePrefix = {arXiv},
       eprint = {1411.7412},
 primaryClass = {astro-ph.EP},
       adsurl = {https://ui.adsabs.harvard.edu/abs/2015AsBio..15..119L}
}

@ARTICLE{2017ApJ...843..122Z,
       author = {{Zahnle}, Kevin J. and {Catling}, David C.},
        title = "{The Cosmic Shoreline: The Evidence that Escape Determines which Planets Have Atmospheres, and what this May Mean for Proxima Centauri B}",
      journal = {\apj},
         year = 2017,
        month = jul,
       volume = {843},
       number = {2},
          eid = {122},
        pages = {122},
          doi = {10.3847/1538-4357/aa7846},
archivePrefix = {arXiv},
       eprint = {1702.03386},
 primaryClass = {astro-ph.EP},
       adsurl = {https://ui.adsabs.harvard.edu/abs/2017ApJ...843..122Z}
}

@ARTICLE{2019PhRvL.122o8701L,
       author = {{Lucarini}, Valerio and {B{\'o}dai}, Tam{\'a}s},
        title = "{Transitions across Melancholia States in a Climate Model: Reconciling the Deterministic and Stochastic Points of View}",
      journal = {\prl},
         year = 2019,
        month = apr,
       volume = {122},
       number = {15},
          eid = {158701},
        pages = {158701},
          doi = {10.1103/PhysRevLett.122.158701},
archivePrefix = {arXiv},
       eprint = {1808.05098},
 primaryClass = {physics.ao-ph},
       adsurl = {https://ui.adsabs.harvard.edu/abs/2019PhRvL.122o8701L}
}

@ARTICLE{2017ApJ...848...33P,
       author = {{Paradise}, Adiv and {Menou}, Kristen},
        title = "{GCM Simulations of Unstable Climates in the Habitable Zone}",
      journal = {\apj},
         year = 2017,
        month = oct,
       volume = {848},
       number = {1},
          eid = {33},
        pages = {33},
          doi = {10.3847/1538-4357/aa8b1c},
archivePrefix = {arXiv},
       eprint = {1704.04535},
 primaryClass = {astro-ph.EP},
       adsurl = {https://ui.adsabs.harvard.edu/abs/2017ApJ...848...33P}
}

@ARTICLE{2013ApJ...765..131K,
       author = {{Kopparapu}, Ravi Kumar and {Ramirez}, Ramses and {Kasting}, James F. and {Eymet}, Vincent and {Robinson}, Tyler D. and {Mahadevan}, Suvrath and {Terrien}, Ryan C. and {Domagal-Goldman}, Shawn and {Meadows}, Victoria and {Deshpande}, Rohit},
        title = "{Habitable Zones around Main-sequence Stars: New Estimates}",
      journal = {\apj},
         year = 2013,
        month = mar,
       volume = {765},
       number = {2},
          eid = {131},
        pages = {131},
          doi = {10.1088/0004-637X/765/2/131},
archivePrefix = {arXiv},
       eprint = {1301.6674},
 primaryClass = {astro-ph.EP},
       adsurl = {https://ui.adsabs.harvard.edu/abs/2013ApJ...765..131K}
}

@BOOK{2010ppc..book.....P,
       author = {{Pierrehumbert}, Raymond T.},
        title = "{Principles of Planetary Climate}",
         year = 2010,
       adsurl = {https://ui.adsabs.harvard.edu/abs/2010ppc..book.....P}
}

@ARTICLE{1981JGR....86.9776W,
       author = {{Walker}, J.~C.~G. and {Hays}, P.~B. and {Kasting}, J.~F.},
        title = "{A negative feedback mechanism for the long-term stabilization of the earth's surface temperature}",
      journal = {\jgr},
         year = 1981,
        month = oct,
       volume = {86},
        pages = {9776-9782},
          doi = {10.1029/JC086iC10p09776},
       adsurl = {https://ui.adsabs.harvard.edu/abs/1981JGR....86.9776W}
}

@ARTICLE{1997Icar..129..254W,
       author = {{Williams}, Darren M. and {Kasting}, James F.},
        title = "{Habitable Planets with High Obliquities}",
      journal = {\icarus},
         year = 1997,
        month = sep,
       volume = {129},
       number = {1},
        pages = {254-267},
          doi = {10.1006/icar.1997.5759},
       adsurl = {https://ui.adsabs.harvard.edu/abs/1997Icar..129..254W}
}

@ARTICLE{2000AREPS..28..611K,
       author = {{Kump}, Lee R. and {Brantley}, Susan L. and {Arthur}, Michael A.},
        title = "{Chemical Weathering, Atmospheric CO$_{2}$, and Climate}",
      journal = {Annual Review of Earth and Planetary Sciences},
         year = 2000,
        month = jan,
       volume = {28},
        pages = {611-667},
          doi = {10.1146/annurev.earth.28.1.611},
       adsurl = {https://ui.adsabs.harvard.edu/abs/2000AREPS..28..611K}
}

@ARTICLE{1998Sci...281.1342H,
       author = {{Hoffman}, Paul F. and {Kaufman}, Alan J. and {Halverson}, Galen P. and {Schrag}, Daniel P.},
        title = "{A Neoproterozoic Snowball Earth}",
      journal = {Science},
         year = 1998,
        month = aug,
       volume = {281},
        pages = {1342},
          doi = {10.1126/science.281.5381.1342},
       adsurl = {https://ui.adsabs.harvard.edu/abs/1998Sci...281.1342H}
}

@ARTICLE{2005JGRD..110.1111P,
       author = {{Pierrehumbert}, R.~T.},
        title = "{Climate dynamics of a hard snowball Earth}",
      journal = {Journal of Geophysical Research (Atmospheres)},
         year = 2005,
        month = jan,
       volume = {110},
       number = {D1},
          eid = {D01111},
        pages = {D01111},
          doi = {10.1029/2004JD005162},
       adsurl = {https://ui.adsabs.harvard.edu/abs/2005JGRD..110.1111P}
}

@ARTICLE{1969Tell...21..611B,
       author = {{Budyko}, M.~I.},
        title = "{The effect of solar radiation variations on the climate of the earth.}",
      journal = {Tellus},
         year = 1969,
        month = jan,
       volume = {21},
        pages = {611-619},
          doi = {10.3402/tellusa.v21i5.10109},
       adsurl = {https://ui.adsabs.harvard.edu/abs/1969Tell...21..611B}
}

@ARTICLE{2007EP&S...59..293T,
       author = {{Tajika}, Eiichi},
        title = "{Long-term stability of climate and global glaciations throughout the evolution of the Earth}",
      journal = {Earth, Planets and Space},
         year = 2007,
        month = apr,
       volume = {59},
       number = {4},
        pages = {293-299},
          doi = {10.1186/BF03353107},
       adsurl = {https://ui.adsabs.harvard.edu/abs/2007EP&S...59..293T}
}

@ARTICLE{2011CliPa...7..249V,
       author = {{Voigt}, A. and {Abbot}, D.~S. and {Pierrehumbert}, R.~T. and {Marotzke}, J.},
        title = "{Initiation of a Marinoan Snowball Earth in a state-of-the-art atmosphere-ocean general circulation model}",
      journal = {Climate of the Past},
         year = 2011,
        month = mar,
       volume = {7},
       number = {1},
        pages = {249-263},
          doi = {10.5194/cp-7-249-201110.5194/cpd-6-1853-2010},
       adsurl = {https://ui.adsabs.harvard.edu/abs/2011CliPa...7..249V}
}

@ARTICLE{2010QJRMS.136....2L,
       author = {{Lucarini}, Valerio and {Fraedrich}, Klaus and {Lunkeit}, Frank},
        title = "{Thermodynamic analysis of snowball Earth hysteresis experiment: Efficiency, entropy production and irreversibility}",
      journal = {Quarterly Journal of the Royal Meteorological Society},
         year = 2010,
        month = jan,
       volume = {136},
       number = {646},
        pages = {2-11},
          doi = {10.1002/qj.543},
archivePrefix = {arXiv},
       eprint = {0905.3669},
 primaryClass = {physics.ao-ph},
       adsurl = {https://ui.adsabs.harvard.edu/abs/2010QJRMS.136....2L}
}

@ARTICLE{2016ApJ...827..120H,
       author = {{Haqq-Misra}, Jacob and {Kopparapu}, Ravi Kumar and {Batalha}, Natasha E. and {Harman}, Chester E. and {Kasting}, James F.},
        title = "{Limit Cycles Can Reduce the Width of the Habitable Zone}",
      journal = {\apj},
         year = 2016,
        month = aug,
       volume = {827},
       number = {2},
          eid = {120},
        pages = {120},
          doi = {10.3847/0004-637X/827/2/120},
archivePrefix = {arXiv},
       eprint = {1605.07130},
 primaryClass = {astro-ph.EP},
       adsurl = {https://ui.adsabs.harvard.edu/abs/2016ApJ...827..120H}
}

@ARTICLE{2011ApJ...731...76C,
       author = {{Cowan}, Nicolas B. and {Robinson}, Tyler and {Livengood}, Timothy A. and {Deming}, Drake and {Agol}, Eric and {A'Hearn}, Michael F. and {Charbonneau}, David and {Lisse}, Carey M. and {Meadows}, Victoria S. and {Seager}, Sara and {Shields}, Aomawa L. and {Wellnitz}, Dennis D.},
        title = "{Rotational Variability of Earth's Polar Regions: Implications for Detecting Snowball Planets}",
      journal = {\apj},
         year = 2011,
        month = apr,
       volume = {731},
       number = {1},
          eid = {76},
        pages = {76},
          doi = {10.1088/0004-637X/731/1/76},
archivePrefix = {arXiv},
       eprint = {1102.4345},
 primaryClass = {astro-ph.EP},
       adsurl = {https://ui-adsabs-harvard-edu.ezproxy1.library.arizona.edu/abs/2011ApJ...731...76C}
}

@ARTICLE{2018AJ....155..230G,
       author = {{Guimond}, Claire Marie and {Cowan}, Nicolas B.},
        title = "{The Direct Imaging Search for Earth 2.0: Quantifying Biases and Planetary False Positives}",
      journal = {\aj},
         year = 2018,
        month = jun,
       volume = {155},
       number = {6},
          eid = {230},
        pages = {230},
          doi = {10.3847/1538-3881/aabb02},
archivePrefix = {arXiv},
       eprint = {1804.00699},
 primaryClass = {astro-ph.EP},
       adsurl = {https://ui-adsabs-harvard-edu.ezproxy1.library.arizona.edu/abs/2018AJ....155..230G}
}

@ARTICLE{2017ApJ...841L..24B,
       author = {{Bean}, Jacob L. and {Abbot}, Dorian S. and {Kempton}, Eliza M. -R.},
        title = "{A Statistical Comparative Planetology Approach to the Hunt for Habitable Exoplanets and Life Beyond the Solar System}",
      journal = {\apjl},
         year = 2017,
        month = jun,
       volume = {841},
       number = {2},
          eid = {L24},
        pages = {L24},
          doi = {10.3847/2041-8213/aa738a},
archivePrefix = {arXiv},
       eprint = {1705.06288},
 primaryClass = {astro-ph.EP},
       adsurl = {https://ui-adsabs-harvard-edu.ezproxy1.library.arizona.edu/abs/2017ApJ...841L..24B}
}

@ARTICLE{1997BAMS...78..197K,
       author = {{Kiehl}, J.~T. and {Trenberth}, Kevin E.},
        title = "{Earth's Annual Global Mean Energy Budget.}",
      journal = {Bulletin of the American Meteorological Society},
         year = 1997,
        month = feb,
       volume = {78},
       number = {2},
        pages = {197-197},
          doi = {10.1175/1520-0477(1997)078<0197:EAGMEB>2.0.CO;2},
       adsurl = {https://ui.adsabs.harvard.edu/abs/1997BAMS...78..197K}
}

@ARTICLE{2006JCli...19.3354S,
       author = {{Soden}, Brian J. and {Held}, Isaac M.},
        title = "{An Assessment of Climate Feedbacks in Coupled Ocean Atmosphere Models}",
      journal = {Journal of Climate},
         year = 2006,
        month = jan,
       volume = {19},
       number = {14},
        pages = {3354},
          doi = {10.1175/JCLI3799.1},
       adsurl = {https://ui.adsabs.harvard.edu/abs/2006JCli...19.3354S}
}

@ARTICLE{2022AJ....164..190B,
       author = {{Bergsten}, Galen J. and {Pascucci}, Ilaria and {Mulders}, Gijs D. and {Fernandes}, Rachel B. and {Koskinen}, Tommi T.},
        title = "{The Demographics of Kepler's Earths and Super-Earths into the Habitable Zone}",
      journal = {Astron.},
         year = 2022,
        month = nov,
       volume = {164},
       number = {5},
          eid = {190},
        pages = {190},
          doi = {10.3847/1538-3881/ac8fea},
archivePrefix = {arXiv},
       eprint = {2209.04047},
 primaryClass = {astro-ph.EP},
       adsurl = {https://ui.adsabs.harvard.edu/abs/2022AJ....164..190B}
}

@ARTICLE{Tuchow2024,
       author = {{Tuchow}, Noah W. and {Stark}, Christopher C. and {Mamajek}, Eric},
        title = "{HPIC: The Habitable Worlds Observatory Preliminary Input Catalog}",
      journal = {Astron. J.},
         year = 2024,
        month = mar,
       volume = {167},
       number = {3},
          eid = {139},
        pages = {139},
          doi = {10.3847/1538-3881/ad25ec},
       adsurl = {https://ui.adsabs.harvard.edu/abs/2024AJ....167..139T}
}

@ARTICLE{Quanz2022,
       author = {{Quanz}, S.~P. and {Ottiger}, M. and {Fontanet}, E. and {Kammerer}, J. and {Menti}, F. and {Dannert}, F. and {Gheorghe}, A. and {Absil}, O. and {Airapetian}, V.~S. and {Alei}, E. and {Allart}, R. and {Angerhausen}, D. and {Blumenthal}, S. and {Buchhave}, L.~A. and {Cabrera}, J. and {Carri{\'o}n-Gonz{\'a}lez}, {\'O}. and {Chauvin}, G. and {Danchi}, W.~C. and {Dandumont}, C. and {Defr{\'e}re}, D. and {Dorn}, C. and {Ehrenreich}, D. and {Ertel}, S. and {Fridlund}, M. and {Garc{\'\i}a Mu{\~n}oz}, A. and {Gasc{\'o}n}, C. and {Girard}, J.~H. and {Glauser}, A. and {Grenfell}, J.~L. and {Guidi}, G. and {Hagelberg}, J. and {Helled}, R. and {Ireland}, M.~J. and {Janson}, M. and {Kopparapu}, R.~K. and {Korth}, J. and {Kozakis}, T. and {Kraus}, S. and {L{\'e}ger}, A. and {Leedj{\"a}rv}, L. and {Lichtenberg}, T. and {Lillo-Box}, J. and {Linz}, H. and {Liseau}, R. and {Loicq}, J. and {Mahendra}, V. and {Malbet}, F. and {Mathew}, J. and {Mennesson}, B. and {Meyer}, M.~R. and {Mishra}, L. and {Molaverdikhani}, K. and {Noack}, L. and {Oza}, A.~V. and {Pall{\'e}}, E. and {Parviainen}, H. and {Quirrenbach}, A. and {Rauer}, H. and {Ribas}, I. and {Rice}, M. and {Romagnolo}, A. and {Rugheimer}, S. and {Schwieterman}, E.~W. and {Serabyn}, E. and {Sharma}, S. and {Stassun}, K.~G. and {Szul{\'a}gyi}, J. and {Wang}, H.~S. and {Wunderlich}, F. and {Wyatt}, M.~C. and {LIFE Collaboration}},
        title = "{Large Interferometer For Exoplanets (LIFE). I. Improved exoplanet detection yield estimates for a large mid-infrared space-interferometer mission}",
      journal = {\aap},
         year = 2022,
        month = aug,
       volume = {664},
          eid = {A21},
        pages = {A21},
          doi = {10.1051/0004-6361/202140366},
archivePrefix = {arXiv},
       eprint = {2101.07500},
 primaryClass = {astro-ph.EP},
       adsurl = {https://ui.adsabs.harvard.edu/abs/2022A&A...664A..21Q}
}

@ARTICLE{Tyrrell2020,
       author = {{Tyrrell}, Toby},
        title = "{Chance played a role in determining whether Earth stayed habitable}",
      journal = {Communications Earth and Environment},
         year = 2020,
        month = dec,
       volume = {1},
       number = {1},
          eid = {61},
        pages = {61},
          doi = {10.1038/s43247-020-00057-8},
       adsurl = {https://ui.adsabs.harvard.edu/abs/2020ComEE...1...61T}
}

@ARTICLE{Foley2016,
       author = {{Foley}, Bradford J. and {Driscoll}, Peter E.},
        title = "{Whole planet coupling between climate, mantle, and core: Implications for rocky planet evolution}",
      journal = {Geochemistry, Geophysics, Geosystems},
         year = 2016,
        month = may,
       volume = {17},
       number = {5},
        pages = {1885-1914},
          doi = {10.1002/2015GC006210},
archivePrefix = {arXiv},
       eprint = {1711.06801},
 primaryClass = {astro-ph.EP},
       adsurl = {https://ui.adsabs.harvard.edu/abs/2016GGG....17.1885F}
}

@ARTICLE{Muller2024,
       author = {{M{\"u}ller}, R. Dietmar and {Dutkiewicz}, Adriana and {Zahirovic}, Sabin and {Merdith}, Andrew S. and {Scotese}, Christopher R. and {Mills}, Benjamin J.~W. and {Ilano}, Lauren and {Mather}, Ben},
        title = "{Solid Earth Carbon Degassing and Sequestration Since 1 Billion Years Ago}",
      journal = {Geochemistry, Geophysics, Geosystems},
         year = 2024,
        month = nov,
       volume = {25},
       number = {11},
        pages = {2024GC011713},
          doi = {10.1029/2024GC011713},
       adsurl = {https://ui.adsabs.harvard.edu/abs/2024GGG....2511713M}
}

@BOOK{Ward2009,
       author = {{Ward}, Peter},
        title = "{The Medea Hypothesis. Is Life on Earth Ultimately Self-Destructive?}",
         year = 2009,
       adsurl = {https://ui.adsabs.harvard.edu/abs/2009mhil.book.....W}
}

@ARTICLE{Kasting1987,
       author = {{Kasting}, James F.},
        title = "{Theoretical constraints on oxygen and carbon dioxide concentrations in the Precambrian atmosphere}",
      journal = {Precambrian Research},
         year = 1987,
        month = jan,
       volume = {34},
       number = {3-4},
        pages = {205-229},
          doi = {10.1016/0301-9268(87)90001-5},
       adsurl = {https://ui.adsabs.harvard.edu/abs/1987PreR...34..205K}
}

@BOOK{NationalAcademiesofSciences2021,
       author = {{National Academies of Sciences, Engineering, and Medicine}},
        title = "{Pathways to Discovery in Astronomy and Astrophysics for the 2020s}",
         year = 2021,
          doi = {10.17226/26141},
       adsurl = {https://ui.adsabs.harvard.edu/abs/2021pdaa.book.....N}
}

@INPROCEEDINGS{Bender2022,
       author = {{Bender}, Chad F. and {Angel}, J. Roger and {Berkson}, Joel and {Gray}, Peter and {Halverson}, Samuel and {Kang}, Hyukmo and {Kim}, Daewood and {Monson}, Andy and {Oh}, Chang Jin and {Rademacher}, Matthew and {Schwab}, Christian and {Young}, Andrew and {Zaritsky}, Dennis},
        title = "{The Large Fiber Array Spectroscopic Telescope: fiber feed and spectrometer conceptual design}",
    booktitle = {Ground-based and Airborne Instrumentation for Astronomy IX},
         year = 2022,
       editor = {{Evans}, Christopher J. and {Bryant}, Julia J. and {Motohara}, Kentaro},
       series = {Society of Photo-Optical Instrumentation Engineers (SPIE) Conference Series},
       volume = {12184},
        month = aug,
          eid = {121844J},
        pages = {121844J},
          doi = {10.1117/12.2628710},
       adsurl = {https://ui.adsabs.harvard.edu/abs/2022SPIE12184E..4JB}
}

@ARTICLE{Lucarini2013,
       author = {{Lucarini}, V. and {Pascale}, S. and {Boschi}, R. and {Kirk}, E. and {Iro}, N.},
        title = "{Habitability and Multistability in Earth-like Planets}",
      journal = {Astronomische Nachrichten},
         year = 2013,
        month = jun,
       volume = {334},
       number = {6},
        pages = {576},
          doi = {10.1002/asna.201311903},
archivePrefix = {arXiv},
       eprint = {1303.5937},
 primaryClass = {astro-ph.EP},
       adsurl = {https://ui.adsabs.harvard.edu/abs/2013AN....334..576L}
}

@ARTICLE{Linsenmeier2015,
       author = {{Linsenmeier}, Manuel and {Pascale}, Salvatore and {Lucarini}, Valerio},
        title = "{Climate of Earth-like planets with high obliquity and eccentric orbits: Implications for habitability conditions}",
      journal = {\planss},
         year = 2015,
        month = jan,
       volume = {105},
        pages = {43-59},
          doi = {10.1016/j.pss.2014.11.003},
archivePrefix = {arXiv},
       eprint = {1401.5323},
 primaryClass = {astro-ph.EP},
       adsurl = {https://ui.adsabs.harvard.edu/abs/2015P&SS..105...43L}
}

@ARTICLE{Checlair2017,
       author = {{Checlair}, Jade and {Menou}, Kristen and {Abbot}, Dorian S.},
        title = "{No Snowball on Habitable Tidally Locked Planets}",
      journal = {\apj},
         year = 2017,
        month = aug,
       volume = {845},
       number = {2},
          eid = {132},
        pages = {132},
          doi = {10.3847/1538-4357/aa80e1},
archivePrefix = {arXiv},
       eprint = {1705.08904},
 primaryClass = {astro-ph.EP},
       adsurl = {https://ui.adsabs.harvard.edu/abs/2017ApJ...845..132C}
}

@ARTICLE{Checlair2019,
       author = {{Checlair}, Jade H. and {Salazar}, Andrea M. and {Paradise}, Adiv and {Menou}, Kristen and {Abbot}, Dorian S.},
        title = "{No Snowball Cycles at the Outer Edge of the Habitable Zone for Habitable Tidally Locked Planets}",
      journal = {\apjl},
         year = 2019,
        month = dec,
       volume = {887},
       number = {1},
          eid = {L3},
        pages = {L3},
          doi = {10.3847/2041-8213/ab5957},
       adsurl = {https://ui.adsabs.harvard.edu/abs/2019ApJ...887L...3C}
}

@incollection{Forster2021,
  author       = {Forster, P. and Storelvmo, T. and Armour, K. and Collins, W. and Dufresne, J.-L. and Frame, D. and Lunt, D. J. and Mauritsen, T. and Palmer, M. D. and Watanabe, M. and Wild, M. and Zhang, H.},
  title        = {The Earth's Energy Budget, Climate Feedbacks, and Climate Sensitivity},
  booktitle    = {Climate Change 2021: The Physical Science Basis. Contribution of Working Group I to the Sixth Assessment Report of the Intergovernmental Panel on Climate Change},
  editor       = {Masson-Delmotte, V. and Zhai, P. and Pirani, A. and Connors, S. L. and P{\'e}an, C. and Berger, S. and Caud, N. and Chen, Y. and Goldfarb, L. and Gomis, M. I. and Huang, M. and Leitzell, K. and Lonnoy, E. and Matthews, J. B. R. and Maycock, T. K. and Waterfield, T. and Yelek{\c c}i, O. and Yu, R. and Zhou, B.},
  publisher    = {Cambridge University Press},
  address      = {Cambridge, United Kingdom and New York, NY, USA},
  year         = {2021},
  pages        = {923--1054},
  doi          = {10.1017/9781009157896.009}
}

@ARTICLE{Underwood2025,
       author = {{Underwood}, Morgan and {Lenardic}, Adrian and {Seales}, Johnny and {Kwait-Gonchar}, Benjamin},
        title = "{Observational Tests of Terrestrial Planet Buffering Feedbacks and the Habitable Zone Concept}",
      journal = {arXiv e-prints},
         year = 2025,
        month = sep,
          eid = {arXiv:2509.02848},
        pages = {arXiv:2509.02848},
          doi = {10.48550/arXiv.2509.02848},
archivePrefix = {arXiv},
       eprint = {2509.02848},
 primaryClass = {astro-ph.EP},
       adsurl = {https://ui.adsabs.harvard.edu/abs/2025arXiv250902848U}
}

@ARTICLE{North1983,
       author = {{North}, G.~R. and {Short}, D.~A. and {Mengel}, J.~G.},
        title = "{Simple energy balance model resolving the seasons and the continents - Application to the astronomical theory of the ice ages}",
      journal = {\jgr},
         year = 1983,
        month = aug,
       volume = {88},
        pages = {6576-6586},
          doi = {10.1029/JC088iC11p06576},
       adsurl = {https://ui.adsabs.harvard.edu/abs/1983JGR....88.6576N}
}

@ARTICLE{Mills2011,
       author = {{Mills}, Benjamin and {Watson}, Andrew J. and {Goldblatt}, Colin and {Boyle}, Richard and {Lenton}, Timothy M.},
        title = "{Timing of Neoproterozoic glaciations linked to transport-limited global weathering}",
      journal = {Nature Geoscience},
         year = 2011,
        month = dec,
       volume = {4},
       number = {12},
        pages = {861-864},
          doi = {10.1038/ngeo1305},
       adsurl = {https://ui.adsabs.harvard.edu/abs/2011NatGe...4..861M}
}

@ARTICLE{Krissansen-Totton2017,
       author = {{Krissansen-Totton}, Joshua and {Catling}, David C.},
        title = "{Constraining climate sensitivity and continental versus seafloor weathering using an inverse geological carbon cycle model}",
      journal = {Nature Communications},
         year = 2017,
        month = may,
       volume = {8},
          eid = {15423},
        pages = {15423},
          doi = {10.1038/ncomms15423},
       adsurl = {https://ui.adsabs.harvard.edu/abs/2017NatCo...815423K}
}

@book{Lenton2011,
  author    = {Timothy M. Lenton and Andrew J. Watson},
  title     = {Revolutions that Made the Earth},
  publisher = {Oxford University Press},
  year      = {2011},
  isbn      = {978-0-19-958704-9},
  pages     = {423},
  note      = {Illustrated edition}
}

@ARTICLE{Haqq-Misra2022,
       author = {{Haqq-Misra}, Jacob and {Wolf}, Eric T. and {Fauchez}, Thomas J. and {Shields}, Aomawa L. and {Kopparapu}, Ravi K.},
        title = "{The Sparse Atmospheric Model Sampling Analysis (SAMOSA) Intercomparison: Motivations and Protocol Version 1.0: A CUISINES Model Intercomparison Project}",
      journal = {\psj},
         year = 2022,
        month = nov,
       volume = {3},
       number = {11},
          eid = {260},
        pages = {260},
          doi = {10.3847/PSJ/ac9479},
archivePrefix = {arXiv},
       eprint = {2209.10480},
 primaryClass = {astro-ph.EP},
       adsurl = {https://ui.adsabs.harvard.edu/abs/2022PSJ.....3..260H}
}

@ARTICLE{Berner1991,
  author  = {Berner, Robert A.},
  title   = {A model for atmospheric {CO2} over Phanerozoic time},
  journal = {American Journal of Science},
  year    = {1991},
  volume  = {291},
  pages   = {339--376},
  doi     = {10.2475/ajs.291.4.339}
}

@ARTICLE{Berner2001,
  author  = {Berner, Robert A. and Kothavala, Zofia},
  title   = {GEOCARB III: A Revised Model of Atmospheric {CO2} over Phanerozoic Time},
  journal = {American Journal of Science},
  year    = {2001},
  volume  = {301},
  pages   = {182--204},
  doi     = {10.2475/ajs.301.2.182}
}

@ARTICLE{Sleep2001,
  author  = {Sleep, Norman H. and Zahnle, Kevin},
  title   = {Carbon dioxide cycling and implications for climate},
  journal = {Journal of Geophysical Research},
  year    = {2001},
  volume  = {106},
  number  = {E1},
  pages   = {1373--1399},
  doi     = {10.1029/2000JE001247}
}

@ARTICLE{Lenton2008,
  author  = {Lenton, Timothy M. and others},
  title   = {Tipping elements in the Earth’s climate system},
  journal = {Proceedings of the National Academy of Sciences},
  year    = {2008},
  volume  = {105},
  pages   = {1786--1793},
  doi     = {10.1073/pnas.0705414105}
}

@ARTICLE{Roe2006,
  author  = {Roe, Gerard H.},
  title   = {In defense of Milankovitch},
  journal = {Science},
  year    = {2006},
  volume  = {314},
  pages   = {793--794},
  doi     = {10.1126/science.1133783}
}

@INPROCEEDINGS{Kirschvink1992,
  author    = {Kirschvink, Joseph L.},
  booktitle = {The Proterozoic Biosphere},
  title     = {Late Proterozoic low-latitude global glaciation: the Snowball Earth},
  editor    = {Schopf, J. William and Klein, Cornelis},
  publisher = {Cambridge University Press},
  year      = {1992},
  pages     = {51--52}
}

@ARTICLE{Hyde2000,
  author  = {Hyde, William T. and others},
  title   = {Neoproterozoic Snowball Earth simulations with a fully coupled climate model},
  journal = {Nature},
  year    = {2000},
  volume  = {405},
  pages   = {425--429},
  doi     = {10.1038/35013031}
}

@ARTICLE{Evans2000,
  author  = {Evans, David A. D.},
  title   = {The palaeomagnetically viable latitudes of Snowball Earth glaciation},
  journal = {Nature},
  year    = {2000},
  volume  = {405},
  pages   = {490--493},
  doi     = {10.1038/35013030}
}

@ARTICLE{Westerhold2020,
       author = {{Westerhold}, Thomas and {Marwan}, Norbert and {Drury}, Anna Joy and {Liebrand}, Diederik and {Agnini}, Claudia and {Anagnostou}, Eleni and {Barnet}, James S.~K. and {Bohaty}, Steven M. and {De Vleeschouwer}, David and {Florindo}, Fabio and {Frederichs}, Thomas and {Hodell}, David A. and {Holbourn}, Ann E. and {Kroon}, Dick and {Lauretano}, Vittoria and {Littler}, Kate and {Lourens}, Lucas J. and {Lyle}, Mitchell and {P{\"a}like}, Heiko and {R{\"o}hl}, Ursula and {Tian}, Jun and {Wilkens}, Roy H. and {Wilson}, Paul A. and {Zachos}, James C.},
        title = "{An astronomically dated record of Earth{\textquoteright}s climate and its predictability over the last 66 million years}",
      journal = {Science},
         year = 2020,
        month = sep,
       volume = {369},
       number = {6509},
        pages = {1383-1387},
          doi = {10.1126/science.aba6853},
       adsurl = {https://ui.adsabs.harvard.edu/abs/2020Sci...369.1383W}
}
\bibliographystyle{aasjournal}

\end{document}